\documentclass{aa}  

\usepackage{graphicx}
\usepackage{txfonts}

\bibpunct{(}{)}{;}{a}{}{,}
\usepackage{natbib,subcaption}
\usepackage{xcolor}
\usepackage{gensymb}

\begin{document}

   \title{Precise dynamical masses of new directly imaged companions from combining relative astrometry, radial velocities, and Hipparcos-Gaia eDR3 accelerations \thanks{Based on observations collected with SPHERE mounted on the VLT at Paranal Observatory (ESO, Chile) under programs 0102.C-0236(A) (PI: Rickman), 0104.C-0724(A) (PI: Rickman), and 105.20SZ.001 (PI: Rickman) as well as observations collected with the CORALIE spectrograph mounted on the 1.2~m Swiss telescope at La Silla Observatory and with the HARPS spectrograph on the ESO 3.6~m telescope at La Silla (ESO, Chile).} \thanks{The radial-velocity measurements, reduced images, and additional data products discussed in this paper are available on the DACE web platform at \url{https://dace.unige.ch/} and the links to individual targets are listed in Appendix~\ref{appendix2}.}}

   \author{E. L. Rickman \inst{1,2} \thanks{European Space Agency Research Fellow}, E. Matthews \inst{2}, W. Ceva \inst{2}, D. S\'{e}gransan \inst{2}, G. M. Brandt \inst{3}, H. Zhang \inst{3}, T. D. Brandt \inst{3}, T. Forveille \inst{4}, J. Hagelberg \inst{2}, S. Udry \inst{2}}

   \institute{European Space Agency (ESA), ESA Office, Space Telescope Science Institute, 3700 San Martin Drive, Baltimore 21218, MD, USA \\ \email{erickman@stsci.edu} 
   \and
   D\'{e}partment d'astronomie de l’Universit\'{e} de Gen\`{e}ve, Chemin Pegasi 51, 1290 Versoix, Switzerland
   \and
   Department of Physics, University of California, Santa Barbara, Santa Barbara, CA 93106, USA
   \and
   Univ. Grenoble Alpes, CNRS, IPAG, 38000 Grenoble, France
   }

    \date{Received; accepted }
    \authorrunning{Rickman et al.}
    \titlerunning{Precise dynamical masses of new directly imaged companions}

% \abstract{}{}{}{}{} 
% 5 {} token are mandatory
 
  \abstract
   {}
   {With an observing time span of more than 20 years, the CORALIE radial-velocity survey is able to detect long-term trends in data corresponding to companions with masses and separations accessible to direct imaging. Combining exoplanet detection techniques, such as  radial velocities from the CORALIE survey, astrometric accelerations from \emph{Hipparcos} and \emph{Gaia} eDR3, and relative astrometry from direct imaging, removes the degeneracy of unknown orbital parameters. This allows precise model-independent masses of detected companions to be derived, which provides a powerful tool to test models of stellar and substellar mass-luminosity relations.}
% methods
{Long-term precise Doppler measurements with the CORALIE spectrograph reveal radial-velocity signatures of companions on long-period orbits. The long baseline of radial-velocity data allows the detectability of the companion candidates to be assessed with direct imaging. We combine long-period radial-velocity data with absolute astrometry from \emph{Hipparcos} and \emph{Gaia} eDR3 and relative astrometry derived from new direct imaging detections with VLT/SPHERE to fit orbital parameters and derive precise dynamical masses of these companions.}
   {In this paper we report the discovery of new companions orbiting HD~142234, HD~143616, and HIP~22059, as well as the first direct detection of HD~92987~B, and update the dynamical masses of two previously directly imaged companions: HD~157338~B and HD~195010~B. The companions span a period range of 32 to 279 years and are all very low-mass stellar companions, ranging from 218 to 487~$M_{\mathrm{Jup}}$. 
   We compare the derived dynamical masses to mass-luminosity relations of very low-mass stars ($<0.5~M_{\odot}$), and discuss the importance of using precursor radial-velocity and astrometric information to inform the future of high-contrast imaging of exoplanets and brown dwarfs.}
   {}


   \maketitle
%
%-------------------------------------------------------------------

\section{Introduction}

Dynamical masses of companions to main-sequence stars provide a powerful tool for testing mass-luminosity relations and  evolutionary models of stellar and substellar objects. They can act as benchmark objects for models that are used to infer masses and radii of isolated objects using their luminosities alone. While there are a number of dynamical masses in the literature for solar mass and higher stars, the known dynamical masses of very low-mass stars ($<0.5~M_{\odot}$) with individual mass components are still limited (e.g., \citealt{2022A&A...658A.145B}). In recent years updated mass-luminosity relationships have been constructed \citep{10.1093/mnras/stu1605,2016AJ....152..141B,2019ApJ...871...63M} that build upon earlier work \citep[e.g.,][]{2000A&A...364..217D}. We  note the importance of continuing to place observational constraints on these relations in order to constrain them.

A powerful, yet underused, method to determine precise dynamical masses of individual components in stellar binary systems is by combining different detection and measurement techniques that are typical of exoplanet and brown dwarf searches. Radial velocities provide constraints on the minimum mass ($m\sin i$), with an unknown orbital inclination ($i$), of a companion orbiting a host star, as well as the orbital period.   Direct imaging provides the  astrometry of a companion relative to a host star.  Proper motions from the combination of \emph{Gaia} \citep{2016A&A...595A...1G} and \emph{Hipparcos} \citep{1997ESASP1200.....E} place additional absolute astrometric constraints on the orbit of the system.

Using one of these techniques alone is not sufficient to reveal the dynamical masses of the individual components of a stellar system. However, long-period radial-velocity data, like those from the CORALIE spectrograph \citep{2000fepc.conf..571U} on the 1.2~m Swiss/EULER telescope and from HARPS \citep{2003Msngr.114...20M} on the ESO 3.6~m telescope, both of which are situated at La Silla Observatory, Chile, can be combined  with relative astrometry from direct imaging with instruments like VLT/SPHERE \citep{2019A&A...631A.155B} and astrometric accelerations \citep[e.g.,][]{2021ApJS..254...42B,2022A&A...657A...7K}. This removes the degeneracy of orbital parameters, such as the orbital inclination of the observed system, and reveals  model-independent masses. In  this way  we are able to detect and characterize a number of low-mass companions to bright primary stars, which can serve as benchmark objects to test mass-luminosity relations, as well as as stellar and substellar evolutionary models. It is only recently that the power of combining direct and indirect techniques has been utilized to derive the  precise dynamical masses of stellar and substellar objects, as well as exoplanets \citep[e.g.,][]{2011A&A...525A..95S,2012ApJ...761...39C,2017ApJS..231...15D,2018AJ....155..159B,2018A&A...614A..16C,2019AJ....158..140B,2020A&A...639A..47M,2020A&A...635A.203R,2021AJ....162..301B}.

We use radial-velocity data taken from the  CORALIE survey as part of the ongoing survey for extra-solar planets \citep{2000fepc.conf..571U} in the southern hemisphere since 1999. This survey  contains a volume-limited sample of  solar-type stars within 50~pc. With more than 20~years  of radial-velocity data the CORALIE survey provides a wealth of data for understanding the occurrence rate of planetary, brown dwarf, and stellar companions in the solar neighborhood. From the CORALIE survey, we selected targets that present long-term linear or quadratic drifts in their radial-velocity data, that indicate massive companion candidates amenable to direct imaging observations, and that could potentially be masquerading as exoplanets or brown dwarfs.

We directly imaged selected promising candidates using VLT/SPHERE to reveal their true masses and therefore companion nature. In order to further refine derived orbital parameters and precise dynamical masses,  we   combined astrometric information from long baseline proper motion anomalies from \emph{Hipparcos} \citep{1997ESASP1200.....E,2007A&A...474..653V,Hipparcos..JAVA} and \emph{Gaia} eDR3 \citep{2021A&A...649A...1G} from the Hipparcos-Gaia catalog  of accelerations \citep[HGCA;][]{2021ApJS..254...42B} to provide additional constraints on fitted orbital parameters. In order to determine the  precise orbital parameters we use the orbital fitting code \texttt{orvara} \citep{2021AJ....162..186B}, which is an efficient code written specifically for the purpose of combining radial-velocity measurements, absolute astrometry, and relative astrometry.

In this paper we report the discovery of new very low-mass stellar companions orbiting HD~142234, HD~143616, and HIP~22059. We also present the first direct detection of HD~92987~B, which was previously detected via radial velocities \citep{2019A&A...625A..71R}, and update the orbital parameters and dynamical masses of two already known very low-mass stellar companions: HD~157338 and HD~195010, both previously imaged with VLT/NACO by \citet{montagnier:tel-00714874}. For each of these systems we present precise dynamical mass determinations combining radial velocities, relative astrometry from direct imaging, and Hipparcos-Gaia eDR3 accelerations.

The paper is  organized as follows. The properties of the host stars are summarized in Sect. 2. In Sect. 3 we present our RV and direct imaging observations and data reduction. In Sect. 4 we discuss the process of determining the astrometry and photometry from the high-contrast images. In Sect. 5 we present the orbit fitting and solution of each of the companions. In Sect. 6 we make a comparison of our derived dynamical masses with low-mass stellar $M_H$-mass relations. The results are discussed in Sect. 7, and some concluding remarks are presented.

\section{Stellar characteristics of the primary stars} \label{sec:stellar_params}

For each of the systems, the primary star spectral types are taken from the \emph{Hipparcos} catalog \citep{1997A&A...323L..49P}, while the $V_T$ band magnitudes and color indices are taken from the Tycho-2 catalog \citep{2000A&A...355L..27H}. The astrometric parallaxes ($\pi$) are taken from the \emph{Gaia} early data release 3 \citep[eDR3;][]{2021A&A...649A...1G}, and the luminosities ($L$) and effective temperatures ($T_{\rm{eff}}$) are taken from the second \emph{Gaia} data release \citep{2018A&A...616A...1G}.

The $v\sin(i)$ for each primary star is computed using the calibration of the width of CORALIE’s cross-correlation function \citep[CCF;][]{2001A&A...373.1019S,marmier_phd_thesis}. The values for the stellar surface gravities ($\log g$) and metallicities ([Fe/H]) are gathered from literature values across several catalogs \citep{2011A&A...530A.138C,2018A&A...615A..76S,2021A&A...656A..53S}.

Isochronal masses, ages, and radii of the primary stars, as well as their uncertainties, are derived using the Geneva stellar-evolution models \citep{2012A&A...537A.146E,2013A&A...558A.103G}. The interpolation in the model grid was made through a Bayesian formalism using
observational Gaussian priors on $T_{\mathrm{eff}}$, $M_V$, and [Fe/H] following the procedure outlined in \citet{marmier_phd_thesis}. The stellar parameters for each primary star are summarized in Table~\ref{tab:stellar_params}.

\begin{table*}
\begin{center}
\caption{Observed and inferred stellar parameters for host stars  HD~92987, HD~142234, HD~143616, HD~157338, HD~195010, and HIP~22059.}
\label{tab:stellar_params}
% used for centering table
\begin{tabular}{c c c c c c c c c}        % centered columns (4 columns)
\hline\hline                    % inserts double horizontal lines
Parameters & Units & HD~92987 & HD~142234 & HD~143616 & HD~157338 & HD~195010 & HIP~22059 \\    % table heading
\hline    
Sp. Type\tablefoottext{a} & & G2/3V & G5V & G6/8V & G0/G1V & G8/K0V & K5V \\
$V_T$\tablefoottext{b} & & 7.10 & 8.62 & 8.34 & 6.99 & 8.91 & 9.69 \\
$B_T-V_T$\tablefoottext{b} & & 0.709 & 0.687 & 0.809 & 0.633 & 0.898 & 1.251 \\ 
$\pi$\tablefoottext{c} & $[\text{mas}]$ & $22.98\pm0.05$ & $21.00\pm0.03$ & $23.81\pm0.03$ & $30.26\pm0.04$ & $19.84\pm0.02$ & $32.38\pm0.02$ \\
$L$\tablefoottext{d} & $[\text{L}_{\odot}]$ & $2.55\pm0.006$ & $0.76\pm0.003$ & $0.77\pm0.002$ & $1.58\pm0.004$ & $0.67\pm0.002$ & $0.19\pm0.0003$ \\
$T_{\text{eff}}$\tablefoottext{d} & $[\text{K}]$ & $5808^{+56}_{-80}$ & $5592^{+213}_{-192}$ & $5530^{+34}_{-47}$ & $6055^{+63}_{-54}$ & $5352^{+84}_{-74}$ & $4699^{+165}_{-97}$ \\
$\log g$ & $[\text{cgs}]$ & $4.14$ %\pm0.03 
\tablefoottext{e} & $4.5$ \tablefoottext{d} & $4.47$ \tablefoottext{f} & $4.41$
% \pm0.26
\tablefoottext{g} & $4.45$ \tablefoottext{f} & $4.60$
% \pm0.02
\tablefoottext{g} \\
$[\text{Fe/H}]$ & $[\text{dex}]$ & $0.07$ % \pm0.01 
\tablefoottext{e} & $-0.47$ \tablefoottext{f} & $-0.02$ \tablefoottext{f} & $-0.06$ % \pm0.06 
\tablefoottext{g} & $0.02$ \tablefoottext{f} & $-0.35$
% \pm0.17$ 
\tablefoottext{g} \\
$v \sin i$\tablefoottext{h} & $[\text{km s}^{-1}]$ & $2.616$ & $1.903$ & $2.111$ & $2.729$ & $1.623$ & $1.699$ \\
\hline
$M_{*}$ & $[\text{M}_{\odot}]$ & $1.10\pm0.01$ & $0.83\pm0.02$ & $0.92\pm0.02$ & $1.07\pm0.03$ & $0.90\pm0.02$ & $0.67\pm0.03$ \\
$R_{*}$ & $[\text{R}_{\odot}]$ & $1.56\pm0.05$ & $0.92\pm0.04$ & $0.97\pm0.02$ & $1.12\pm0.03$ & $0.96\pm0.03$ & $0.63\pm0.03$ \\
Age & $[\text{Gyr}]$ & $7.98\pm0.41$ & $10.74\pm2.36$ & $8.39\pm1.78$ & $4.31\pm1.29$ & $10.56\pm2.39$ & $7.30\pm4.54$ \\
\hline
\end{tabular}
\tablebib{(1)~\citet{1997A&A...323L..49P}; (2)~\citet{2000A&A...355L..27H}; (3)~\citet{2021A&A...649A...1G}; (4)~\citet{2018A&A...616A...1G}; (5)~\citet{2021A&A...656A..53S}; (6)~\citet{2011A&A...530A.138C}; (7)~\citet{2018A&A...615A..76S}}
\tablefoot{
\tablefoottext{a}{Parameters taken from the \emph{Hipparcos} catalog \citep{1997A&A...323L..49P}.}
\tablefoottext{b}{Parameters taken from the Tycho-2 catalog \citep{2000A&A...355L..27H}.}
\tablefoottext{c}{Parameters taken from \emph{Gaia} early data release 3 \citep{2021A&A...649A...1G}.}
\tablefoottext{d}{Parameters taken from \emph{Gaia} data release 2 \citep{2018A&A...616A...1G}}
\tablefoottext{e}{Parameters taken from \citet{2021A&A...656A..53S}.}
\tablefoottext{f}{Parameters taken from \citet{2011A&A...530A.138C}.}
\tablefoottext{g}{Parameters taken from \citet{2018A&A...615A..76S}.}
\tablefoottext{h}{Parameters derived using CORALIE CCF.}}
\end{center}
\end{table*}

\section{Observations and data reduction}

\subsection{Absolute astrometry}

Our absolute astrometry comes from the Hipparcos-Gaia Catalog of Accelerations \citep[HGCA,][]{2018ApJS..239...31B,2021ApJS..254...42B}.  The HGCA is a cross-calibration of Hipparcos \citep{1997ESASP1200.....E,2007A&A...474..653V} and Gaia EDR3 \citep{2016A&A...595A...1G,2021A&A...649A...1G,2021A&A...649A...2L} that places both on a common reference frame with calibrated uncertainties.  Each star in the HGCA has three proper motions: a Hipparcos proper motion near 1991.25, a Gaia proper motion near 2016.0, and the position difference between Hipparcos and Gaia scaled by the time baseline between the missions.

Table \ref{tab:hgca} lists each of our target star's absolute astrometry from the HGCA.  The listed Gaia proper motions match the EDR3 catalog values, but the epochs listed are those that minimize the uncertainty in each star's position.  The epochs differ from 2016.0 because of covariance between position and proper motion at the catalog epoch adopted by Gaia.  The Hipparcos proper motions are a linear combination of the ESA \citep{1997ESASP1200.....E} and Hipparcos-2 \citep{2007A&A...474..653V} reductions, corrected for a locally variable frame rotation as discussed in \cite{2018ApJS..239...31B}.  All uncertainties are inflated from the catalog values to achieve statistical agreement between the three proper motions for a sample of stars that show no radial-velocity acceleration \citep{2021ApJS..254...42B}.

All stars listed in Table \ref{tab:hgca} show highly significant discrepancies between the proper motion measurements, indicative of a tug by an unseen companion.  The formal significance of this astrometric acceleration ranges from almost $50\sigma$ for HD~142234 to nearly $300\sigma$ for HD~92987.

\begin{table*}
\centering
\caption{HGCA astrometry}
\begin{tabular}{ccccccccc}
\hline
\hline
Star & Data Source & $\mu_{\alpha*}$ (mas\,yr$^{-1}$) & $\sigma[\mu_{\alpha*}]$ & 
$\mu_{\delta}$ (mas\,yr$^{-1}$) & $\sigma[\mu_{\delta}]$ & 
Corr$[\mu_{\alpha*},\mu_\delta]$ & $t_{\alpha*}$ (Jyr) & $t_{\delta}$ (Jyr) \\
\hline
  HD 92987 & Hip & 7.272 & 0.678 & 7.711 & 0.593 & $-$0.207 & 1991.02 & 1991.54 \\
           & Hip-Gaia & 17.875 & 0.020 & 9.420 & 0.018 & $-$0.044 \\
           & Gaia & 14.816 & 0.070 & 23.633 & 0.055 & 0.385 & 2015.96 & 2015.97 \\
 HD 142234 & Hip & 61.504 & 1.095 & 112.270 & 0.879 & $-$0.278 & 1991.38 & 1991.18 \\
           & Hip-Gaia & 57.593 & 0.039 & 111.694 & 0.025 & $-$0.200 \\
           & Gaia & 55.225 & 0.049 & 110.892 & 0.040 & $-$0.328 & 2015.92 & 2015.66 \\
 HD 143616 & Hip & 28.909 & 1.242 & 21.917 & 0.855 & $-$0.053 & 1991.11 & 1991.13 \\
           & Hip-Gaia & 25.063 & 0.041 & 14.123 & 0.027 & $-$0.059 \\
           & Gaia & 18.598 & 0.052 & 6.781 & 0.037 & $-$0.249 & 2015.96 & 2015.39 \\
 HD 157338 & Hip & $-$0.010 & 0.856 & $-$182.868 & 0.515 & $-$0.263 & 1991.02 & 1990.85 \\
           & Hip-Gaia & 3.750 & 0.026 & $-$179.733 & 0.016 & $-$0.134 \\
           & Gaia & 9.194 & 0.051 & $-$176.202 & 0.034 & $-$0.010 & 2015.95 & 2016.27 \\
 HD 195010 & Hip & 68.103 & 1.292 & $-$281.046 & 0.950 & $-$0.065 & 1991.15 & 1991.18 \\
           & Hip-Gaia & 60.710 & 0.036 & $-$280.653 & 0.024 & 0.099 \\
           & Gaia & 62.713 & 0.028 & $-$286.095 & 0.023 & 0.032 & 2015.91 & 2015.95 \\
 HIP 22059 & Hip & $-$50.714 & 1.653 & $-$133.077 & 1.816 & $-$0.045 & 1991.10 & 1991.15 \\
           & Hip-Gaia & $-$41.054 & 0.056 & $-$133.137 & 0.062 & $-$0.063 \\
           & Gaia & $-$34.356 & 0.031 & $-$120.663 & 0.041 & $-$0.211 & 2016.01 & 2016.15 \\
\hline
\end{tabular}
\label{tab:hgca}
\end{table*}

\subsection{Radial velocities}

The radial-velocity data are taken from the CORALIE Survey for Extra-Solar Planets \citep{2000fepc.conf..548Q},  a planet search survey in progress for more than 20 years in the southern hemisphere, whose observations began in June 1998. The survey makes use of the CORALIE spectrograph at La Silla Observatory, and monitors a volume-limited sample of 1647 main-sequence stars spanning from spectral types F8 to K0 within 50~pc of the Sun.

In order to increase the efficiency and accuracy of the CORALIE spectrograph over the years, it underwent two major upgrades in June 2007 \citep{2010A&A...511A..45S} and in November 2014. While these changes improved the overall performance of the CORALIE spectrograph, they also introduced small offsets in the measured radial velocities; therefore, we treat the radial-velocity data from the CORALIE spectrograph as three separate instruments  corresponding to these upgrades. We refer to the original CORALIE as CORALIE-98 (C98), the upgrade in 2007   as CORALIE-07 (C07), and the latest upgrade in 2014  as CORALIE-14 (C14).

We also include some additional data for a few of the targets from other spectrographs, namely HARPS \citep{2003Msngr.114...20M} and HIRES \citep{1994SPIE.2198..362V}. All of the data products presented in this paper are available at the Data and Analysis Center for Exoplanets (DACE)\footnote{The data are available at the Data and Analysis Center for Exoplanets (DACE), which can be accessed at \url{https://dace.unige.ch}.}.

The RV data were reduced using the CORALIE automated pipeline, which also measures the cross-correlation function (CCF), full width at half maximum (FWHM), the bisector, and the $H_{\alpha}$ chromospheric activity indicator. These indicators allow us to pinpoint the origin of observed periodic signals, to ensure that we are not observing, for example, significant stellar activity.

As part of the ongoing CORALIE survey, we have been monitoring for long-term trends in the data that could be indicative of widely separated companions that are amenable to high-contrast imaging observations. To monitor these candidates, we assessed our sample of RV data for long-period linear or quadratic trends over the past $>20$-year baseline.

\subsection{High-contrast imaging}

\begin{figure*}
    \centering
    \includegraphics[width=\textwidth]{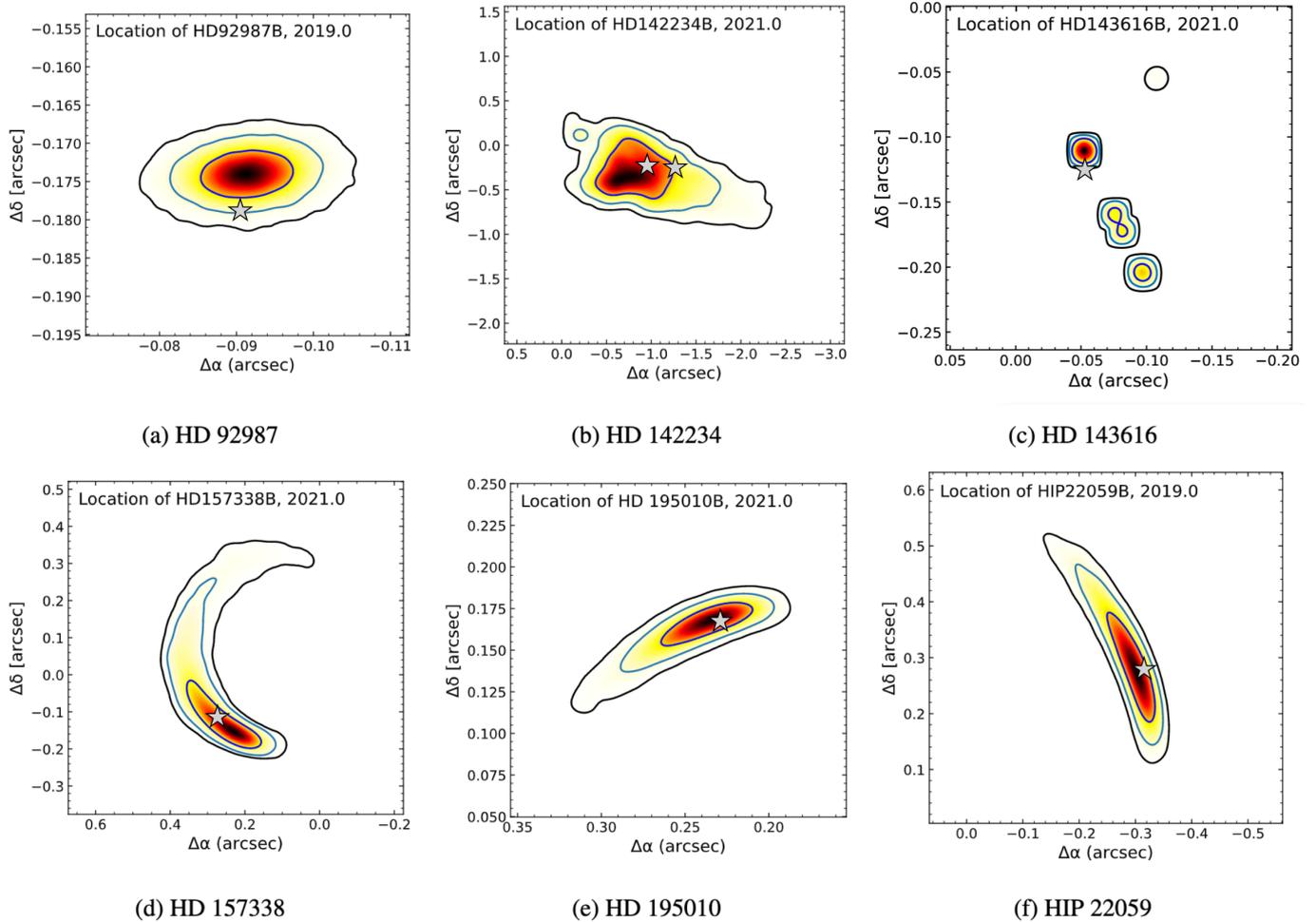}
    \caption{Contours showing the astrometric predictions of each companion relative to the host star in ascension ($\Delta\alpha$) and declination ($\Delta\delta$) using the \texttt{orvara} orbit fitting package \citep{2021AJ....162..186B}, with the measured companion positions shown by the overplotted gray   stars. Relative astrometric positions indicate whether a companion can be directly detected with imaging given its projected separation between the host star and the companion itself. The 1, 2, and 3$\sigma$ contour levels of the predicted positions of each companion are plotted from a full orbit fit using just the radial velocities and the HGCA astrometry, with the astrometric predictions being shown for the year in which each companion was then subsequently directly imaged with VLT/SPHERE in order to visually compare the prior predictions to the direct detections. The measured position of each companion from the VLT/SPHERE imaging are shown by the overplotted gray stars. In the case of HD~142234, the left star corresponds to HD~142234~B and the right star corresponds to HD~142234~C. The measured positions of each companion from direct imaging observations agree well with the predicted positions from the orbit fitting using just the radial velocities and astrometry from the HGCA.}
    \label{fig:astrometric_predictions}
\end{figure*}

The radial-velocity data taken from the CORALIE spectrograph \citep{2000fepc.conf..548Q}, as well as HARPS \citep{2003Msngr.114...20M} and HIRES \citep{1994SPIE.2198..362V}, are used as precursor observations to assess the detectability of any companion candidates through direct imaging. We combined the radial-velocity data with the astrometric information from the HGCA to determine the predicted relative astrometry in ascension and declination ($\Delta\alpha,\Delta\delta$) of each companion on the sky, using the orbit-fitting  code \texttt{orvara} \citep{2021AJ....162..186B}. The results of the predicted positions, which show full orbital fits of each system using the radial velocities and the astrometric accelerations from the HGCA, are shown in Figure~\ref{fig:astrometric_predictions}. The detectability of these objects was assessed by translating these values into a predicted contrast ratio and a projected angular separation for each system. These values were then compared against the on-sky measured performance $5\sigma$ contrast curves from VLT/SPHERE. 

We selected to observe the most promising targets using the  VLT/SPHERE dual-band imaging mode \citep{2010MNRAS.407...71V} using the InfraRed Dual-Band Imager and Spectrograph \citep[IRDIS;][]{2008SPIE.7014E..3LD} in the $H2$ and $H3$ bands ($\lambda_{H2} = 1.593~\mu$m, $\lambda_{H3} = 1.667~\mu$m). Observations were taken with SPHERE \citep{2019A&A...631A.155B} via the extreme adaptive optics system at the VLT, under  programs 0102.C-0236(A) (PI: Rickman), 0104.C-0724(A) (PI: Rickman), and 105.20SZ.001 (PI: Rickman).

\begin{figure*}
    \centering
    \begin{subfigure}[t]{0.3\linewidth}
    \includegraphics[width=\textwidth]{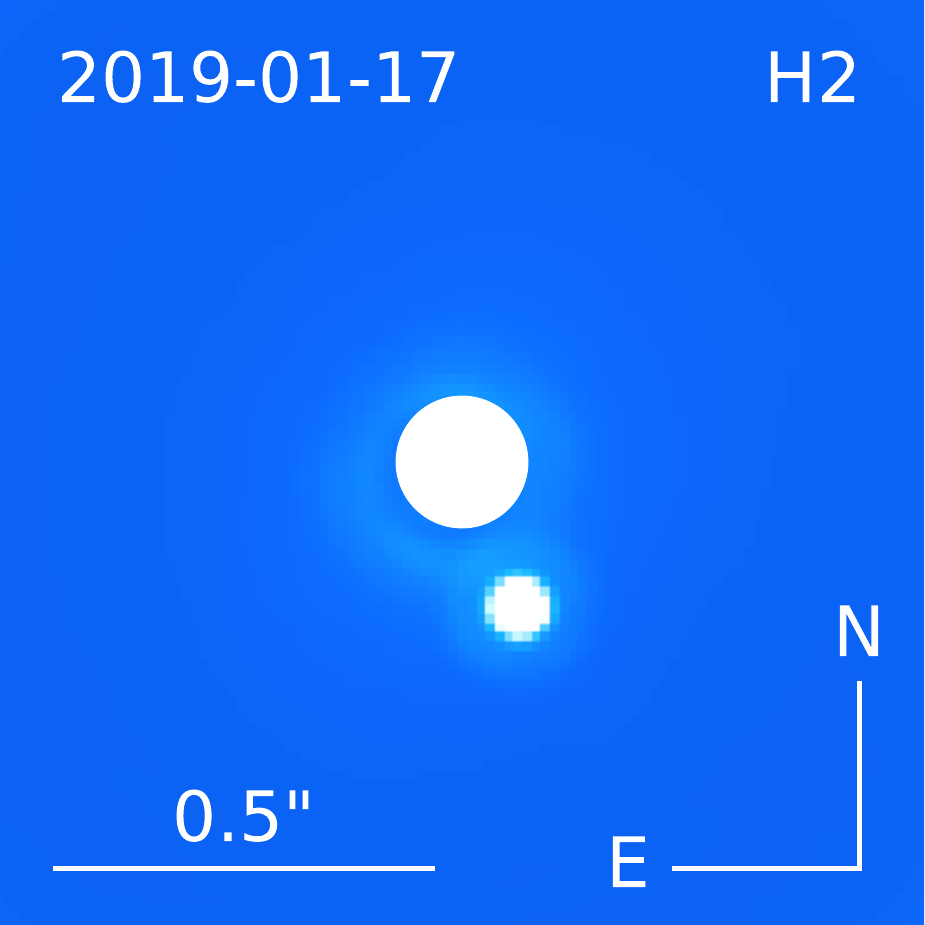}
   \caption{HD~92987}
    \end{subfigure}
    \begin{subfigure}[t]{0.3\linewidth}
    \includegraphics[width=\textwidth]{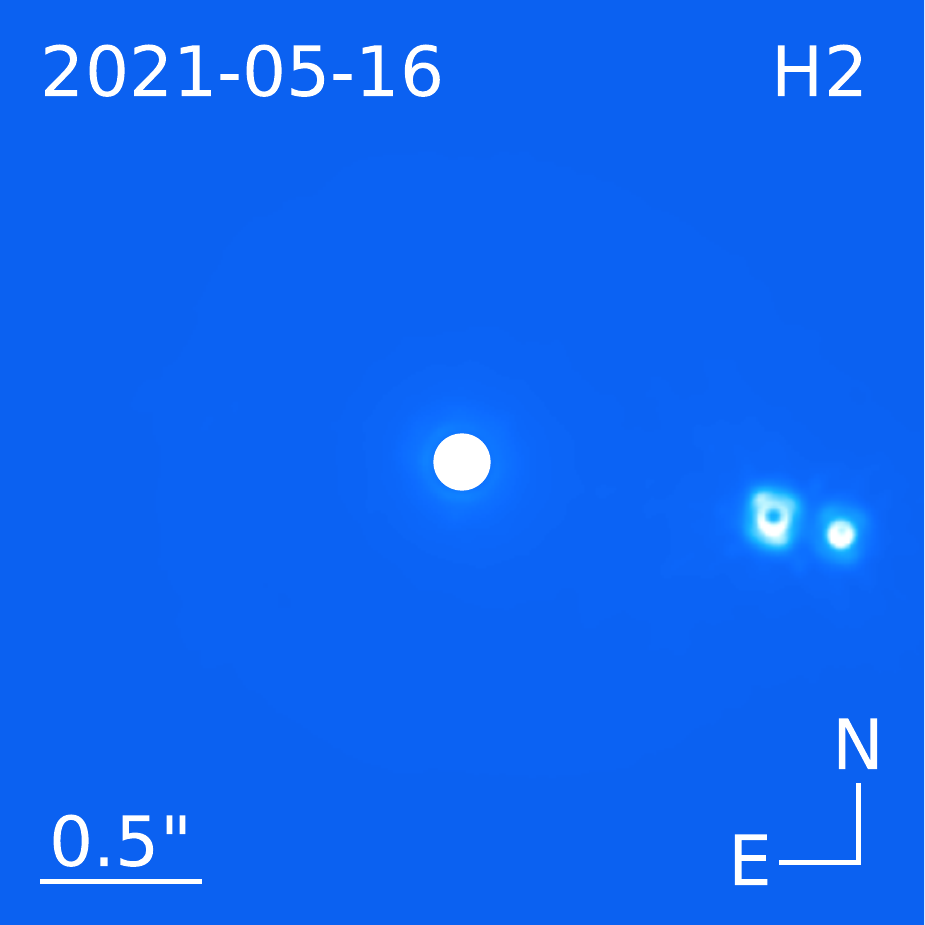}
    \caption{HD~142234}
    \end{subfigure}
    \begin{subfigure}[t]{0.3\linewidth}
    \includegraphics[width=\textwidth]{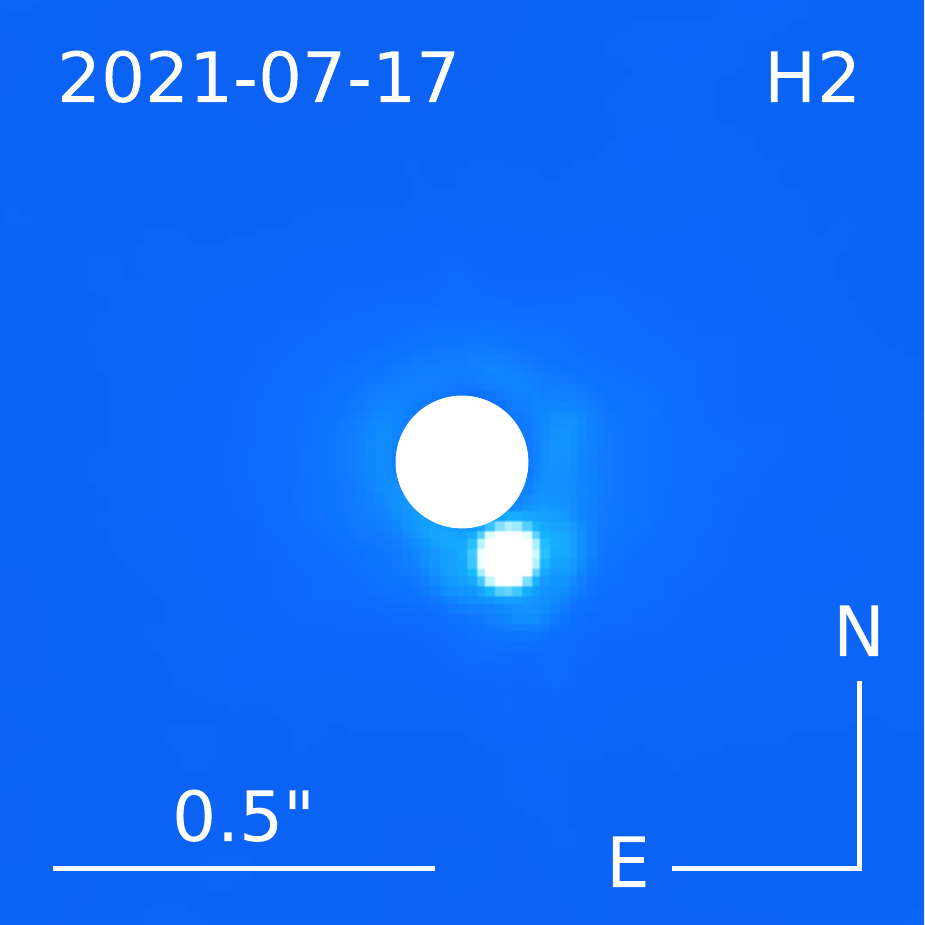}
    \caption{HD~143616}
    \end{subfigure}
    \begin{subfigure}[t]{0.3\linewidth}
    \includegraphics[width=\textwidth]{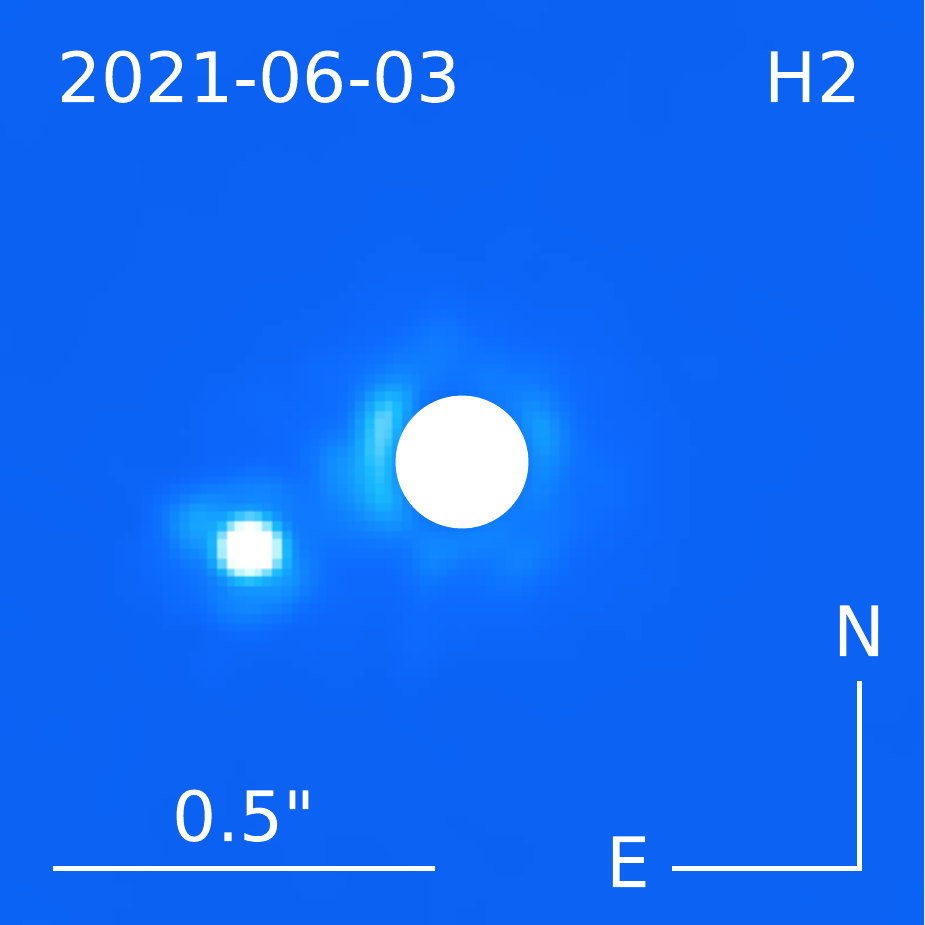}
    \caption{HD~157338}
    \end{subfigure}
    %\begin{subfigure}[t]{0.3\linewidth}
    %\includegraphics[width=\textwidth]{HD157338_figs/HD157338_2_detection_image_H2.pdf}
    %\caption{HD~157338}
    %\end{subfigure}
    \begin{subfigure}[t]{0.3\linewidth}
    \includegraphics[width=\textwidth]{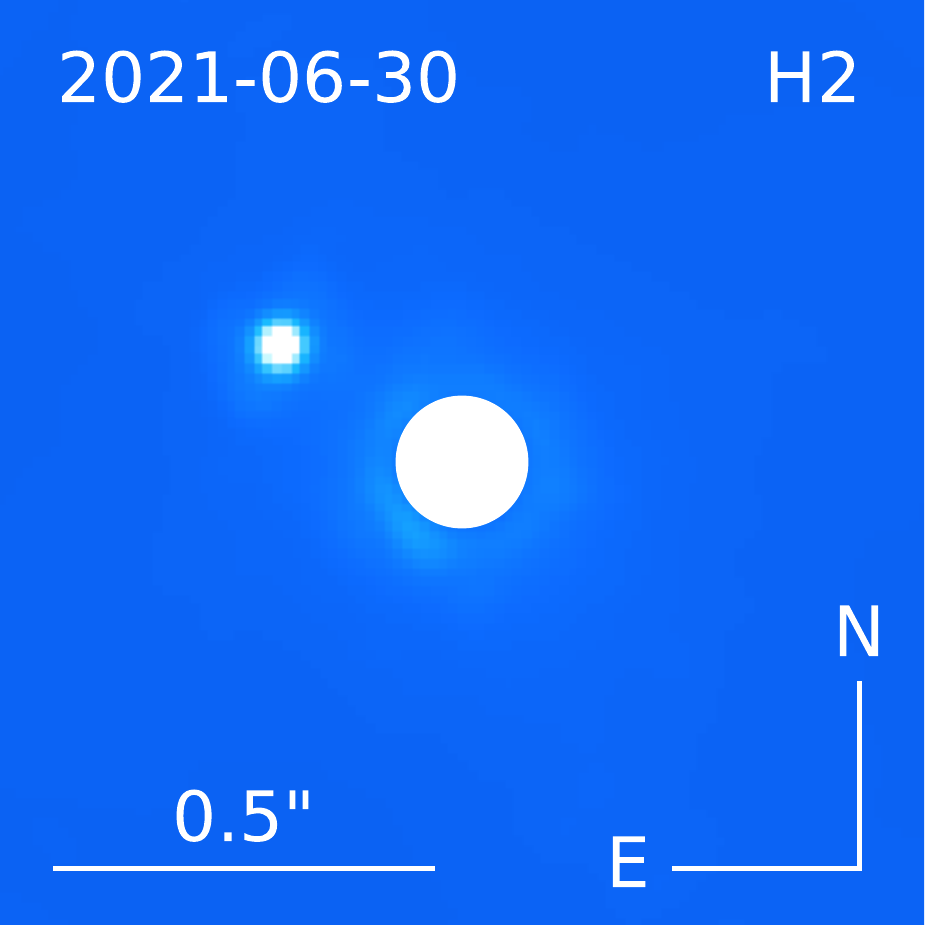}
    \caption{HD~195010}
    \end{subfigure}
    \begin{subfigure}[t]{0.3\linewidth}
    \includegraphics[width=\textwidth]{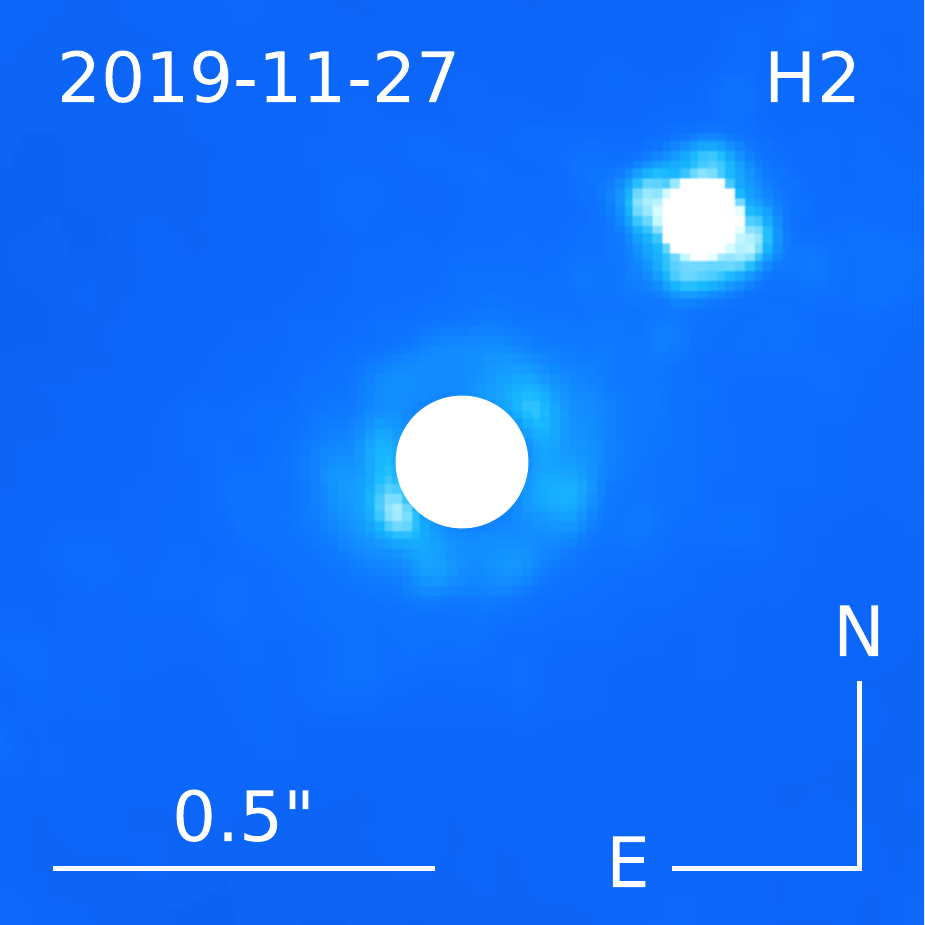}
    \caption{HIP~22059}
    \end{subfigure}
    \caption{High-contrast images taken with VLT/SPHERE using IRDIS coronagraphic imaging in the $H2$ band. In each image the star is masked behind the central white circle. Some of the coronagraphic images shown here show some saturation, which is why the flux frames were used  to obtain the photometry of each companion. The image for HD~142234 is shown on a wider angular scale than the other images.}
    \label{fig:direct_images}
\end{figure*}

The observations were taken using an apodized Lyot coronagraph \citep{2003A&A...397.1161S}, where the length of exposure time was calculated to be optimal for the expected contrast ratio of the detections, with observations ranging from 0.5 to 3 hours. In order to measure the position of the centre of the star behind the coronagraph, several exposures at the beginning and at the end of the observing sequence were taken with two orthogonal sinusoidal modulations applied to the deformable mirror, which generates satellite spots around the star.

The observation sequence also consisted of short exposure images of the primary star moved from behind the coronagraph to obtain the flux and the shape of the point spread function (PSF), making use of a neutral density (ND) filter\footnote{The corrections for the ND filter transmission make use of the
ND filter curves available  at \url{https://www.eso.org/sci/facilities/
paranal/instruments/sphere/inst/filters.html}}. Long-exposure sky frames were also taken to estimate the background flux and correct for any bad pixels on the detector. The observations were carried out in pupil tracking mode, which allows   the observed field of view to rotate during the course of the coronagraphic observations, in order to obtain high-contrast images using the angular differential imaging (ADI) approach. 

The VLT/SPHERE data were reduced using the Geneva Reduction and Analysis Pipeline for High-contrast Imaging of planetary Companions \citep[GRAPHIC,][]{2016MNRAS.455.2178H}. The GRAPHIC pipeline carries out sky subtraction, flat-fielding, bad pixel correction, frame selection, and a correction for the distortion as outlined in \citet{2016SPIE.9908E..34M}. We then performed a classical ADI (cADI) PSF subtraction algorithm before derotating and median combining to produce a final PSF-subtracted image. The resulting images in the $H2$ band are shown in Fig.~\ref{fig:direct_images}.

\section{Astrometry and photometry} \label{sec:astro_photo}

\begin{table*}
\centering
\caption{Measured astrometry and photometry of the very low-mass stellar companions.}
\begin{tabular}{cccccc}
\hline
\hline
Companion & Date (yyyy-mm-dd) & $\rho$ (mas) & $\theta$ (deg) & Contrast ($H2$) & Contrast ($H3$) \\
\hline

HD~92987~B & 2019-01-17 & 200.81 $\pm$ 5.99 & 206.76 $\pm$ 1.86 & $5.26 \pm 0.10$ & $5.21 \pm 0.12$ \\

HD~142234~B & 2021-05-16 & 1007.26 $\pm$ 1.38 & 258.38 $\pm$ 0.20 & $2.30 \pm 0.01$ & $2.22 \pm 0.01$ \\

HD~142234~C & 2021-05-16 & 1227.57 $\pm$ 3.89 & 257.88 $\pm$ 0.25 & $3.17 \pm 0.01$ & $3.09 \pm 0.01$ \\

HD~143616~B & 2021-07-17 & 144.84 $\pm$ 2.72 & 200.34 $\pm$ 1.08 & $3.87 \pm 0.09$ & $3.78 \pm 0.08$ \\

HD~157338~B & 2021-06-03 & 306.25 $\pm$ 3.29 & 111.33 $\pm$ 0.62 & $3.93 \pm 0.02$ & $3.82 \pm 0.01$ \\

HD~157338~B & 2021-06-30 & 303.78 $\pm$ 3.15 & 112.20 $\pm$ 0.61 & $3.92 \pm 0.01$ & $3.85 \pm 0.03$ \\

HD~195010~B & 2021-06-03 & 284.27 $\pm$ 1.73 & $54.70 \pm 0.43$ & $4.76 \pm 0.03$ & $4.69 \pm 0.02$ \\

HD~195010~B & 2021-06-30 & 282.76 $\pm$ 3.53 & 54.42 $\pm$ 0.73 & $4.75 \pm 0.07$ & $4.77 \pm 0.07$ \\

HIP~22059~B & 2019-11-27 & 455.23 $\pm$ 0.58 & 314.49 $\pm$ 0.26 & $4.07 \pm 0.02$ & $3.99 \pm 0.02$ \\

\hline
\end{tabular}
\tablefoot{The astrometric measurements were calculated using the average of the IRDIS $H2$ and $H3$ channels SPHERE flux frame data. The photometry for  the $H2$ and $H3$ bands are shown, with HD~157338 and HD~195010 both having two epochs of data taken for each system.}
\label{tab:astrometry}
\end{table*}

Due to saturation in some of the coronagraphic images caused by the brightness of the companions, we opted to use the flux frames of each data set to determine the astrometry and photometry of the companions. These frames use a neutral density filter with no coronagraph and a shorter exposure time than a normal science sequence.  Our use of unsaturated flux frames enabled us to perform PSF photometry and astrometry, using the primary star itself as the PSF template.  

We first bias-correct and flat-field the images, interpolating over bad pixels.  We then perform reference differential imaging (RDI), subtracting a scaled IRDIS stellar PSF taken under the same observing conditions to remove the primary star's light.  For this purpose we use HD~142234 (which has no companions within $1''$) and two additional comparison stars without companions from the ESO archive: HIP~82154 and HIP~102979.  HD~142234 itself has companions sufficiently far away that RDI is unnecessary.  For all other stars, we have six image pairs: three PSF template stars, and two sets of SPHERE/IRDIS coadds for each epoch.  

We optimize the position and relative flux of each PSF reference star within one of our flux frames and use it to remove the light of the primary star.  We then perform PSF photometry on the residual, which contains the light from the companion, together with noise and low-level residuals from the primary.  We use the unsubtracted PSF of the primary as the template for the secondary.  In this way, we fit for an offset between the two stars without needing to centroid the image since the stellar PSF and companion PSF are both present in the same image.  We also obtain a contrast without having to assume anything about the form of the PSF. 

Our approach resulted in six measurements of the contrast and position (three PSF reference images times two sets of coadds) for each epoch.  We report the mean and standard deviation of these six epochs as our measurements and uncertainties; we bias-corrected the standard deviations by a factor of $\sqrt{6/5}$.  We conservatively did not divide the uncertainties by $\sqrt{6}$ as we did not have a large enough data set to verify the statistical independence of our measurements.

The separation and position angle detector positions were converted into on-sky separation and position angles by accounting for an anamorphic distortion of $1.0062\pm0.0002$, plate scale of $12.255\pm0.009$~mas/pix for the $H2$ band observations, and $12.250\pm0.009$~mas/pix for the $H3$ band observations \citep{2016SPIE.9908E..34M}. A true north offset of $-1.75\pm0.08$~deg \citep{2016SPIE.9908E..34M}, and a pupil offset of $135.99 \pm 0.11$~deg were also applied\footnote{Values are taken from the VLT SPHERE User Manual, 16th release}. 

%Additionally, a systematic uncertainty of $\pm3$~mas ($\pm0.245$~pix) for the target stars' positions was included when calculating the positional errors of the companions. 

The resulting astrometry and photometry of the very low-mass stellar companions are given in Table~\ref{tab:astrometry}. We   note that the error on the relative astrometry and photometry of HD~92987 is larger than the other systems. This is due to the  HD~92987 system having a higher contrast ratio of the companion to the primary star than the other observed systems, making it more difficult to obtain the same level of precision for the relative astrometry and photometry. Despite this, we still measure values with reasonable uncertainty that place orbital constraints on the HD~92987 system.

In Fig.~\ref{fig:astrometric_predictions} we compare the predicted and measured relative astrometry of each companion by overplotting the measured astrometry outlined in Table~\ref{tab:astrometry} on  the predicted positions. From this we see that the positions of each observed companion are in good agreement with each other, where each companion is within at least 2$\sigma$ of the prediction.

\begin{figure}
    \centering
    \includegraphics[width=0.48\textwidth]{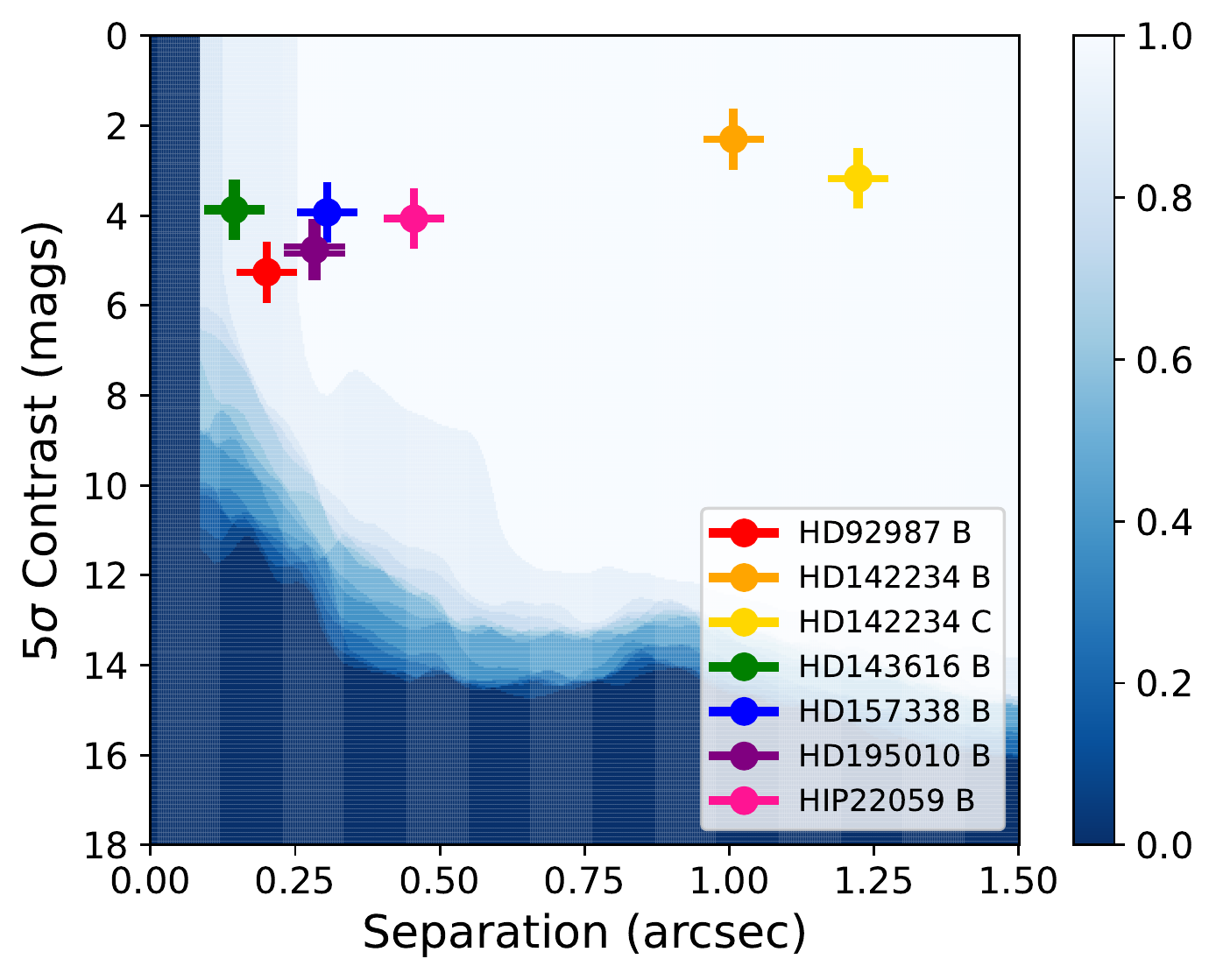}
    \caption{Measured contrast of each companion at the measured separations with stacked $5\sigma$ contrast curves corresponding to measured on-sky performance from VLT/SPHERE data for IRDIS $H2$ bands. The detection probability scale is shown on the right. The darker color corresponds to a lower likelihood of detectability with imaging with VLT/SPHERE IRDIS $H2$ band, and the lighter colors correspond to a higher likelihood of detectability at a given separation and contrast ratio. The astrometry and photometry for HD~157338 and HD~195010 are each shown as an average of the two 2021 SPHERE $H2$ band data points.}
    \label{fig:detection_probability}
\end{figure}

The values from Table~\ref{tab:astrometry} are shown in Fig.~\ref{fig:detection_probability} as a comparison against the measured on-sky performance $5\sigma$ contrast curves from $H2$ VLT/SPHERE observations to assess their direct imaging detectability. The observations were obtained through injection and recovery of PSFs of companions at increasing separation for a number of targeted stars from the CORALIE survey.

%Because the flux calibration frames are not acquired simultaneously with the science frames, accounting for changes in weather or the target star's flux between flux and science observations cannot be easily incorporated in the error on the contrast values of the companions.  Therefore, the uncertainty of the companions' photometric measurments presented here should be regarded as conservative.  

%\begin{table*}
%\centering
%\caption{The measured astrometry and photometry of the very low-mass stellar companions.  All astrometric and photometric measurements were conducted using IRDIS H2 channel SPHERE data.  Note that the errors on the contrast values do not account for changes in weather or the target stars' flux.}
%\begin{tabular}{ccccc}
%\hline
%\hline
%Companion & Date & $\rho$ (mas) & $\theta$ (deg) & Contrast (mag)\\
%\hline

%HD~92987~B & 2019-01-17 & 198.4 $\pm$ 3.0 & 206.82 $\pm$ 0.88 &  5.11 $\pm$ 0.01 \\

%HD~142234~B & 2021-05-16 & 994.8 $\pm$ 3.6 & 255.77 $\pm$ 0.23 & 4.88 $\pm$ 0.03 \\

%HD~142234~C & 2021-05-16 & 1220.8 $\pm$ 3.7 & 256.62 $\pm$ 0.20 & 4.73 $\pm$ 0.01 \\

%HD~143616~B & 2021-07-17 & 149.7 $\pm$ 3.1 & 195.75 $\pm$ 1.18 & 4.82 $\pm$ 0.02 \\

%HD~157338~B & 2021-06-03 & 318.9 $\pm$ 3.1 & 112.55 $\pm$ 0.56 & 4.33 $\pm$ 0.01 \\

%HD~157338~B & 2021-06-30 & 317.7 $\pm$ 3.1 & 114.72 $\pm$ 0.57 & 4.80 $\pm$ 0.02 \\
 
%HD~195010~B & 2021-06-30 & 282.3 $\pm$ 3.0 & 54.38 $\pm$ 0.62 & 4.14 $\pm$ 0.01 \\

%HIP~22059~B & 2019-11-27 & 454.0 $\pm$ 3.1 & 314.80 $\pm$ 0.40 & 3.44 $\pm$ 0.01 \\
%\hline
%\end{tabular}
%\label{tab:astrometry}
%\end{table*}

\section{Orbital solutions} \label{sec:orbital_solutions}

In order to obtain orbital fits of our observed systems, we made use of \texttt{orvara} \citep{2021AJ....162..186B}, which is an orbit-fitting code  specifically designed to combine radial-velocity data, relative astrometry from direct imaging, and absolute astrometric data from the \emph{Hipparcos} and \emph{Gaia}. The \texttt{orvara} code \citep{2021AJ....162..186B} utilizes a comprehensive Markov chain Monte Carlo (MCMC) approach, incorporating astrometric accelerations from Hipparcos-Gaia eDR3 using the Hipparcos-Gaia catalog of accelerations \citep[HGCA;][]{2021ApJS..254...42B} and the intermediate astrometry fitting tool \texttt{htof} \citep{2021AJ....162..230B}.

Using \texttt{orvara}, we were able to obtain robust orbital fits by combining the radial-velocity data with relative astrometry obtained from our VLT/SPHERE observations, as described in Section~\ref{sec:astro_photo}, as well as proper motion anomalies from the long baseline between the \emph{Hipparcos} and \emph{Gaia} missions. In this section we present the orbital solutions for each of the systems. The full list of posteriors of the fitted orbital parameters is given in Table~\ref{tab:orbital_parameters}, and each of the corner plots showing the posterior distributions is shown in Appendix~\ref{appendix}.

\subsection{HD~92987 (HIP~52472)}

\begin{figure*}
    \centering
    \includegraphics[width=0.49\textwidth]{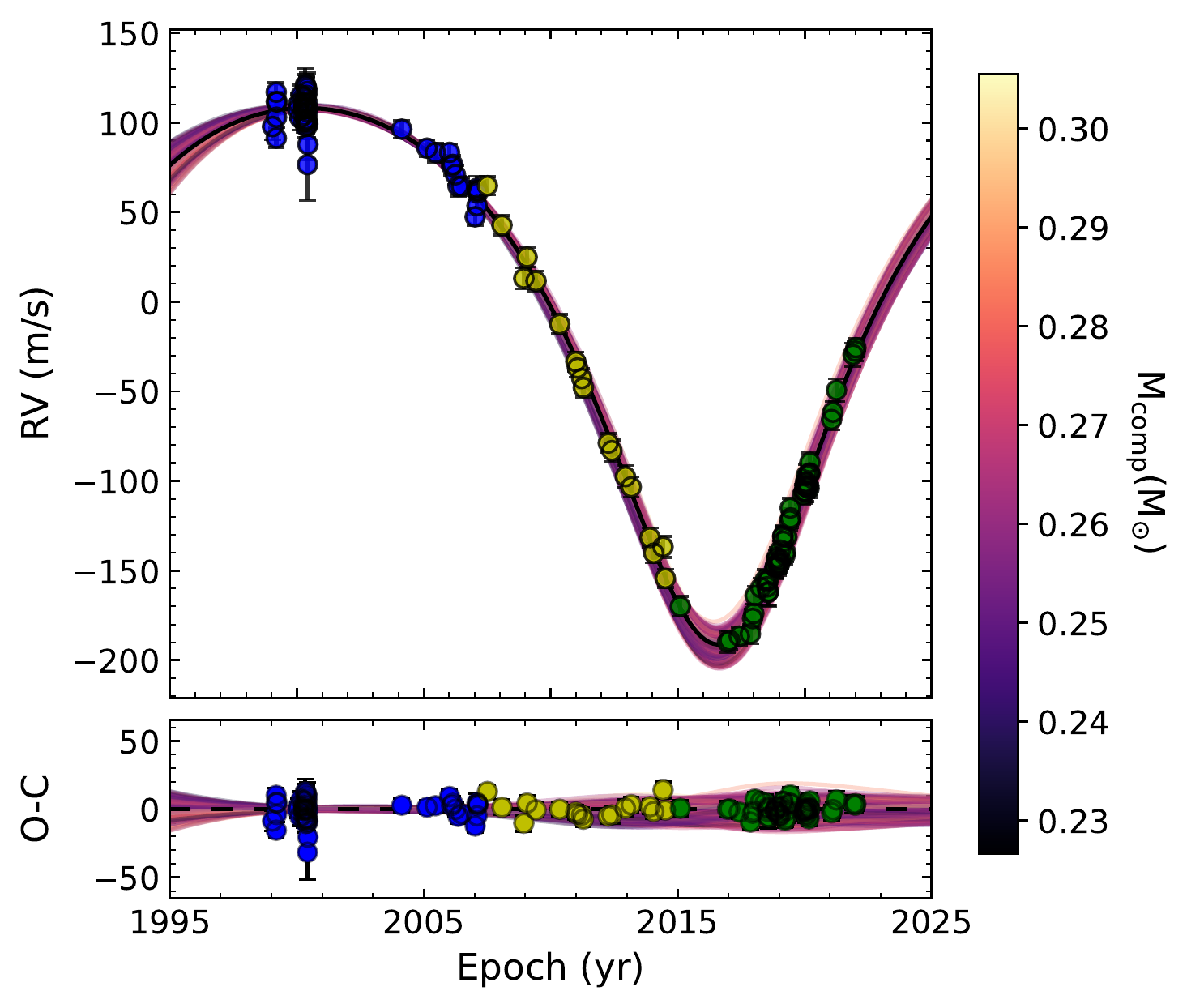}
    \includegraphics[width=0.49\textwidth]{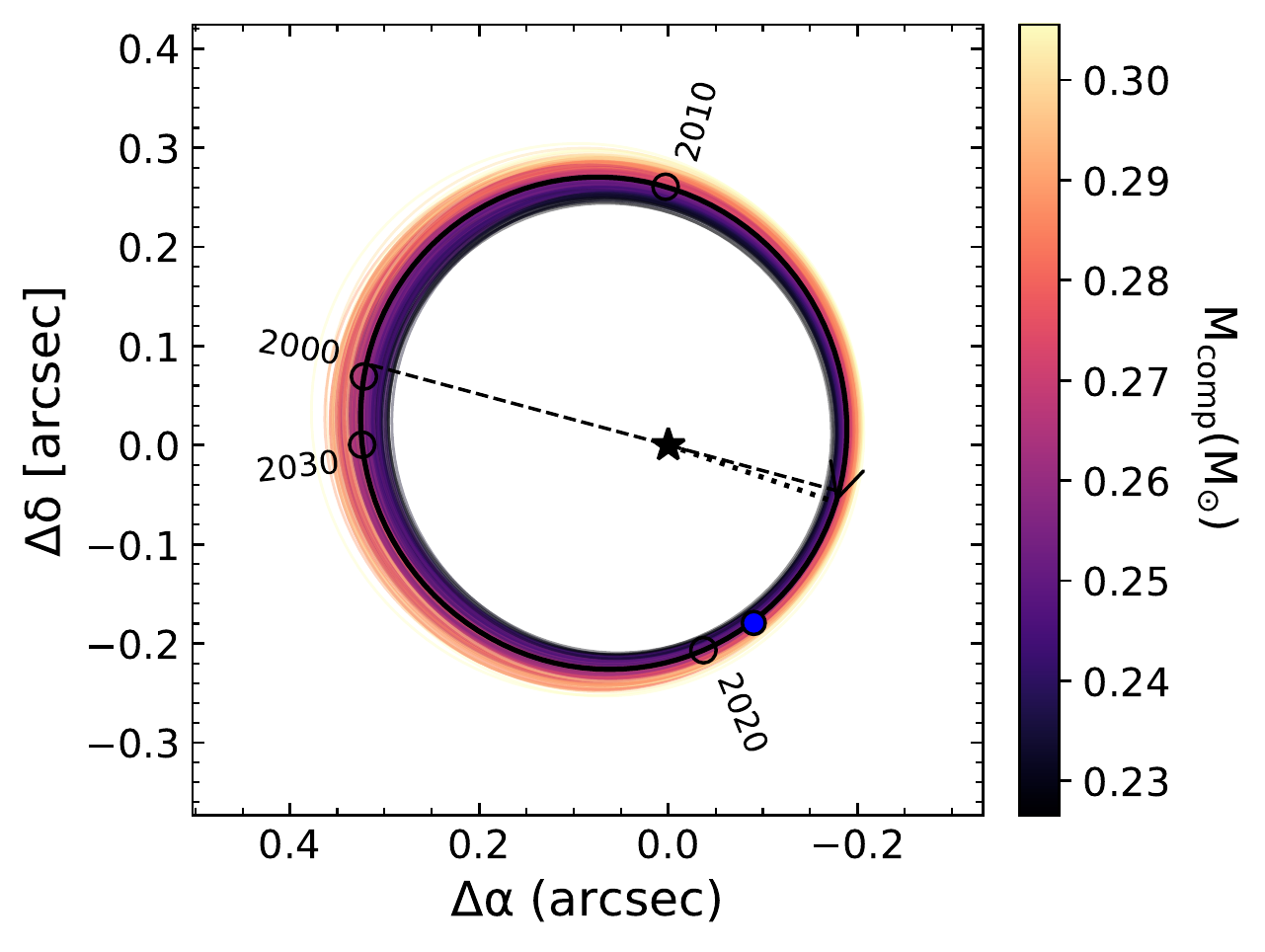}
    \includegraphics[width=\textwidth]{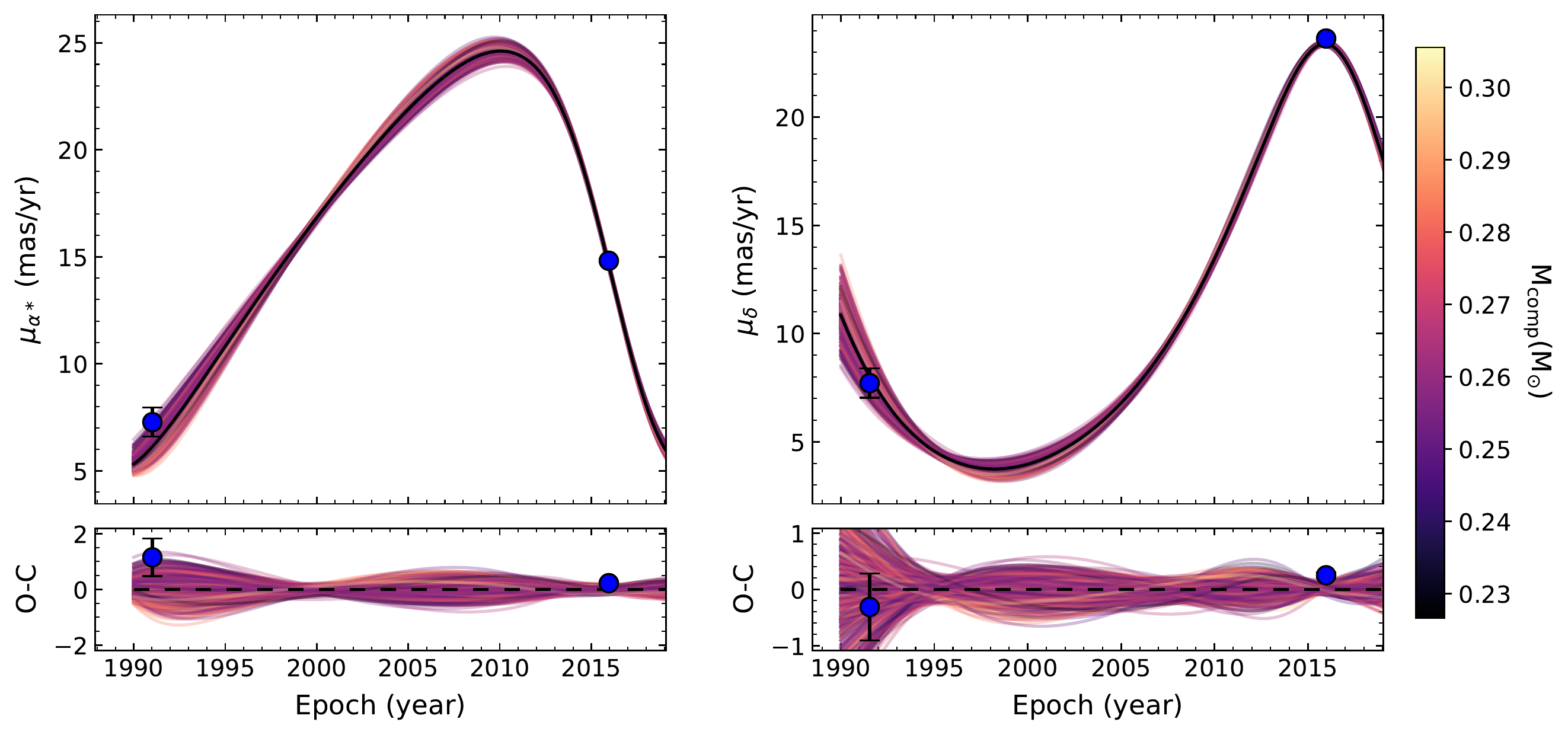}
    \caption{ Orbit fits of HD~92987 using the orbit-fitting code \texttt{orvara}. \emph{Top left.} Radial-velocity orbit induced by HD~92987~B over a full orbital period. Shown are the radial-velocity data of COR-98 (blue points), COR-07 (yellow points), and COR-14 (green points). The thick line shows the highest likelihood fit; the thin colored lines show 500 orbits drawn randomly from the posterior distribution.\emph{Top right.} Relative astrometric orbit of HD~92987~B. The thick black line represents the highest likelihood orbit; the thin colored lines represent 500 orbits drawn randomly from the posterior distribution. Dark purple  corresponds to a low companion mass and  light yellow  corresponds to a high companion mass. The dotted black line shows the periastron passage, and the arrow at the periastron passage shows the direction of the orbit. The dashed line indicates the line of nodes. Predicted past and future relative astrometric points are shown by black circles with their respective years, while the observed relative astrometric point from VLT/SPHERE data is shown by the blue-filled data point, where the measurement error is smaller than the plotted symbol. \emph{Bottom.} Acceleration induced by the companion on the host star as measured from absolute astrometry from \emph{Hipparcos} and \emph{Gaia}. The thick black line represents the highest likelihood orbit;  the thin colored lines are 500 orbits drawn randomly from the posterior distribution. Darker purple represents a lower companion mass and   light yellow  represents a higher companion mass. The residuals of the proper motions are shown in the bottom panels.}
    \label{fig:HD92987_orbits}
\end{figure*}

HD~92987 is a G2/3V star that has been observed with CORALIE at La Silla Observatory since January 1999. Fifty-three measurements were taken with CORALIE-98, 18 additional RV measurements were obtained with CORALIE-07, and 57 additional RV measurements were obtained with CORALIE-14. The detection of HD~92987~B was originally presented in \citet{2019A&A...625A..71R} with a minimum mass derived from the radial velocities of $16.88^{+0.69}_{-0.65}~M_{\rm{Jup}}$, interpreted as being either a massive planetary companion, or a very low-mass brown dwarf. The radial-velocity detection of HD~92987~B was also reported independently by \citet{2019AJ....157..252K}, who measured a minimum mass of $17.9^{+2.4}_{-1.9}~M_{\rm{Jup}}$, in agreement with the interpretation presented in \citet{2019A&A...625A..71R}.

HD~92987 was then observed with the  VLT/SPHERE program 0102.C-0236(A) (PI: Rickman) on 2019 January 17, with a total observation time of three hours, with a 64-second integration time on the coronagraphic frames using IRDIS, as well as several sky, center, and flux frames taken which was necessary for the post-processing of the data. The direct detection of HD~92987~B is shown in Fig.~\ref{fig:direct_images}. Since \citet{2019A&A...625A..71R}, we increased the radial-velocity baseline by about three years, continuing to monitor how the radial-velocity trend changes over time, which enabled us to further constrain its orbital parameters.

Combining the radial-velocity measurements with the relative astrometry in Table~\ref{tab:astrometry} and with the Hipparcos-Gaia eDR3 accelerations, we ran the MCMC orbital fit using \texttt{orvara} \citep{2021AJ....162..186B} with a log-flat prior on the primary mass and with 500,000 steps in each chain. The resulting plots are shown in Fig.~\ref{fig:HD92987_orbits} with a burn-in phase of 700 multiplied by every 250th step that  is saved on the chain with 500 randomly drawn orbits.

We derive a companion mass of $276.6^{+17.8}_{-16.8}~M_{\mathrm{Jup}}$, with an orbital period of $31.76^{+0.54}_{-0.48}$~years. For the primary mass of HD~92987~A we derive a mass of $1.17^{+0.12}_{-0.11}~M_{\odot}$, which is in agreement with the mass we determined using the Geneva stellar-evolution models \citep{2012A&A...537A.146E,2013A&A...558A.103G}, as described in Section~\ref{sec:stellar_params}. Our dynamical mass value of the primary star, HD~92987~A, is also in agreement with a recent value from \citet{2021A&A...656A..53S}, who derived a spectroscopic mass of $1.155 \pm 0.005~M_{\odot}$.

\citet{2019AJ....157..252K} presented imaging observations of HD~92987 using the Differential Speckle Survey Instrument (DSSI) operating on the Gemini South telescope, but did not directly detect HD~92987~B. They did, however, use these observations to constrain the upper mass limit of HD~92987~B as a function of the angular separation, and found the upper mass limit to be in the 200-500~$M_{\rm{Jup}}$ mass range at approximately 200~mas separation. This is in agreement with the mass derived using our SPHERE observations and absolute astrometry from \emph{Hipparcos} and \emph{Gaia}, with a mass of $276.6^{+17.8}_{-16.8}~M_{\mathrm{Jup}}$ at $200.81\pm5.99$~mas in the $H2$ band at $1.593\mu$m.

More recently, the mass of HD~92987~B was determined by \citet{2021AJ....162...12V} through combining published radial velocities with astrometry from \emph{Hipparcos} and \emph{Gaia} eDR3, who derived a mass of $268.4\pm4.7~M_{\rm{Jup}}$. For their orbital fit they placed a Gaussian prior on the primary star based on their model mass of the primary of $1.043\pm0.012~M_{\odot}$. The companion mass found in \citet{2021AJ....162...12V} has a precision on their mass estimate of 1.2\%, but relies on a modeled  mass of the primary star, unlike the primary mass we  derive in this work. The mass of HD~92987~B determined by \citet{2021AJ....162...12V} of $268.4~\pm4.7M_{\rm{Jup}}$ is in good agreement with our determined mass of $276.6^{+17.8}_{-16.8}~M_{\mathrm{Jup}}$, with an expected increase on the error bar as we used an uninformed log-flat prior on the primary mass when doing the orbital fit, leading to a completely model independent derived mass of HD~92987~A and HD~92987~B.

A precise dynamical mass can be well constrained for HD~92987~B thanks to the well-sampled orbital phase coverage from the CORALIE survey, as shown in Fig.~\ref{fig:HD92987_orbits}. The relative astrometric point from the VLT/SPHERE observations is able to underpin the tight fit of the relative astrometric orbit shown in Fig.~\ref{fig:HD92987_orbits}, which could be further constrained with additional direct imaging observations. The fitted orbital parameters are listed in Table~\ref{tab:orbital_parameters}. The resulting corner plots from the probed parameter space from the MCMC fit with \texttt{orvara} and the posterior distributions are shown in Fig.~\ref{fig:HD92987_corner} in Appendix~\ref{appendix}.

The HD~92987 system has an almost face-on orbit with an inclination of $175.76\pm0.08$~degrees. This emphasizes the importance of not taking the mass derived from radial-velocities alone (which only gives  the minimum mass, $m\sin i$)  as the true mass. The companion nature of such detections are undetermined and could be a stellar or substellar companion masquerading as an exoplanetary candidate. This is the case for HD~92987~B, which joins an increasing list of substellar companions discovered using radial velocities alone that are in fact stellar in nature \citep[e.g.,][]{2007AJ....134..749B,2010A&A...509A.103S,2011A&A...528L...8S,2016A&A...588A.144W,2017A&A...602A..87M,2017AJ....153..258B,2019A&A...631A.125K}.

\subsection{HD~142234 (HIP~77931)} \label{sect:HD144234_orbit}

\begin{figure*}
    \centering
    \includegraphics[width=0.49\textwidth]{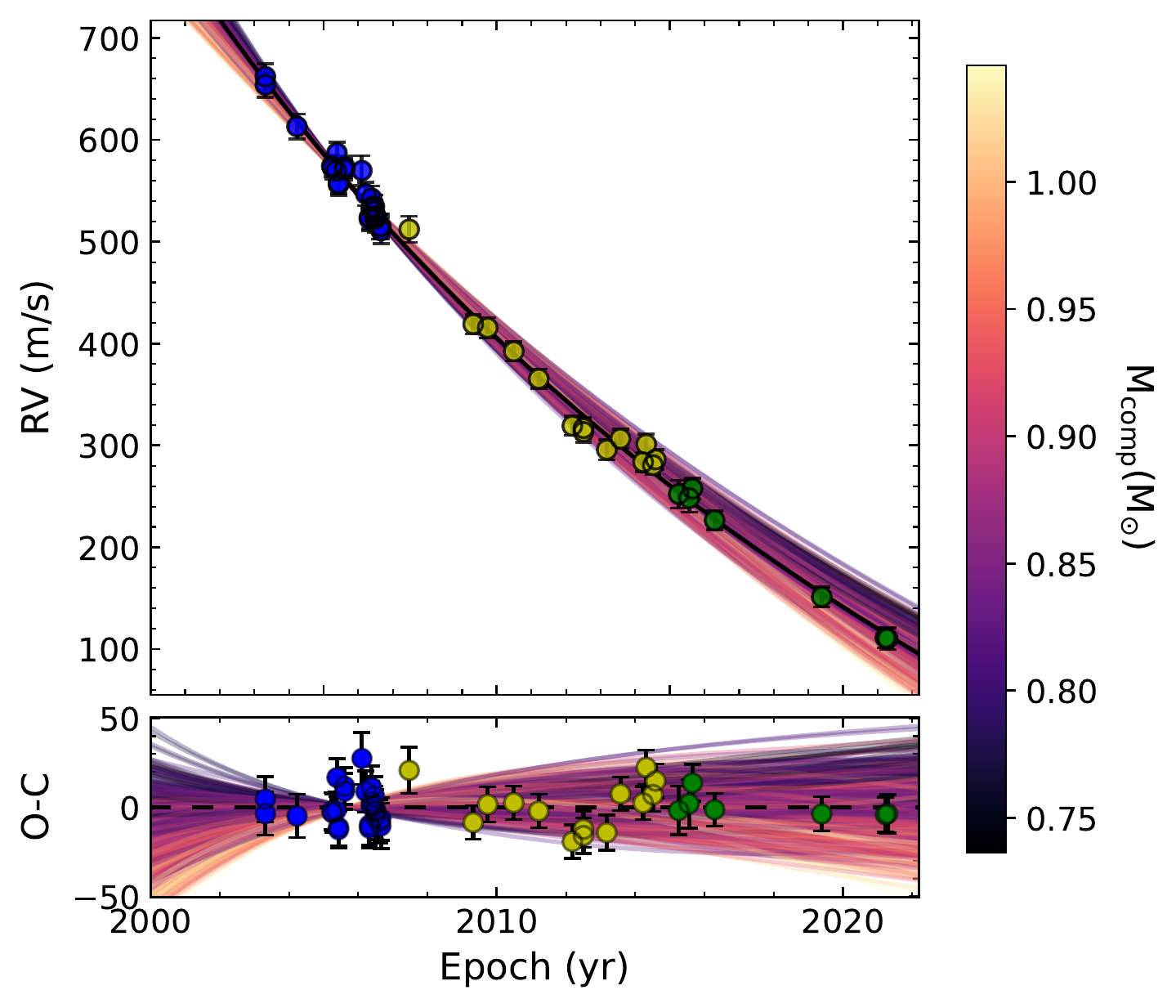}
    \includegraphics[width=0.49\textwidth]{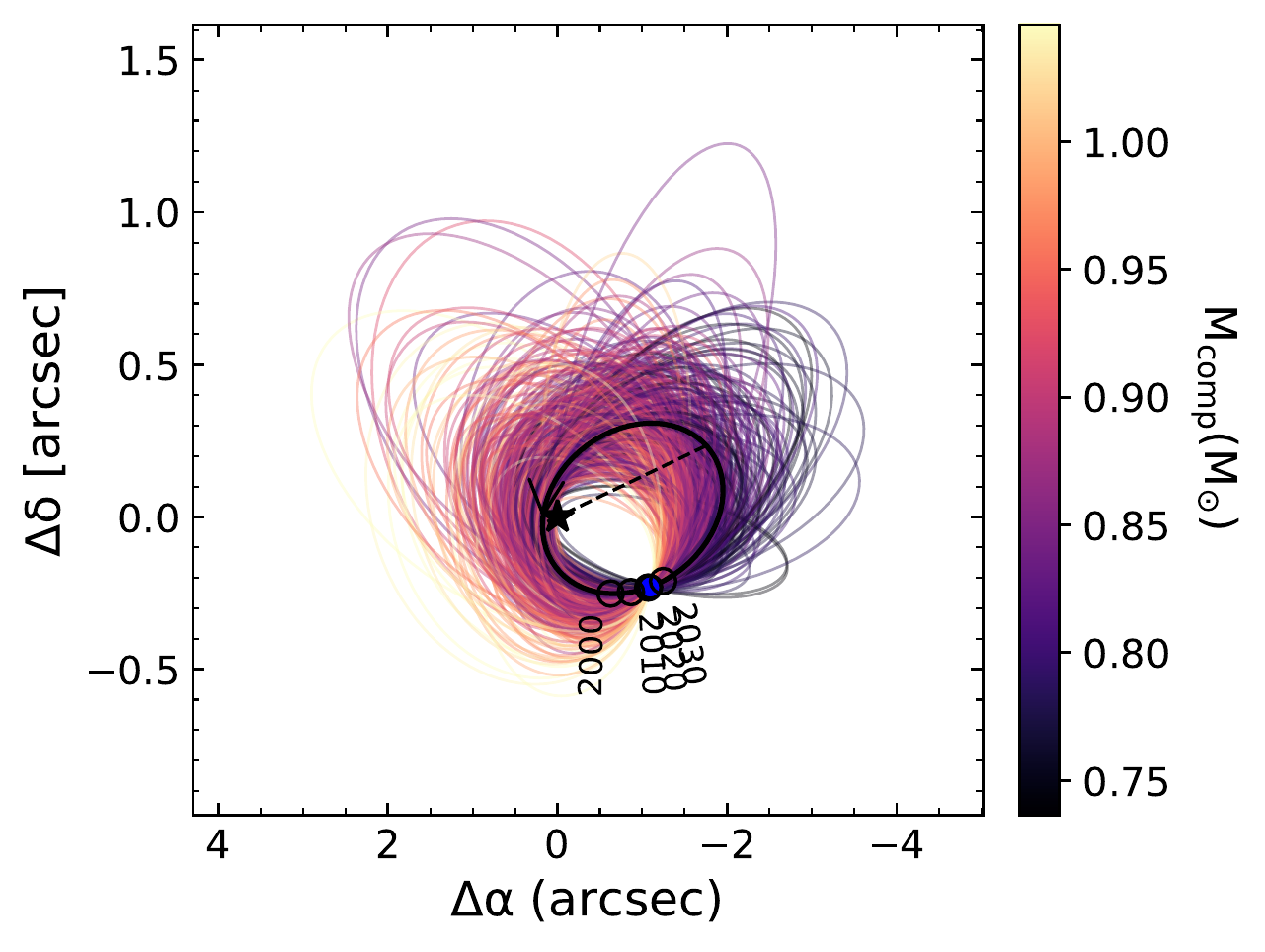}
    \includegraphics[width=\textwidth]{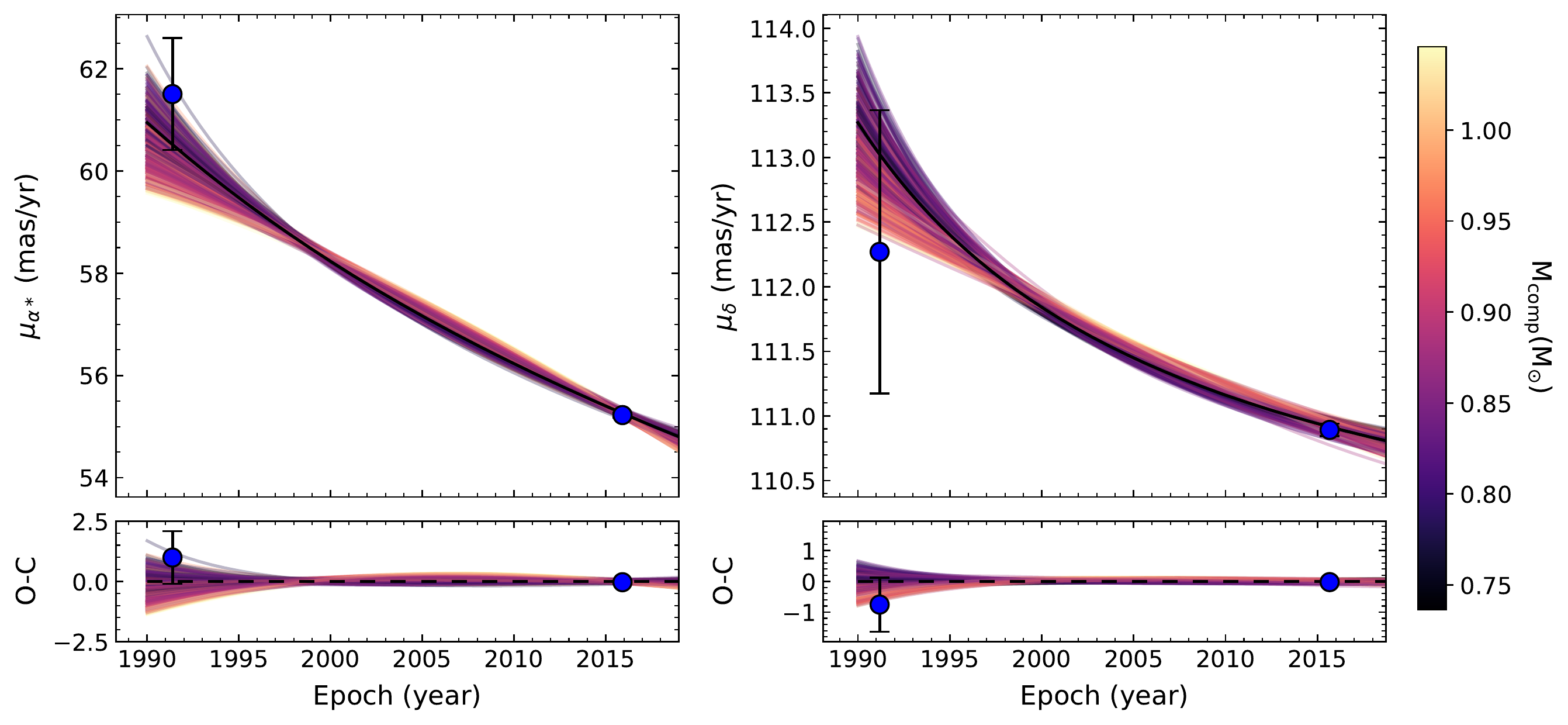}
    \caption{Same as Fig.~\ref{fig:HD92987_orbits}, but for HD~142234~B+C. Here the orbital fit is performed with HD~142234~B+C together.}
    \label{fig:HD142234_orbits}
\end{figure*}

HD~142234 is a G5V star that has been observed with CORALIE at La Silla Observatory since January 2003. Twenty-six measurements were taken with CORALIE-98, an additional 14 measurements were taken with CORALIE-07, and 7 measurements obtained with CORALIE-14.

HD~142234 was then observed with VLT/SPHERE program 105.20SZ.001 (PI: Rickman) on 2021 May 16, with a total observation time of 1 hour 50 minutes, with a 64-second integration time for the coronagraphic frames using IRDIS, as well as several sky, center, and flux frames taken which was necessary for the post-processing of the data. The imaging data from the VLT/SPHERE observations are shown in Fig.~\ref{fig:direct_images}.

From the VLT/SPHERE images we found HD~142234 to be a triple system. Given    the linear trend of the radial velocities (top left plot  in Fig.~\ref{fig:HD142234_orbits}) and the closeness of the B and C components, we conclude that this is a hierarchical triple system with BC in a tight orbit around each other, and a longer orbit around HD~142234~A. As \texttt{orvara} does not yet have the capability to perform orbital fits of this configuration, we performed an orbital fit with the photocenter of the relative astrometry between the two components as the barycenter to estimate the combined mass of HD~142234~B+C as if it were one point source, and then used the photometric ratio of the two resolved point sources to estimate the ratio of the masses of the individual components. 

Only   a small fraction of the orbital phase is sampled by the radial-velocity data, and so we implemented a Gaussian prior on the primary mass  derived using the Geneva stellar-evolution models \citep{2012A&A...537A.146E,2013A&A...558A.103G}, as described in Section~\ref{sec:stellar_params}. We used a value of $0.83\pm0.02~M_{\odot}$ for the primary (see Table~\ref{tab:stellar_params}). We ran \texttt{orvara} using 500,000 steps in each chain with the CORALIE radial-velocity data, the relative astrometry from VLT/SPHERE taken to be a single point as the photocenter between HD~142234~B and HD~142234~C, and the Hipparcos-Gaia astrometry. The resulting orbital fit is shown in Fig.~\ref{fig:HD142234_orbits}, with a burn-in phase of 700 multiplied by every  250th step that is saved on the chain with 500 randomly drawn orbits.

We note that the relative astrometric orbital fit is not as tightly constrained as the other systems in this paper. This is primarily due to the unknown true barycenter of HD~142234~B and HD~142234~C, leaving a large uncertainty on the relative astrometric position. We take the photocenter to be the point that is equidistant between HD~142234~B and HD~142234~C, giving the astrometry to be separation $\rho=1.117\pm0.003~''$ and position angle $\theta=258.13\pm0.23\degree$. Furthermore, we only sample a relatively uninformative portion of its RV orbit leading to an uncertainty on the orbital period. The orbital constraints provide a good first estimate of the companion masses, and in future versions of \texttt{orvara} with the updated capability to incorporate this orbital configuration, an updated orbital fit and comparison can be carried out.

From this orbital fit, we calculate a combined mass of HD~142334~B+C to be ${911.4}_{-63.9}^{+76.5}~M_{\mathrm{Jup}}$ with an orbital period of ${279}_{-62}^{+111}$~years. Using the photometric ratio of HD~142234~B to HD~144234~C from the calculated contrasts shown in Table~\ref{tab:astrometry}, we estimate the masses of each component from the combined orbital fit. In this way  we obtain an estimate of HD~142234~B and HD~142234~C to be $487.0^{+40.9}_{-34.1}~M_{\mathrm{Jup}}$ and $424.4^{+35.6}_{-29.8}~M_{\mathrm{Jup}}$, respectively.

The fitted orbital parameters are listed in Table~\ref{tab:orbital_parameters}. The resulting corner plots from the probed parameter space from the MCMC fit with \texttt{orvara} and the posterior distributions are shown in Fig.~\ref{fig:HD142234_corner} in Appendix~\ref{appendix}. 

As shown in Fig.~\ref{fig:HD142234_corner}, the eccentricity of HD~142234~B+C is not very well constrained, with the posterior distribution spanning  from the lower to the upper bounds, with a skew  toward a higher eccentricity, and a resulting fit of $e=0.73^{+0.18}_{-0.35}$. The large uncertainty is expected in this case, as the orbital phase coverage is not completely covered by the radial velocities. As seen in the top left panel of Fig.~\ref{fig:HD142234_orbits}, the radial velocities mostly  reveal a long-term linear trend, making it difficult to tightly constrain the orbital parameters. This is also due to  the orbital fit of HD~142234~B and HD~142234~C fitted as a single point source; there are large uncertainties in the true relative astrometry of the system. However, the orbital fit does provide good first constraints on this system.

\subsection{HD~143616 (HIP~78551)}

\begin{figure*}
    \centering
    \includegraphics[width=0.49\textwidth]{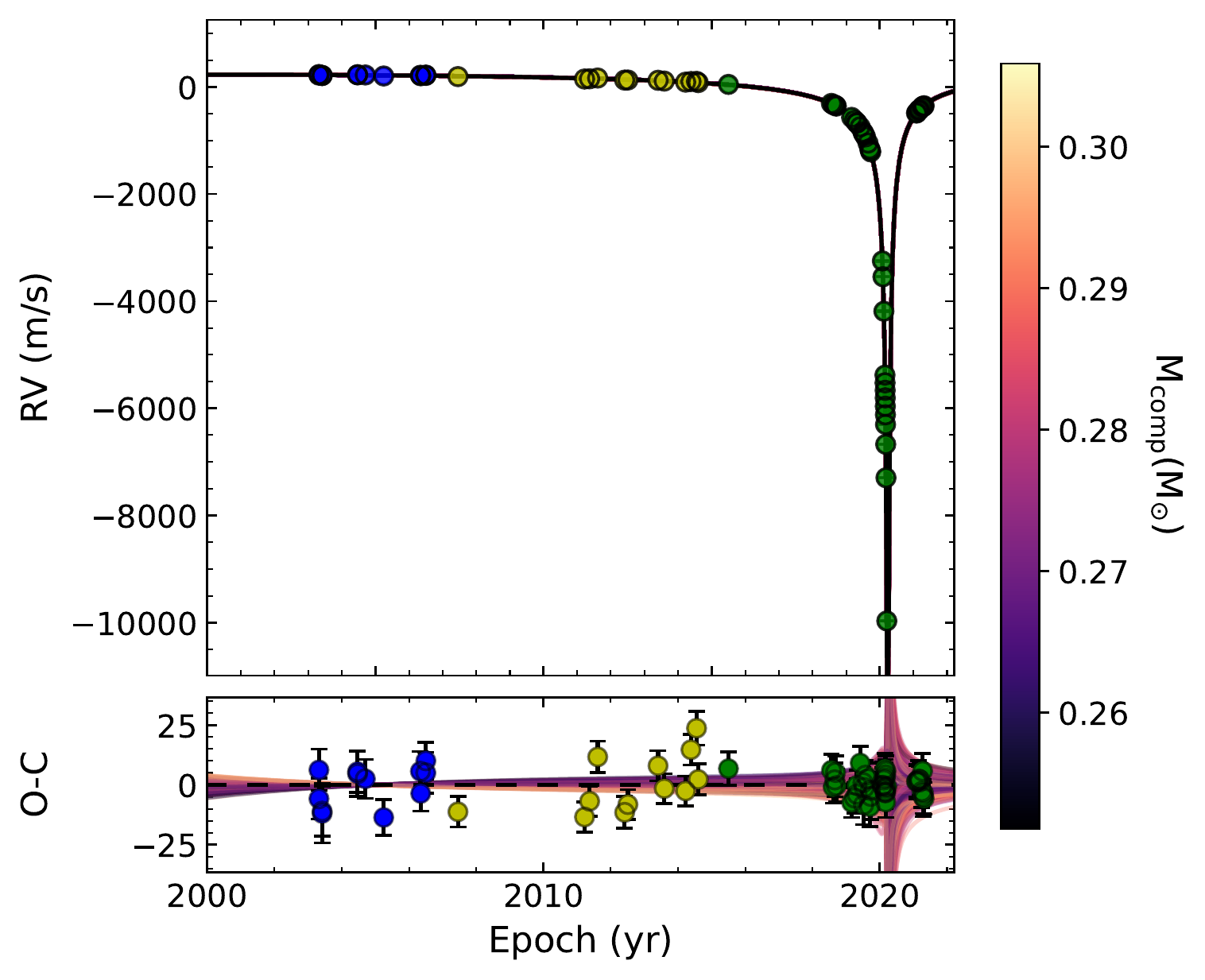}
    \includegraphics[width=0.49\textwidth]{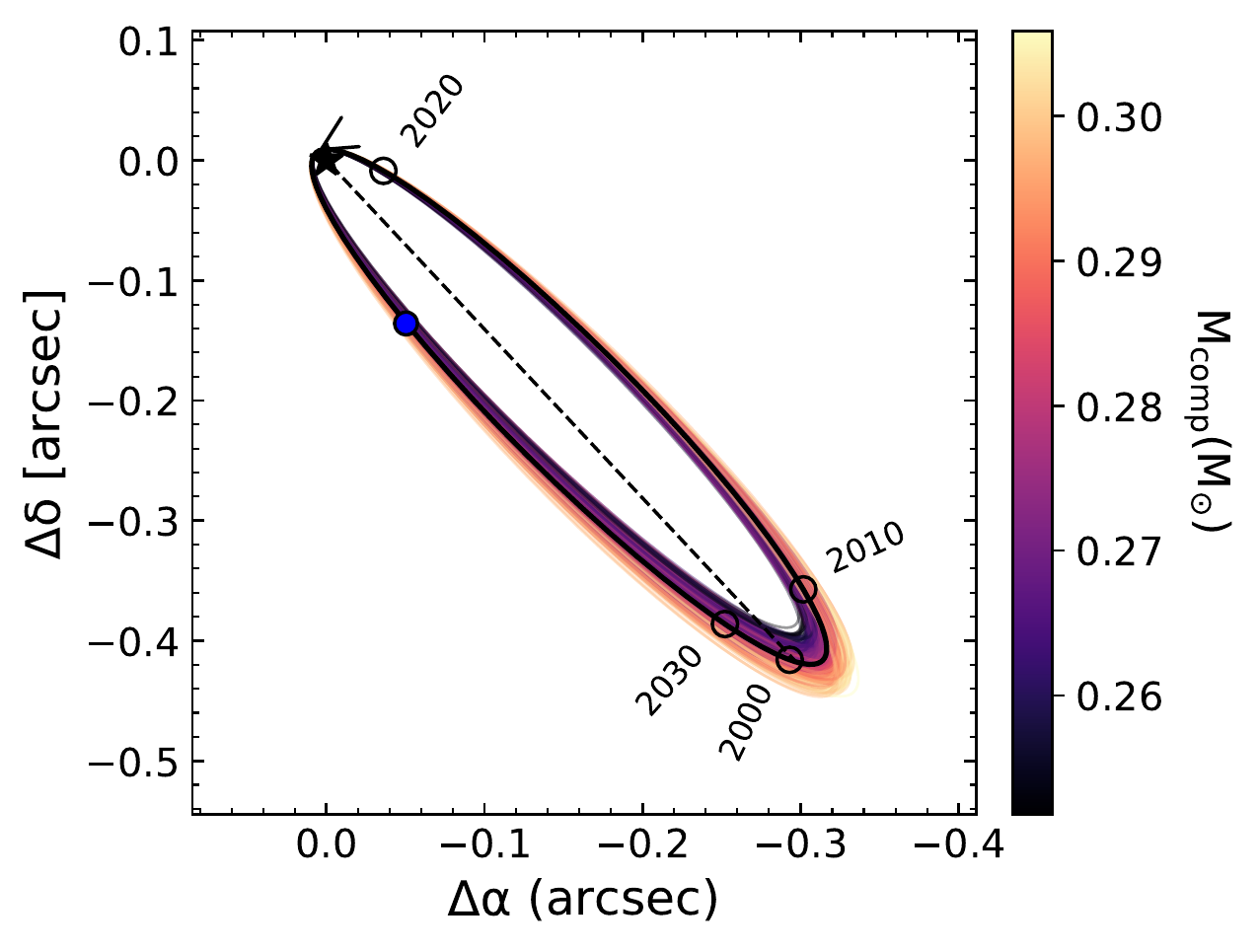}
    \includegraphics[width=\textwidth]{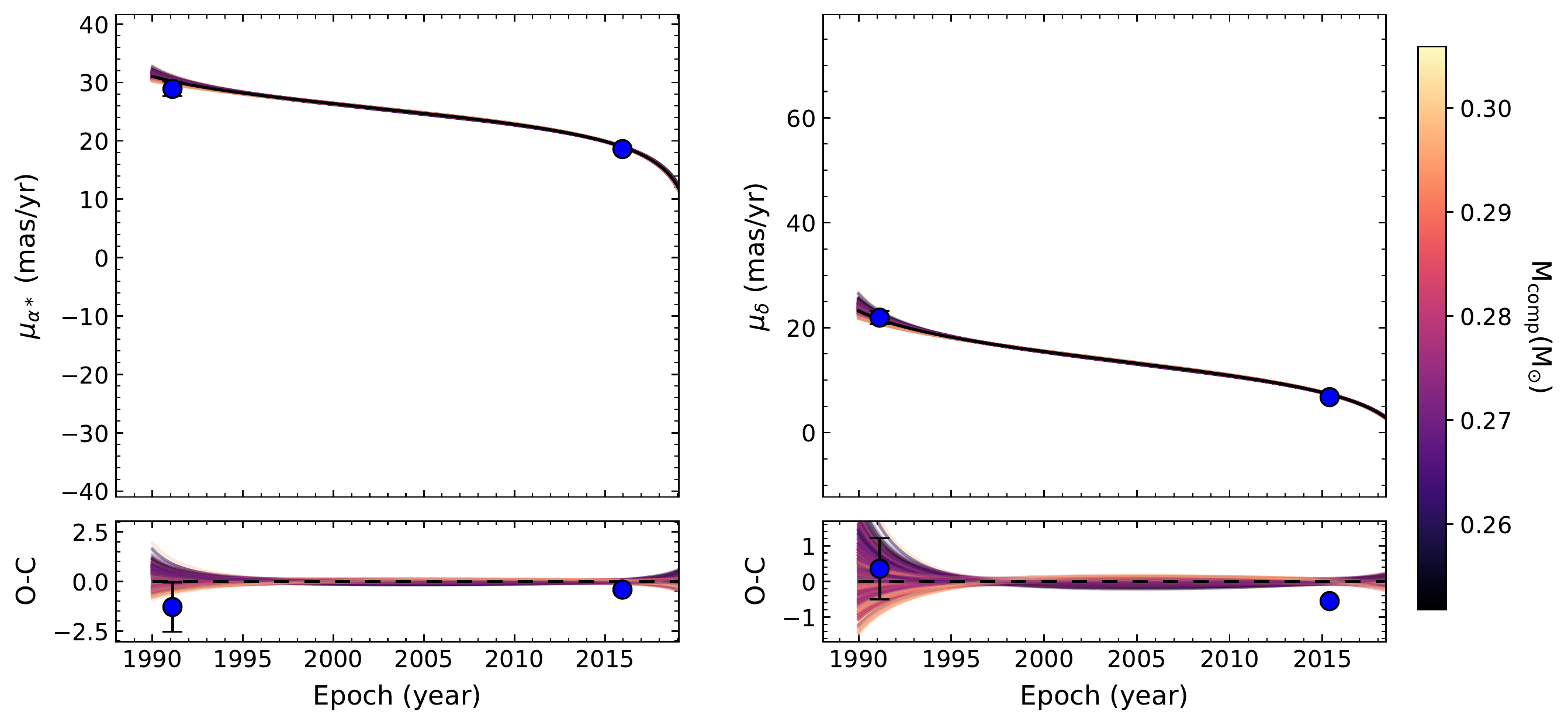}
    \caption{Same as Fig.~\ref{fig:HD92987_orbits}, but for HD~143616~B.}
    \label{fig:HD143616_orbits}
\end{figure*}

HD~143616 is a G6/8V star that has been observed with CORALIE at La Silla Observatory since April 2003. Twelve measurements were taken with CORALIE-98, an additional 12 measurements were taken with CORALIE-07, and 48 measurements were taken with CORALIE-14.

HD~143616 was then observed with VLT/SPHERE program 105.20SZ.001 (PI: Rickman) on 2021 July 17, with a total observation time of 40 minutes, with a 64-second integration time on the coronagraphic frames using IRDIS, as well as several sky, center, and flux frames taken which was necessary for the post-processing of the data. The direct detection of HD~143616~B is shown in Fig.~\ref{fig:direct_images}.

From the radial velocities alone, we already see that the orbit is extremely eccentric (see Fig.~\ref{fig:HD143616_orbits}). Due to its extreme eccentricity, we implemented a starting position for the orbital fit, which we determined from a fit on the radial-velocity data alone.

To perform an initial fit on just the radial-velocity data, we used the Data and Analysis Center for Exoplanets \citep[DACE;][]{2019ASPC..521..757B}\footnote{The Data and Analysis Center for Exoplanets (DACE) can be accessed at \url{https://dace.unige.ch}}. The DACE online tool fits a Keplerian model using the formalism described in \citet{2016A&A...590A.134D}. The modeling of instrumental noise and stellar jitter is described in \citet{2016A&A...585A.134D} and \citet{2018A&A...614A.133D}. We used the parameters derived for the  semimajor axis ($a$), $\sqrt{e}\sin\omega$, and $\sqrt{e}\cos\omega$ as the start points for the orbital fit with \texttt{orvara}, which can be implemented using a start file. The start points we implemented from the DACE radial-velocity fit are  $9.0\pm0.143$~AU for the semimajor axis $a$;  $-0.036460 \pm 0.000330$ for $\sqrt{e}\sin \omega$; and $-0.9790787 \pm  0.0000866$ for  $\sqrt{e}\cos \omega$. 

The choice of these start points was motivated by  the well-sampled periastron passage with the radial-velocity data and the extremity of the eccentricity of the system. Using these start points we fit the orbit combining all of the CORALIE radial-velocity data, with the relative astrometry from the VLT/SPHERE imaging, as well as the proper motion anomalies from the HGCA. We ran \texttt{orvara} with 500,000 steps in each chain, and for the orbital plots shown in Fig.~\ref{fig:HD143616_orbits} we defined a burn-in phase of 700 multiplied by every 250th step that is saved on the chain, with the best fit orbit shown, with 500 orbits randomly selected from the posterior distribution plotted alongside the RVs, and proper motions from \emph{Hipparcos} and \emph{Gaia}. From the combined orbital fit, we derive a dynamical mass of HD~143616~B of $291.2^{+12.6}_{-11.5}~M_{\mathrm{Jup}}$, an orbital period of $33.52^{+0.85}_{-0.70}$~years, and an extreme eccentricity of $e=0.9655^{+0.00058}_{-0.00050}$.

\begin{figure*}
    \centering
    \includegraphics[width=0.49\textwidth]{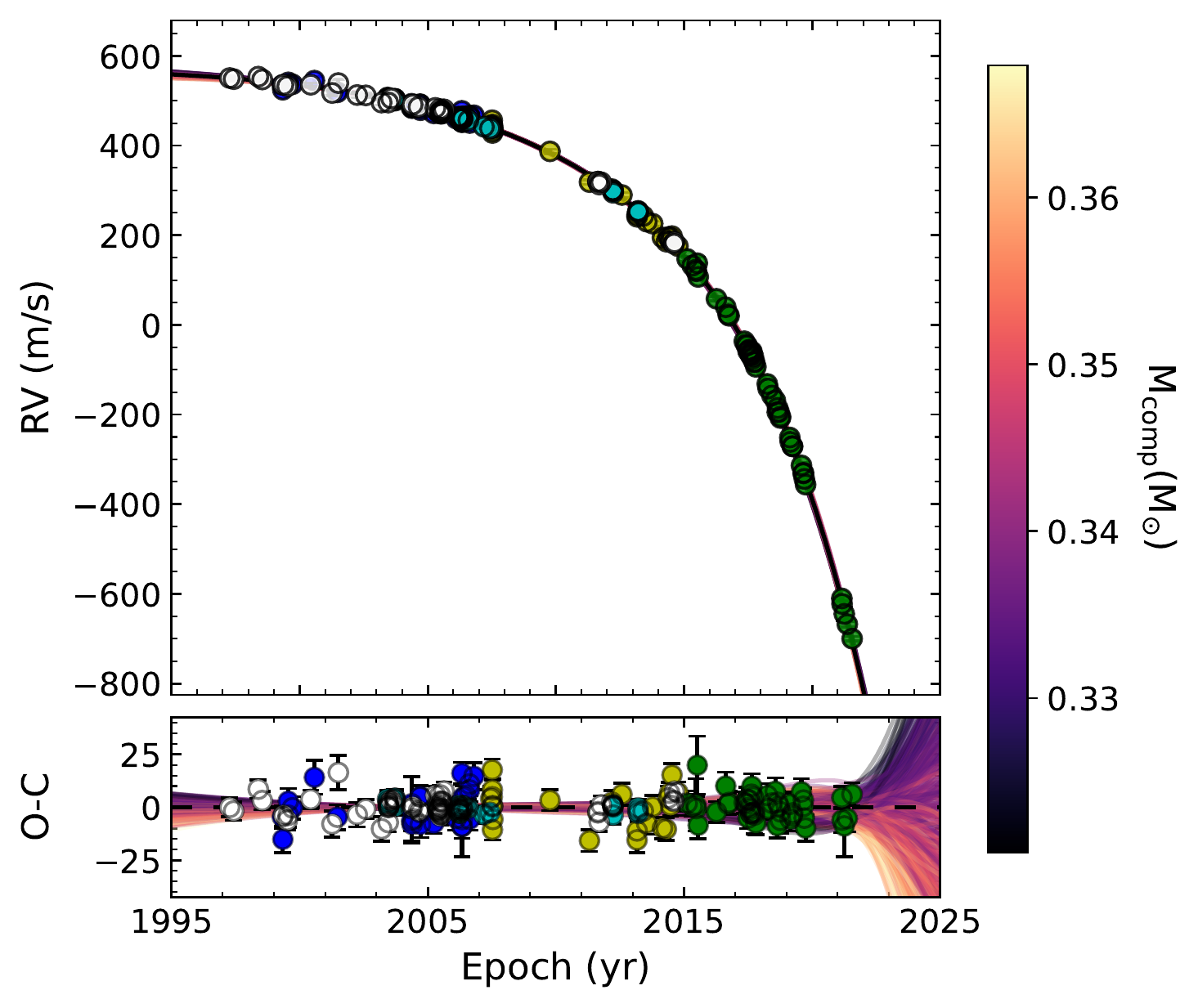}
    \includegraphics[width=0.49\textwidth]{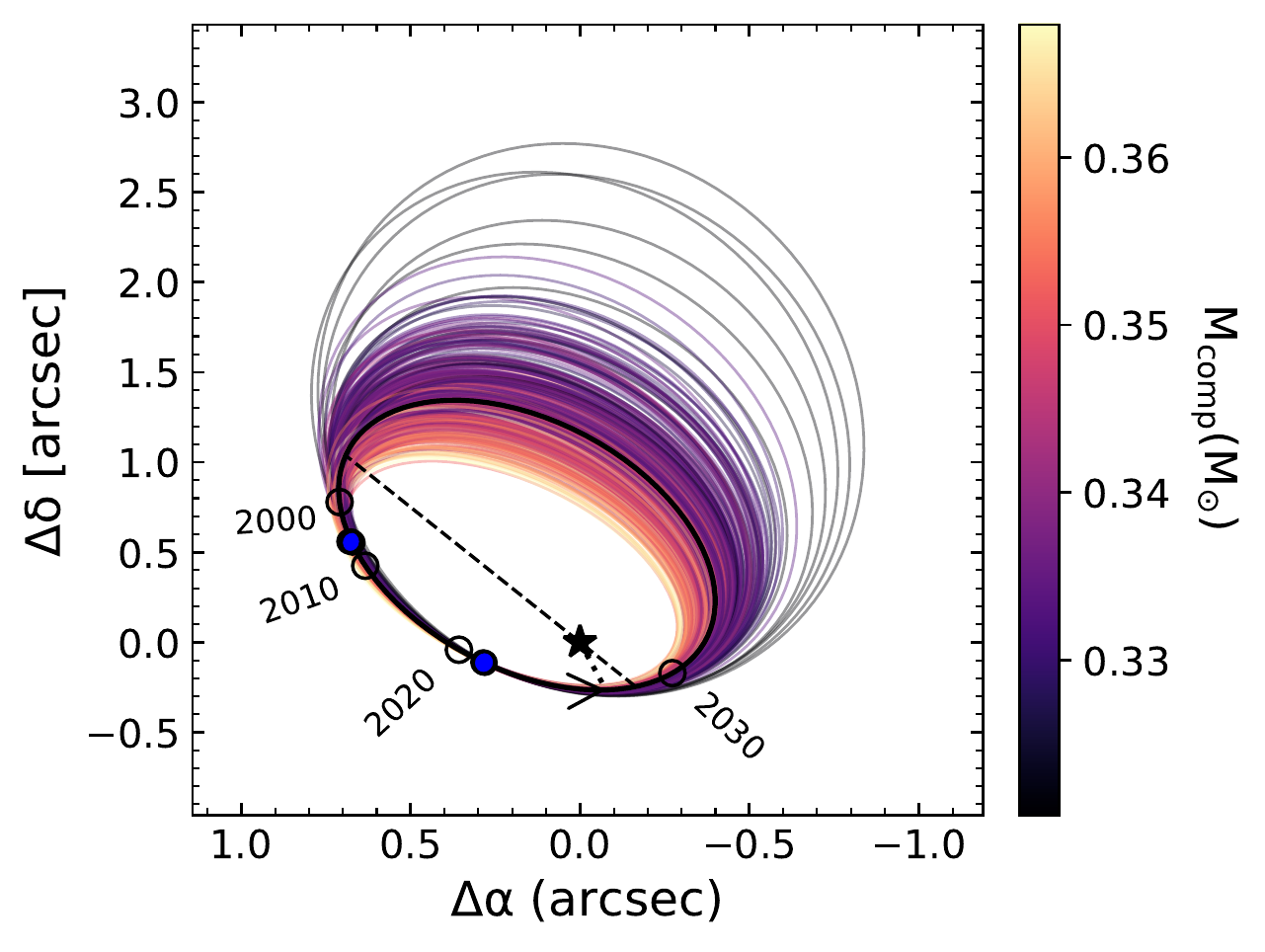}
    \includegraphics[width=\textwidth]{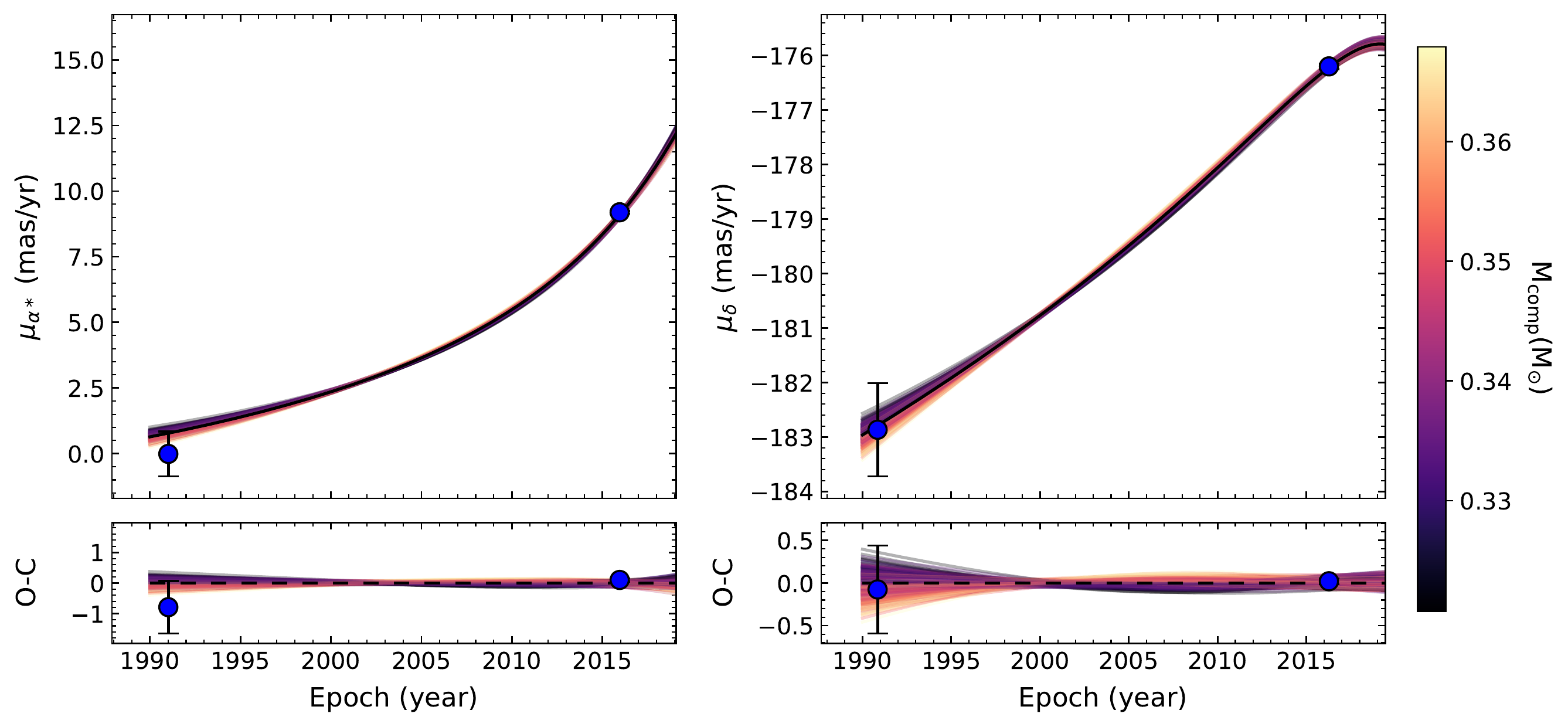}
    \caption{Same as Fig.~\ref{fig:HD92987_orbits}, but for HD~157338~B. \emph{Top left.} Radial-velocity orbit induced by HD~157338~B. The white data points show the HIRES data, the blue data points show the COR-98 data, the cyan data points show the HARPS data, the yellow points show the COR-07 data, and the green points show the COR-14 data.
    \emph{Top right.} Predicted past and future relative astrometric points are shown as black circles with their respective years, while the observed relative astrometric points from the 2006 VLT/NACO data and the 2021 VLT/SPHERE data are shown as blue filled data points, where the measurement error is smaller than the plotted symbol.}
    \label{fig:HD157338_orbits}
\end{figure*}

\begin{figure*}
\centering
\includegraphics[width=0.48\textwidth]{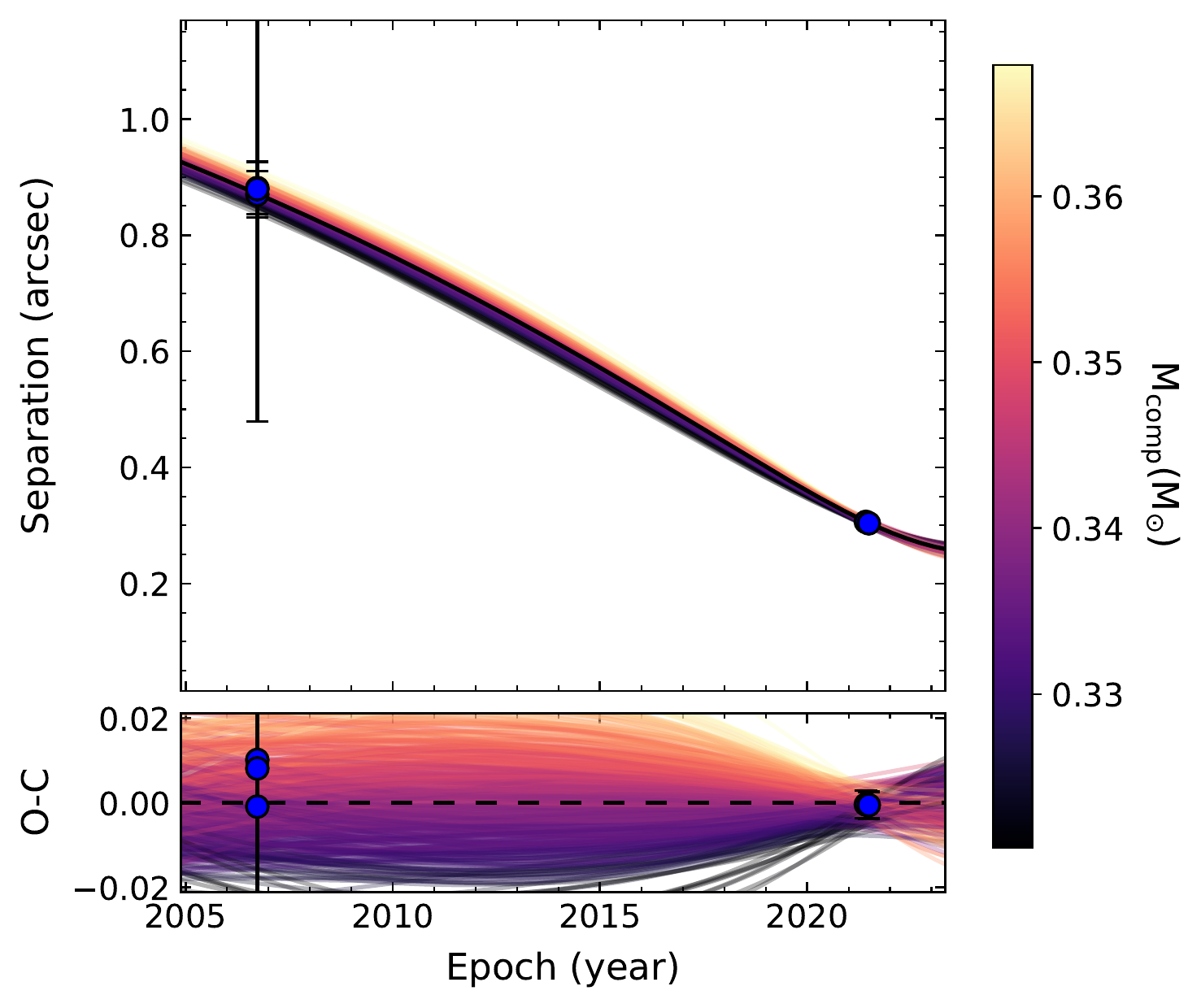}
\includegraphics[width=0.48\textwidth]{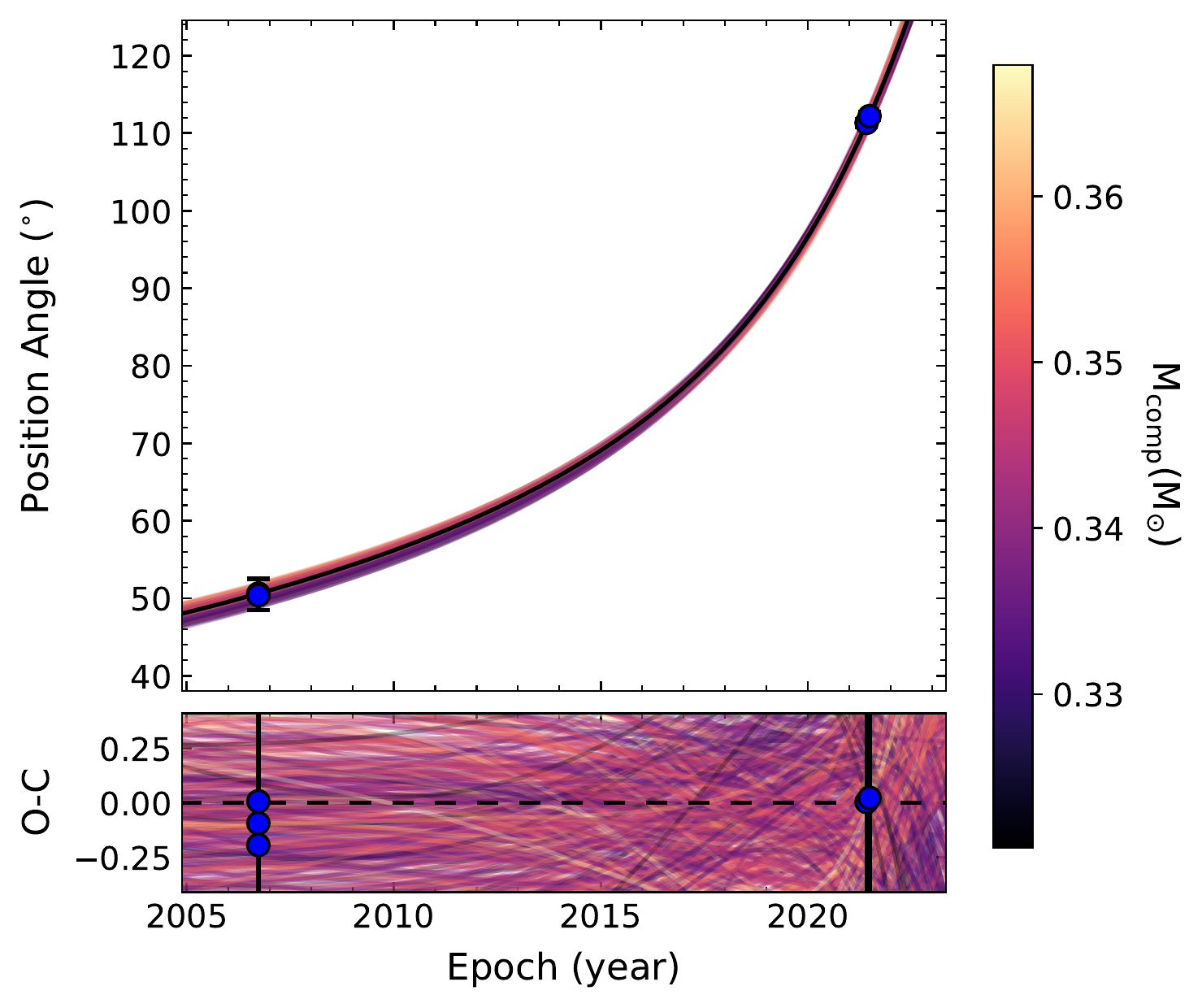}
\caption{Relative separation and position angle of HD~157338~B across three epochs of data. The 2006 data corresponds to VLT/NACO data as reported by \citet{montagnier:tel-00714874}, and the two 2021 data points correspond to the VLT/SPHERE data   presented in this paper. The second epoch of direct imaging observations over a long baseline is able to constrain the orbital parameters significantly, as shown in the residual plots in the corresponding bottom panels.}
\label{fig:HD157338_relative_astrometry}
\end{figure*}

We placed an uninformed log-flat prior on the primary mass when doing the orbital fit, leading to a completely model-independent derived mass of HD~143616~A and HD~143616~B. From the orbital fit, we find the dynamical mass of the primary star to be $0.935\pm0.017~M_{\odot}$, which is in good agreement with the mass derived from the Geneva stellar-evolution models \citep{2012A&A...537A.146E,2013A&A...558A.103G}, as described in Section~\ref{sec:stellar_params}, where we find a value of $0.92\pm0.02~M_{\odot}$, as listed in Table~\ref{tab:stellar_params}.

The fitted orbital parameters are listed in Table~\ref{tab:orbital_parameters}. The resulting corner plots from the probed parameter space from the MCMC fit with \texttt{orvara} and the posterior distributions are shown in Fig.~\ref{fig:HD143616_corner} in Appendix~\ref{appendix}.

\subsection{HD~157338 (HIP~85158)}

HD~157338 is a G0/1V star that has been observed with CORALIE at La Silla Observatory since April 1999. Thirty-one measurements were taken with CORALIE-98, an additional 26 were taken with CORALIE-07, and 53 points with HARPS \citep{2003Msngr.114...20M}. We also include 33 radial-velocity measurements from HIRES \citep{2017AJ....153..208B} that date back to April 1997, increasing the time span of the CORALIE survey sampling by an additional two years.

HD~157338 was first directly imaged by \citet{montagnier:tel-00714874} with VLT/NACO \citep{2003SPIE.4841..944L,2003SPIE.4839..140R} in 2006 across the F1 ($\lambda=1.575\mu$m), F2 ($\lambda=1.600\mu$m), and F3  ($\lambda=1.625\mu$m) bands. From those observations, \citet{montagnier:tel-00714874} reported the detection of HD~157338~B. We used these archival NACO data in our orbital fit, as shown in Fig.~\ref{fig:HD157338_orbits} and Fig.~\ref{fig:HD157338_relative_astrometry}, which helped further refine our derived orbital parameters and dynamical mass.

HD~157338 was then observed with VLT/SPHERE program 105.20SZ.001 (PI: Rickman) on 2021 June 03, with a total observation time of 30 minutes, with a 32-second integration time on the coronagraphic frames using IRDIS, as well as several  sky, center, and flux frames taken which was necessary for the post-processing of the data. A follow-up observation of HD~157338 was taken on 2021 June 30, providing an additional relative astrometric point for our orbital fit. The direct detection of HD~157338~B is shown in Fig.~\ref{fig:direct_images}.

We ran \texttt{orvara} for the orbital fit with 500,000 steps in each chain, using the radial-velocity data from CORALIE, HARPS, and HIRES, along with relative astrometry from VLT/NACO and VLT/SPHERE, and astrometric accelerations from the HGCA. The resulting orbital fits are shown in Fig.~\ref{fig:HD157338_orbits}, which are plotted  with a burn-in phase of 700 multiplied by every 250th step that is saved on the chain with 500 orbits randomly selected from the posterior distribution plotted alongside the RVs, and proper motions from \emph{Hipparcos} and \emph{Gaia}. From the orbital fit we derive a companion mass of $359.3\pm10.5~M_{\mathrm{Jup}}$, and an orbital period of $126^{+34}_{-21}$~years.

The long baseline between direct imaging observations means, despite only having a fraction of the orbital phase covered with radial velocities, that the orbit is well-constrained. The significance of the additional imaging data with VLT/SPHERE is shown in Fig.~\ref{fig:HD157338_relative_astrometry}, where the long baseline between direct imaging observations and the gain in the precision of the relative astrometry contribute greatly to constraining the orbital parameters.

\citet{montagnier:tel-00714874} derived a mass of $0.405\pm0.03~M_{\odot}$ ($424.3\pm31.4~M_{\rm{Jup}}$) for HD~157338~B. The mass we derive using \texttt{orvara} is less massive at $359.3\pm10.5~M_{\mathrm{Jup}}$, but as seen in Fig.~\ref{fig:HD157338_relative_astrometry} and from an additional 16 years  of radial-velocity data, and proper motion data from the HGCA, we are able to determine a much more precise updated dynamical mass value.

We placed an uninformed log-flat prior on the primary mass when doing the orbital fit, leading to a completely model independent derived mass of HD~157338~A and HD~157338~B. From the orbital fit, we find the dynamical mass of the primary star to be $1.07^{+0.16}_{-0.15}~M_{\odot}$, which is in agreement with the primary mass derived using the Geneva stellar-evolution models \citep{2012A&A...537A.146E,2013A&A...558A.103G}, as described in Section~\ref{sec:stellar_params} and  listed in Table~\ref{tab:stellar_params}.

The greatest uncertainty in the orbital fit comes from the lack of full orbital phase coverage from the radial-velocity sampling due to the long orbital period ($>100$~years). However, because the system will approach its periastron passage in late 2024, the change in the orbital motion will be at its greatest, which is an ideal time to gain additional radial velocity or direct imaging astrometry. This would place even tighter constraints on an already relatively well-constrained orbit.

The fitted orbital parameters are listed in Table~\ref{tab:orbital_parameters}. The resulting corner plots from the probed parameter space from the MCMC fit with \texttt{orvara} and the posterior distributions are shown in Fig.~\ref{fig:HD157338_corner} in Appendix~\ref{appendix}.

\subsection{HD~195010 (HIP~101145)}

HD~195010 is a G8/K0V star that has been observed with CORALIE at La Silla Observatory since September 2000. Forty-two measurements were taken with CORALIE-98, 15 were taken with CORALIE-07, and 51 with CORALIE-14.

\begin{figure*}
    \centering
    \includegraphics[width=0.49\textwidth]{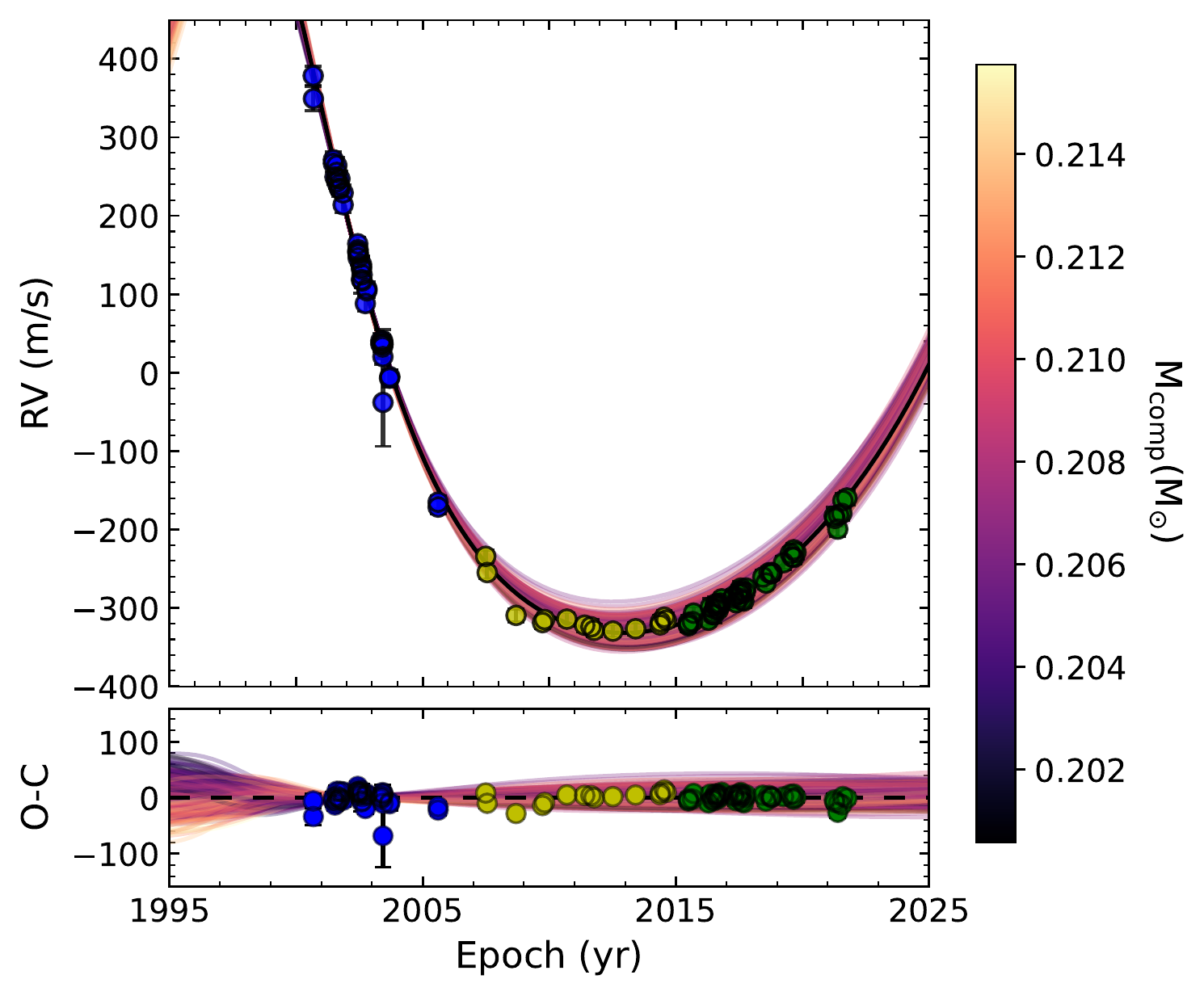}
    \includegraphics[width=0.49\textwidth]{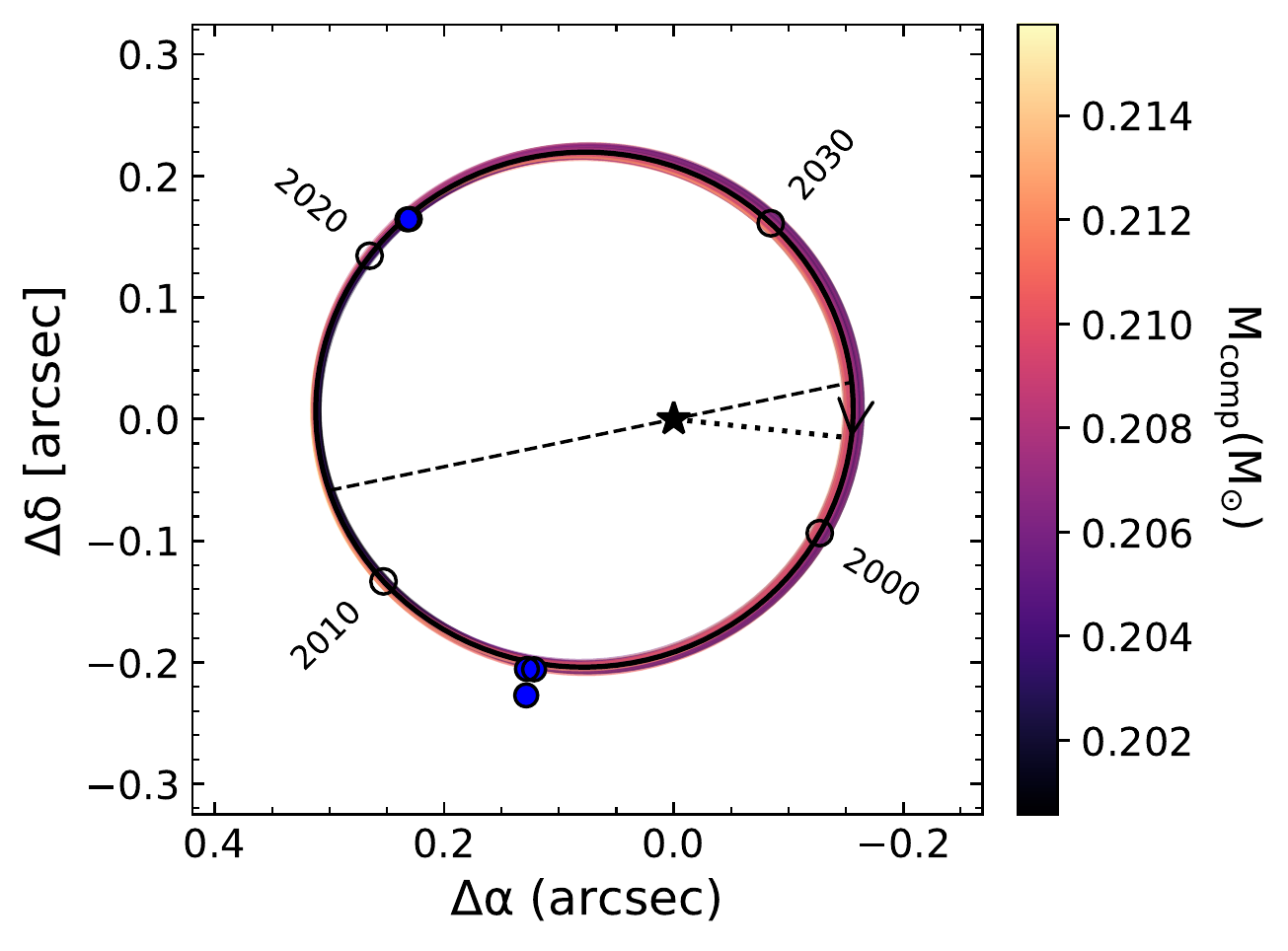}
    \includegraphics[width=\textwidth]{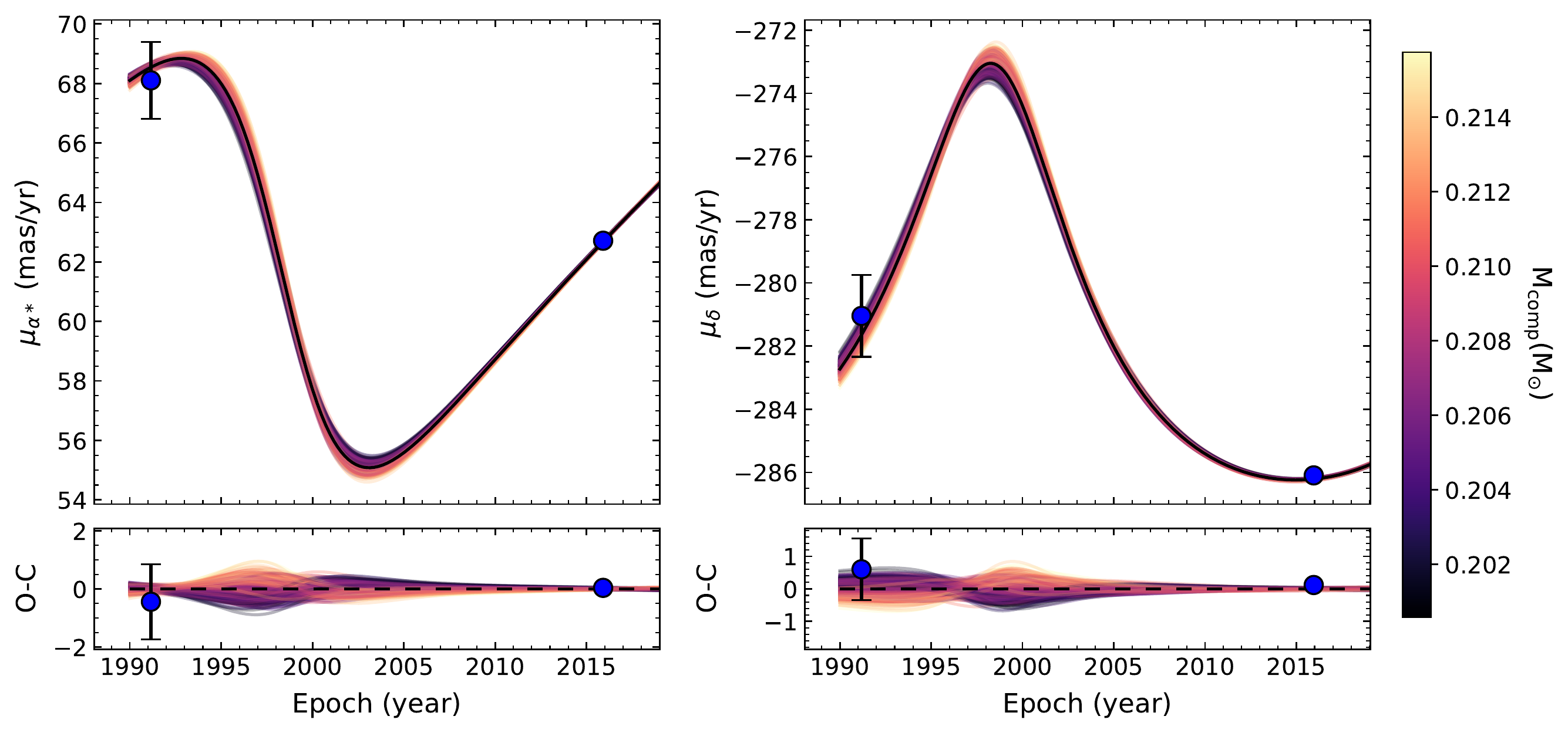}
    \caption{Orbit fits of HD~195010 using the orbit-fiting code \texttt{orvara.} \emph{Top left.} Same as Fig.~\ref{fig:HD92987_orbits}, but for HD~195010~B.
    \emph{Top right.} Predicted past and future relative astrometric points are shown as black circles with their respective years, while the observed relative astrometric points from the 2006 VLT/NACO and the 2021 VLT/SPHERE data is shown as   blue filled data points, where the measurement error is smaller than the plotted symbol.}
    \label{fig:HD195010_orbits}
\end{figure*}

\begin{figure*}
\centering
\includegraphics[width=0.48\textwidth]{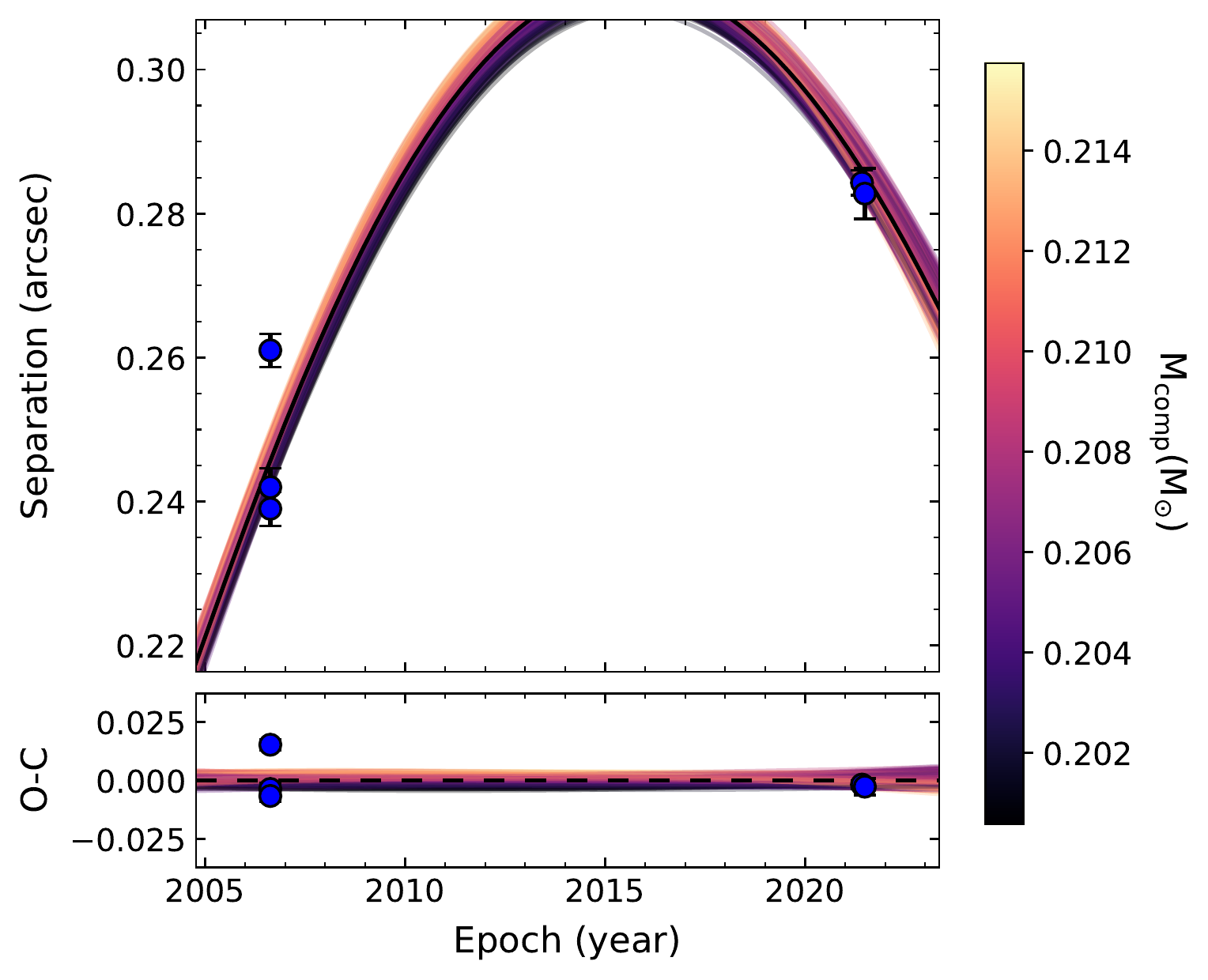}
\includegraphics[width=0.48\textwidth]{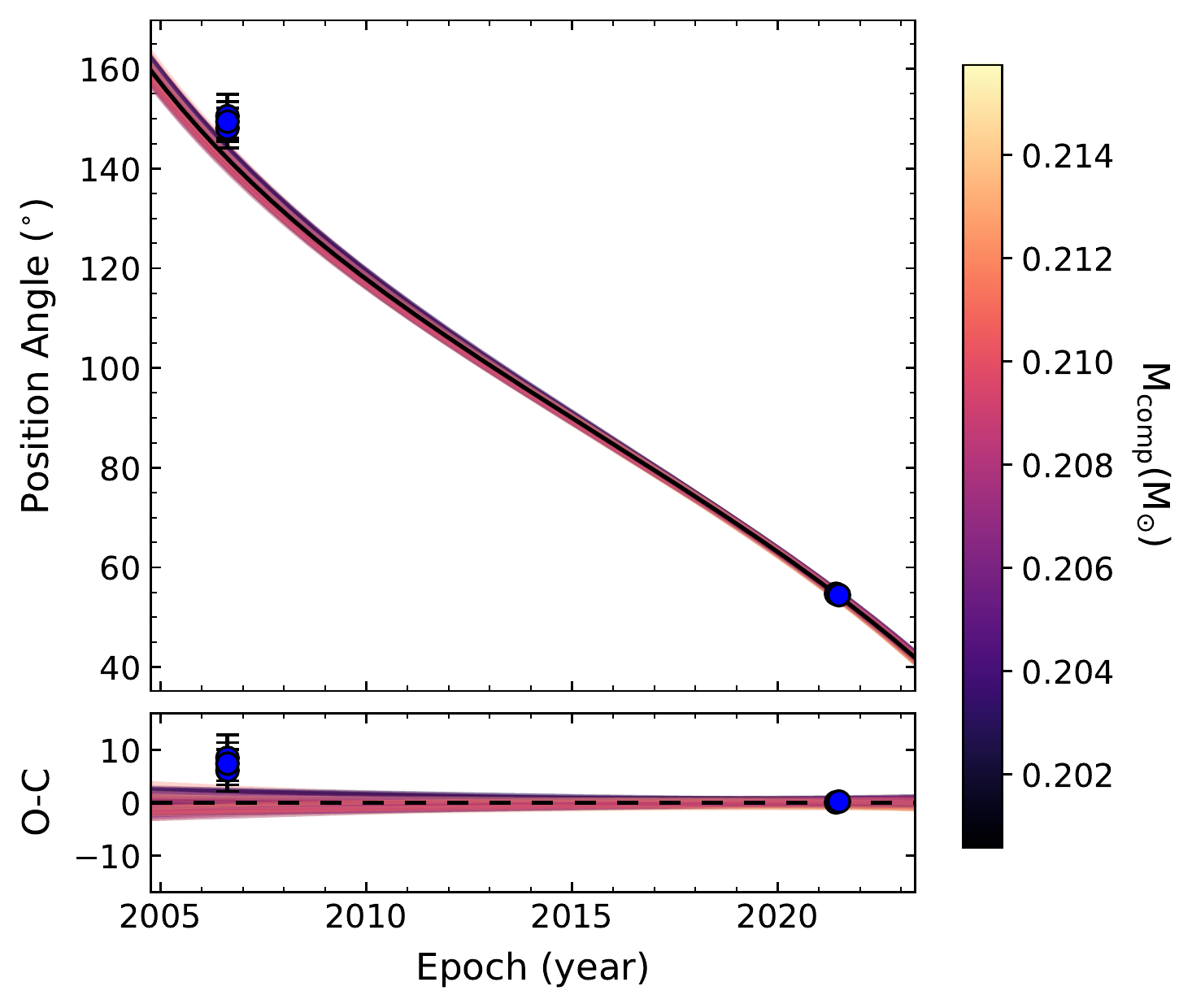}
\caption{Relative separation and position angle of HD~195010~B across three epochs of data. The 2006 data corresponds to the  VLT/NACO data  reported by \citet{montagnier:tel-00714874}, and the 2021 data point correspond to the VLT/SPHERE data  presented in this paper. The residuals are shown in the bottom two panels.}
\label{fig:HD195010_relative_astrometry}
\end{figure*}

HD~195010 was directly imaged by \citet{montagnier:tel-00714874} with VLT/NACO \citep{2003SPIE.4841..944L,2003SPIE.4839..140R} in 2006 across the F1 ($\lambda=1.575\mu$m), F2 ($\lambda=1.600\mu$m), and F3  ($\lambda=1.625\mu$m) bands. From those observations, \citet{montagnier:tel-00714874} reported the detection of HD~195010~B. We used these archival NACO data in our orbital fit, as shown in Fig.~\ref{fig:HD195010_orbits} and Fig.~\ref{fig:HD195010_relative_astrometry}, to help further refine our derived orbital parameters and dynamical mass.

We observed HD~195010 with VLT/SPHERE program 105.20SZ.001 (PI: Rickman) on 2021 June 03 and 2021 June 30, with an observation time of 30 minutes for each epoch, with a 32-second integration time on the coronagraphic frames using IRDIS, as well as several sky, center, and flux frames taken which was necessary for the post-processing of the data. The direct detection of HD~195010~B from the VLT/SPHERE observations is shown in Fig.~\ref{fig:direct_images}.

We ran the orbit-fitting code \texttt{orvara} with 500,000 steps in each chain using the radial-velocity data from the CORALIE survey, the relative astrometry from high-contrast imaging with NACO in 2006, as well as the relative astrometry from high-contrast imaging with VLT/SPHERE in 2021, and the astrometric accelerations from the Hipparcos-Gaia catalog of accelerations \citep[HGCA;][]{2021ApJS..254...42B}.

\begin{figure*}
    \centering
    \includegraphics[width=0.49\textwidth]{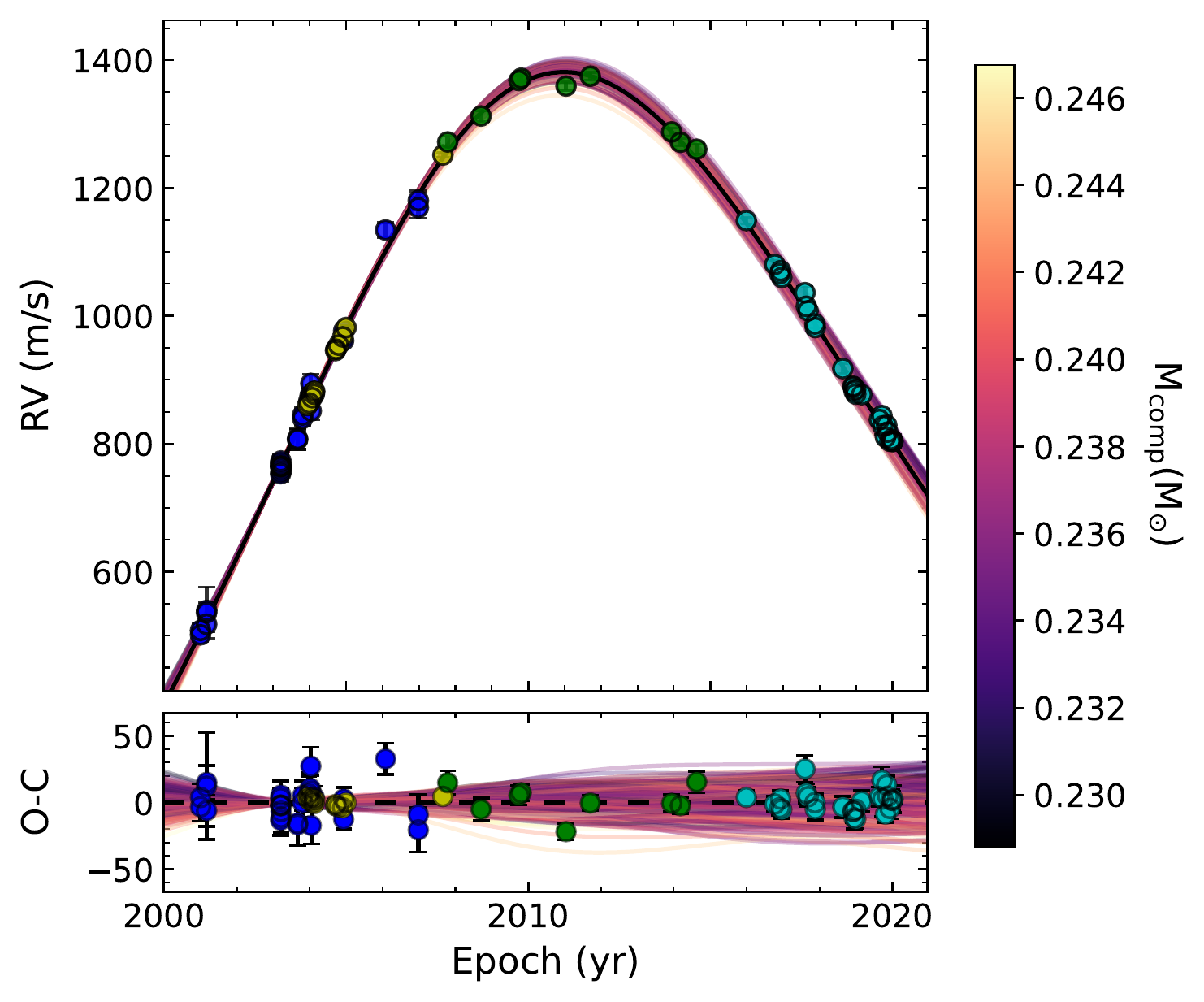}
    \includegraphics[width=0.49\textwidth]{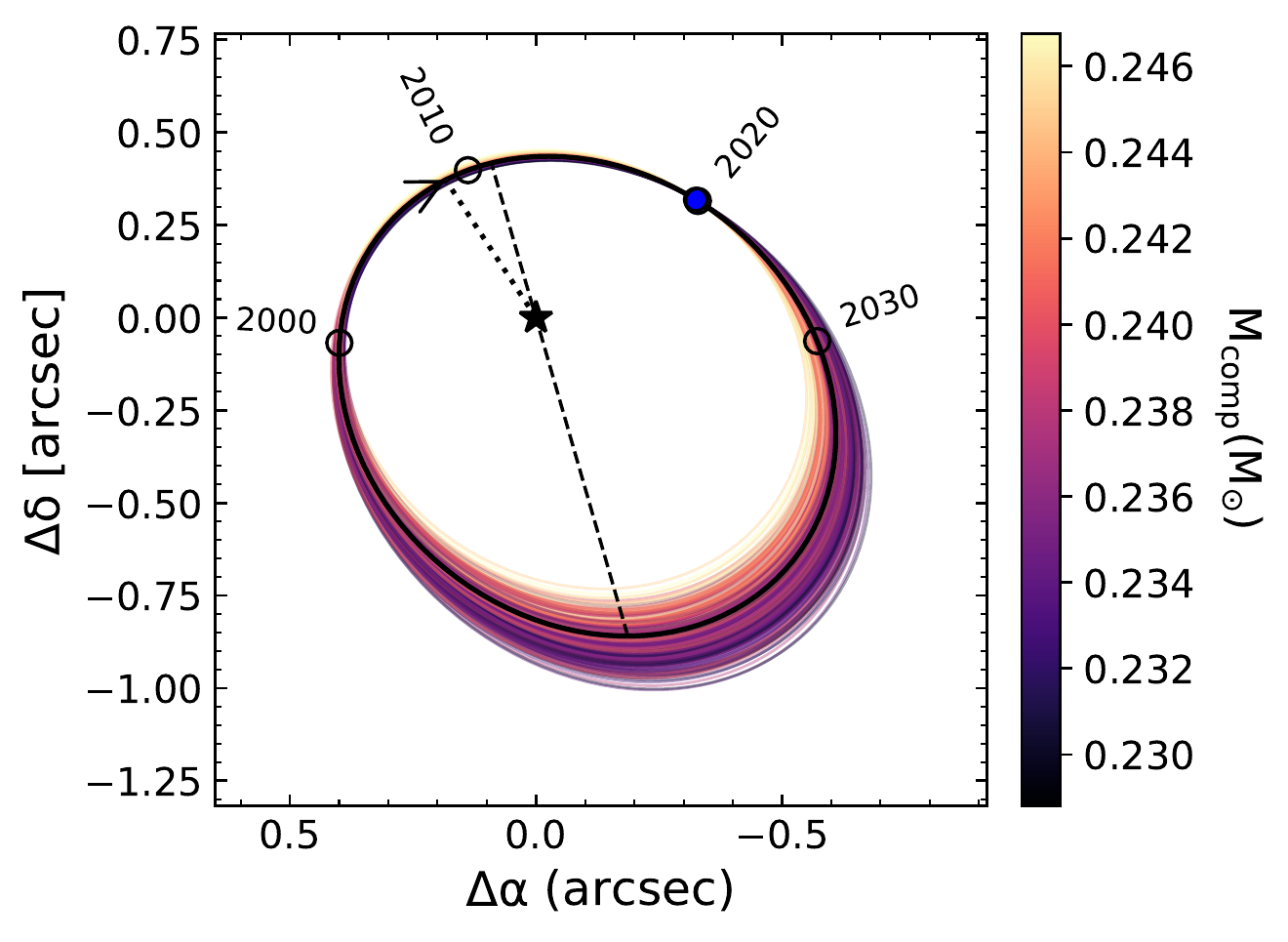}
    \includegraphics[width=\textwidth]{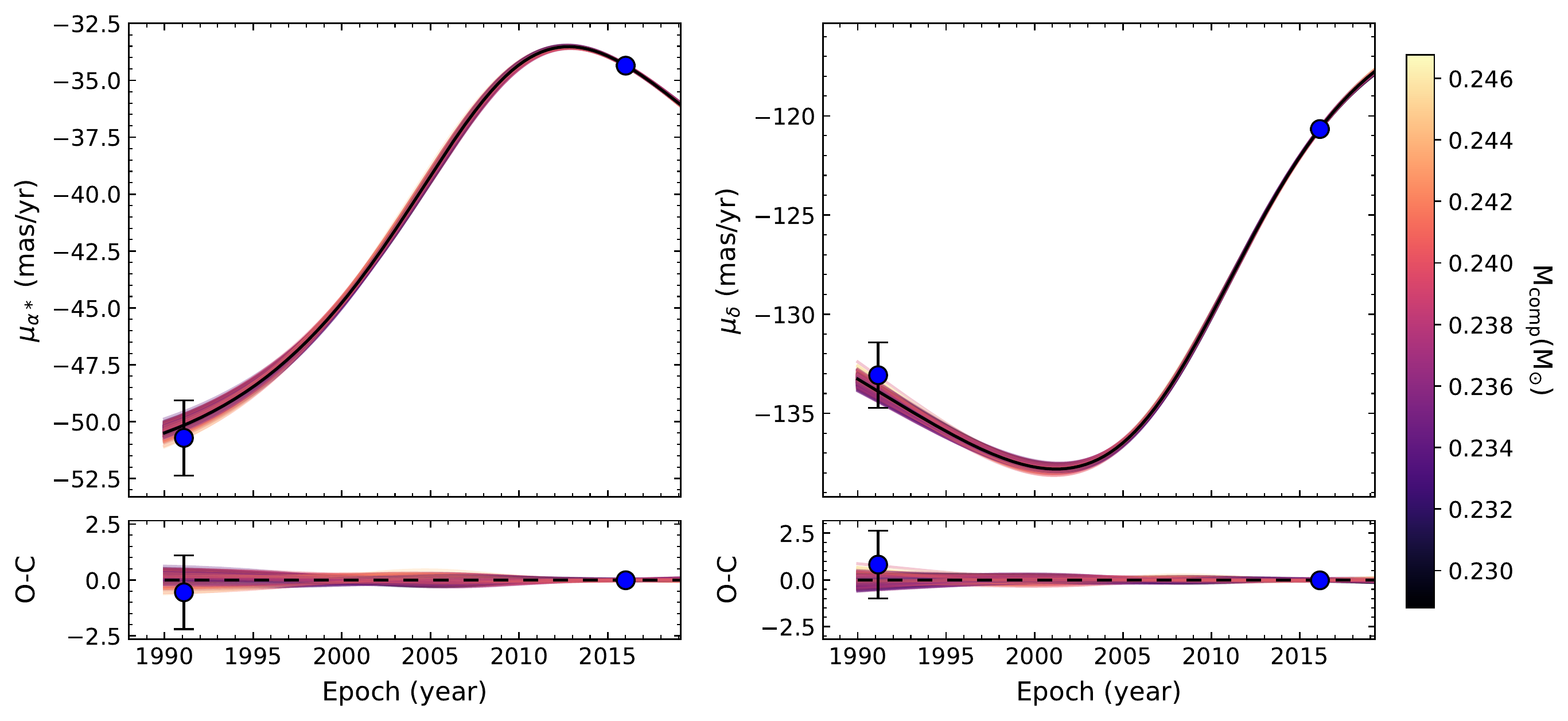}
    \caption{Same as Fig.~\ref{fig:HD92987_orbits}, but for HIP~22059~B. \emph{Top left.} Radial-velocity orbit induced by HIP~22059~B. The blue data points show the COR-98 data, the yellow points show the HARPS data, the green points show the COR-07 data, and the cyan points show the COR-14 data.}
    \label{fig:HIP22059_orbits}
\end{figure*}

The best orbital fits, with 500 orbits randomly selected from the posterior distribution and a burn-in phase of 700 multiplied by every 250th step that is saved on the chain, are plotted alongside the RVs, and proper motions from \emph{Hipparcos} and \emph{Gaia} as shown in Fig.~\ref{fig:HD195010_orbits}. From this we derived a companion mass of $217.8^{+3.5}_{-3.4}~M_{\mathrm{Jup}}$ and an orbital period of $35.12^{+0.32}_{-0.31}$~years. We compare this to the value obtained in \citet{montagnier:tel-00714874}, who find a mass from combining the VLT/NACO imaging data with CORALIE radial velocities of $0.26\pm0.03~M_{\odot}$ or $272.4\pm31.4~M_{\rm{Jup}}$.

The dynamical mass that we determine for HD~195010~B is lower, but we now have far more significant constraints than the previously reported mass from \citet{montagnier:tel-00714874}. This is due to an increase in the baseline of CORALIE radial velocities of 16 years, which is a significant increase in orbital phase coverage with the radial-velocity data, especially throughout its periastron passage, which is key in fitting such a well-constrained orbit. Furthermore, our new data increases the baseline of the relative astrometry, with constraints from both the VLT/NACO and VLT/SPHERE data, as shown in Fig.~\ref{fig:HD195010_relative_astrometry}. In addition, the proper motion data from the HGCA helps place additional constraints on the orbital parameters, as shown in Fig.~\ref{fig:HD195010_orbits}.

We placed a log-flat prior on the primary mass when doing the orbital fit, leading to a completely model independent derived mass of HD~195010~A and HD~195010~B. Much like in the case for HD~92987~B, the system HD~195010 has a highly inclined orbit that is close to face-on, with an inclination of  $i=162.79\pm0.43$~degrees. This once again demonstrates the importance of obtaining dynamical masses by combining these mass measuring techniques to confirm the companion type of detected objects, which could indeed be stellar objects masquerading as brown dwarfs or exoplanetary candidates. From fitting the radial velocities alone, with no additional astrometric information, we obtain the minimum mass ($m\sin i$) of HD~195010~B to be $M_B\approx51~M_{\rm{Jup}}$. Without the additional relative astrometry or proper motion information, this would be interpreted as a brown dwarf, leading to a different understanding of this system and emphasizing the importance of deriving these dynamical masses.

The fitted orbital parameters of HD~195010~B are listed in Table~\ref{tab:orbital_parameters}. The resulting corner plots from the probed parameter space from the MCMC fit with \texttt{orvara} and the posterior distributions are shown in Fig.~\ref{fig:HD195010_corner} in Appendix~\ref{appendix}.

\subsection{HIP~22059}

HIP~22059 is a K5V star that has been observed with CORALIE at La Silla Observatory since January 2001. Twenty-nine measurements were taken with CORALIE-98, 9 additional RV measurements were obtained with CORALIE-07, and 30 additional RV measurements were obtained with CORALIE-14. HIP~22059 was also   observed with HARPS \citep{2003Msngr.114...20M} with 11 RV measurements from December 2003 to August 2007.

HIP~22059 was then observed with VLT/SPHERE program 0104.C-0724(A) (PI: Rickman) on 2019 November 27, with a total observation time of 1 hour 30 minutes, with a 32-second integration time on the coronagraphic frames using IRDIS, as well as several sky, center, and flux frames taken which was necessary for the post-processing of the data. The direct detection of HIP~22059~B is shown in Fig.~\ref{fig:direct_images}.

The orbital fit was run using \texttt{orvara} with 500,000 steps in each chain by combining CORALIE and HARPS radial velocities, with the relative astrometry outlined in Table~\ref{tab:astrometry}, and epoch astrometry from \emph{Hipparcos} and \emph{Gaia} eDR3. We imposed an uninformative log-flat prior on the primary and secondary mass and uniform priors on the remaining orbital parameters, meaning that the tight constraints on the primary mass come from the orbital fit alone, and leads to precise dynamical masses of the two components in the system. From this we determine the mass of the primary star to be $0.769^{+0.028}_{-0.027}~M_{\odot}$.

For the plots shown in Fig.~\ref{fig:HIP22059_orbits}, we implemented a burn-in phase of 700 multiplied by every 250th step that is saved on the chain, and randomly drew 500 orbits showing the posterior distribution. The highest likelihood fit is shown by the thick black line for the RVs, proper motions from \emph{Hipparcos} and \emph{Gaia}, as well as the relative astrometric orbit. From the orbital fit we find a very low-mass stellar companion with a companion mass of $248.7^{+4.1}_{-3.9}~M_{\mathrm{Jup}}$, with an orbital period of $93.1^{+5.9}_{-5.4}$~years.

The largest uncertainty in the mass is from only having partial orbital phase coverage from the radial-velocity data, as shown in the top left panel of Fig.~\ref{fig:HIP22059_orbits}. Full phase coverage would take more than 90~years to collect, which is more than a scientific lifetime. However, given the dispersion of the uncertainty in the orbit in just a few years from now, an additional relative astrometric point in five to ten years would tightly constrain the orbital parameters of this system.

The fitted orbital parameters are listed in Table~\ref{tab:orbital_parameters}. The resulting corner plots from the probed parameter space from the MCMC fit with \texttt{orvara} and the posterior distributions are shown in Fig.~\ref{fig:HIP22059_corner} in Appendix~\ref{appendix}.

\section{Dynamical vs. isochronal masses} \label{sec:masses}

\begin{figure}
    \centering
    \includegraphics[width=0.48\textwidth]{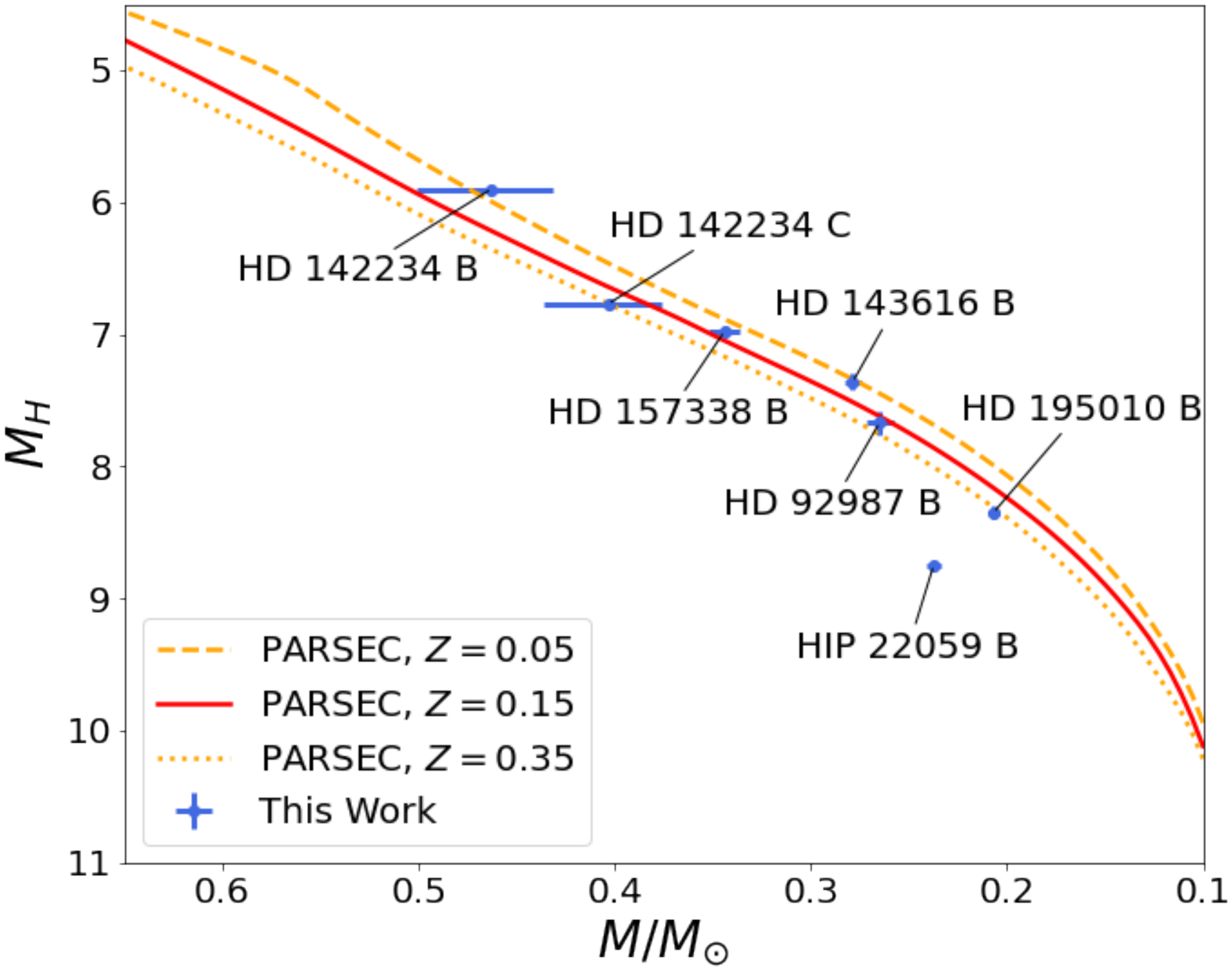}
    \caption{Comparison of the dynamical masses and the $H$-band magnitudes of the companions to theoretical mass-luminosity relations. We adopted the mass-luminosity relations from PARSEC models for low-mass stars \citep{2012MNRAS.427..127B,10.1093/mnras/stu1605} at low ($Z=0.05$, dotted orange line), intermediate ($Z=0.15$, solid red line), and high ($Z=0.35$, dashed orange line) metallicities. Most of our companions (blue dots with error bars and labels) agree reasonably well with the theoretical mass-luminosity relations, except for HIP~22059~B, as explained in Section~\ref{sec:masses}.}
    \label{fig:MLR_comp}
\end{figure}

We compare the companions' dynamical masses to their luminosity masses inferred from stellar isochrones. We extracted the narrow-band $H2$ and $H3$ absolute magnitudes for each companion using the BT-NextGen models \citep{2012RSPTA.370.2765A}, and applying our measured contrast values listed in Table~\ref{tab:astrometry} and \emph{Gaia} eDR3 parallaxes listed in Table~\ref{tab:stellar_params}. We then convert the extracted $H2$ and $H3$ magnitudes into a $H_{\rm{2MASS}}$ magnitude for each companion, making use of the 2MASS filter system \citep{2006AJ....131.1163S}, to be able to directly compare with the mass-luminosity relation in the $H$ band shown in Fig.~\ref{fig:MLR_comp}. The results are summarized in Table \ref{tab:masses}. 

In Figure \ref{fig:MLR_comp}, we compare the masses and magnitudes of the companions to the mass-luminosity relation near the low-mass end of PARSEC isochrones \citep{2012MNRAS.427..127B,10.1093/mnras/stu1605}. We adopted three different isochrones at low ($Z=0.05$, dotted orange line), intermediate ($Z=0.15$, solid red line), and high ($Z=0.35$, dashed orange line) metallicities, accounting for variations in the mass-luminosity relation due to compositional differences. The effect of age on the mass-luminosity relation is negligible for low-mass stars. We take an age of 10 Gyr for all three isochrones. Most of our companions agree reasonably well with the theoretical mass-luminosity relations, except HIP~22059~B.

In the case of HIP~22059, which is under-luminous for its derived dynamical mass, this could be explained by being an unresolved binary. One or more unseen components would contribute up to $\sim0.07~M_{\odot}$ to the dynamical mass, but only negligibly to the flux that we measure. This is especially true if the unresolved components are brown dwarf companions, where the flux drops off dramatically with age, and would be difficult to detect such a low-mass component in the presence of a brighter component, as well as the flux from the primary star. Alternatively, we could be observing two equal-mass components, but with a blended flux. Given that the PSF of HIP~22059~B is unresolved, these two objects must have a projected separation of less than $\sim$40~mas, which is 1.2~AU at the distance of the host star. This is compatible with observed distributions of field M-dwarf binaries \citep{Shan2015,Winters2019}. In either case, taking into account the additional mass component would place HIP~22059 nicely on the mass-luminosity relation in Fig.~\ref{fig:MLR_comp}.

We also note that the large uncertainty in the derived dynamical masses of HD~142234~B and HD~142234~C are mostly due to  unconstrained orbits, as described in Section~\ref{sect:HD144234_orbit}, the result of  treating HD~142234~B and HD~142234~C as a combined mass, with their photocenter taken to be their center of mass. Furthermore, we placed an informed prior on the primary star for the orbital fit, and therefore these two comparisons with the mass-luminosity relation shown in Fig.~\ref{fig:MLR_comp} should be taken with caution.

\section{Discussion and conclusions}

\begin{sidewaystable*}
    \caption{MCMC Orbital posteriors for the orbital fits of each system using \texttt{orvara} \citep{2021AJ....162..186B}.}
    \centering
    \begin{tabular}{ccccccccc}
    \hline\hline
         Parameters & Units & HD~92987 & HD~142234 & HD~143616 & HD~157338 & HD~195010 & HIP~22059 \\
         \hline
         Companion mass $M_{\rm{comp}}$ & $M_{\rm{Jup}}$ & $276.6^{+17.8}_{-16.8}$ & $911.4^{+76.5}_{-63.9}$\tablefoottext{1} & $291.2^{+12.6}_{-11.5}$ & $359.3\pm10.5$ & $217.8^{+3.5}_{-3.4}$ & $248.7\pm^{+4.1}_{-3.9}$ \\
         Companion mass $M_{\rm{comp}}$ & $M_{\odot}$ & $0.26\pm0.02$ & ${0.87}_{-0.06}^{+0.07}$ \tablefoottext{1} & $0.28\pm0.01$ & $0.34\pm0.01$ & $0.21\pm0.003$ & $0.24\pm0.004$ \\
         Host-star mass $M_{\rm{host}}$ & 
         $M_{\odot}$ & ${1.17}_{-0.11}^{+0.12}$ & $0.83\pm0.02$ \tablefoottext{2} & $0.94\pm0.06$ & ${1.07}_{-0.15}^{+0.16}$ & $1.14\pm0.02$ & $0.77\pm0.03$ \\
         Parallax & mas & ${22.983}_{-0.024}^{+0.023}$ & $21.003\pm0.013$ & $23.805\pm0.010$ & ${30.2640}_{-0.0093}^{+0.0091}$ & ${19.8432}_{-0.0068}^{+0.0069}$ & $32.379\pm0.013$ \\
         Semimajor axis $a$ & AU & ${11.31}_{-0.36}^{+0.37}$ & ${50.8}_{-7.6}^{+12}$ & ${11.11}_{-0.25}^{+0.26}$ & ${28.2}_{-2.3}^{+3.6}$ & $235.0\pm1.5$ & ${20.58}_{-0.66}^{+0.71}$ \\
         Inclination $i$ & $\degree$ & ${175.755}_{-0.083}^{+0.082}$ & ${65.6}_{-14}^{+7.0}$ & $49.2\pm1.5$ & ${43.0}_{-1.4}^{+1.3}$ & $162.79\pm0.43$ & ${142.06}_{-0.35}^{+0.36}$ \\
         Orbital Period $P$ & years & ${31.76}_{-0.48}^{+0.54}$ & ${279}_{-62}^{+111}$ & ${33.52}_{-0.70}^{+0.85}$ & ${126}_{-21}^{+34}$ & ${35.12}_{-0.31}^{+0.32}$ & ${93.1}_{-5.4}^{+5.9}$ \\
         Semimajor axis & mas & ${259.9}_{-8.4}^{+8.5}$ & ${1068}_{-160}^{+262}$ & ${264.6}_{-5.9}^{+6.2}$ & ${852}_{-70}^{+108}$ & $235.0\pm1.5$ & ${666}_{-21}^{+23}$ \\
         Eccentricity $e$ & & ${0.276}_{-0.011}^{+0.012}$ & ${0.73}_{-0.35}^{+0.18}$ & ${0.96549}_{-0.00050}^{+0.00058}$ & ${0.664}_{-0.021}^{+0.031}$ & $0.3341\pm0.0087$ & $0.368\pm0.026$ \\
         Jitter $\sigma$ & & ${4.37}_{-0.45}^{+0.49}$ & ${9.0}_{-1.3}^{+1.5}$ & ${5.61}_{-0.79}^{+0.87}$ & ${4.11}_{-0.35}^{+0.38}$ & ${6.96}_{-0.79}^{+0.86}$ & ${3.6}_{-1.0}^{+1.2}$ \\
         $\sqrt{e}\sin\omega$ & & ${-0.033}_{-0.016}^{+0.017}$ & $-0.12\pm0.35$ & ${-0.03574}_{-0.00050}^{+0.00051}$ & ${0.311}_{-0.065}^{+0.071}$ & $0.178\pm0.012$ & $-0.182\pm0.010$ \\
         $\sqrt{e}\cos\omega$ & & ${-0.525}_{-0.012}^{+0.011}$ & ${0.78}_{-0.35}^{+0.10}$ & ${-0.98195}_{-0.00030}^{+0.00026}$ & ${-0.753}_{-0.012}^{+0.015}$ & ${0.5499}_{-0.010}^{+0.0094}$ & ${0.578}_{-0.025}^{+0.023}$ \\
         PA of the ascending node $\Omega$ & $\degree$ & ${75.6}_{-1.2}^{+1.3}$ & ${93.3}_{-6.7}^{+19}$ & ${215.47}_{-0.47}^{+0.46}$ & ${33.9}_{-1.3}^{+1.2}$ & ${281.39}_{-0.82}^{+0.81}$ & ${12.29}_{-0.51}^{+0.50}$ \\
         Time of Periastron $T_0$ & JD & ${2457669}_{-51}^{+50}$ & ${2537894}_{-23911}^{+43441}$ & $2458936.59\pm0.13$ & ${2461296}_{-43}^{+36}$ & ${2463838}_{-102}^{+105}$ & ${2488846}_{-2021}^{+2198}$ \\
         Argument of periastron $\omega$ & $\degree$ & $183.5\pm1.8$ & ${325}_{-309}^{+25}$ & ${182.084}_{-0.030}^{+0.029}$ & ${157.6}_{-4.9}^{+4.6}$ & $17.9\pm1.3$ & ${342.5}_{-1.4}^{+1.3}$ \\
         Mass ratio $q$ & $M_{\rm{comp}}/M_{\rm{host}}$ & ${0.226}_{-0.009}^{+0.010}$ & ${1.049}_{-0.079}^{+0.091}$ \tablefoottext{3} & ${0.295}_{-0.011}^{+0.013}$ & ${0.320}_{-0.035}^{+0.044}$ & ${0.1827}_{-0.0027}^{+0.0028}$ & $0.309\pm-0.007$ \\
         \hline \\
    \end{tabular}
    \tablefoot{\tablefoottext{1}{This is the total mass of HD~142234~B+C.} \\
    \tablefoottext{2}{This is the Gaussian prior implemented on the orbital fit taken from Table~\ref{tab:stellar_params}.} \\
    \tablefoottext{3}{This is the mass ratio of the primary to the total combined mass of the companions: $\frac{M_{\rm{B}} + M_{\rm{C}}}{M_{\rm{A}}}$}.}
    \label{tab:orbital_parameters}
\end{sidewaystable*}

\begin{table*}[t]
\centering
\caption{Contrast and dynamical mass values for each stellar companion.}
\begin{tabular}{ccccccc}
\hline
\hline
Companion & Contrast ($H2$) & Contrast ($H3$) & App Mag ($H_{\rm{2MASS}}$) & Abs Mag ($H_{\rm{2MASS}}$) & $M_{\rm{dyn}}$ ($M_\odot$) & $M_{\rm{dyn}}$ ($M_{\rm{Jup}}$) \\
\hline

HD~92987~B & $5.26 \pm 0.10$ & $5.21 \pm 0.12$ & $10.87 \pm 0.08$ & $7.67 \pm 0.09$ & $0.26\pm0.02$ & $276.6^{+17.8}_{-16.8}$ \\

HD~142234~B & $2.30 \pm 0.01$ & $2.22 \pm 0.01$ & $9.28 \pm 0.02$ & $5.90 \pm 0.02$ & $0.46\pm0.02$ & $487.0^{+40.9}_{-34.1}$ \\

HD~142234~C & $3.17 \pm 0.01$ & $3.09 \pm 0.01$ & $10.15 \pm 0.02$ & $6.77 \pm 0.02$ & $0.41\pm0.03$ & $424.4^{+35.6}_{-29.8}$ \\

HD~143616~B & $3.87 \pm 0.09$ & $3.78 \pm 0.08$ & $10.48 \pm 0.06$ & $7.36 \pm 0.06$ & $0.28\pm0.01$ & $291.2^{+12.6}_{-11.5}$ \\

HD~157338~B & $3.93 \pm 0.02$ \tablefoottext{a} & $3.84 \pm 0.03$ \tablefoottext{b} & $9.57 \pm 0.03$ & $6.98 \pm 0.03$ & $0.34\pm0.01$ & $359.3\pm10.5$ \\
 
HD~195010~B & $4.76 \pm 0.08$ \tablefoottext{a} & $4.73 \pm 0.07$ \tablefoottext{b} & $11.86 \pm 0.06$ & $8.35 \pm 0.05$ & $0.21\pm0.003$ & $217.8^{+3.5}_{-3.4}$ \\

HIP~22059~B & $4.07 \pm 0.02$ & $3.99 \pm 0.02$ & $11.19 \pm 0.04$ & $8.75 \pm 0.03$ & $0.24\pm0.004$ & $248.7^{+4.1}_{-3.9}$ \\

\hline
\end{tabular}
\tablefoot{Contrast values measured in the $H2$ and $H3$ bands as described in Section~\ref{sec:astro_photo}, along with calculated apparent and absolute magnitudes in the $H_{\rm{2MASS}}$ band using a combination of the extracted values from the $H2$ and $H3$ contrasts (where the full contrast values for $H2$ and $H3$ are listed in Table~\ref{tab:astrometry}), and each companion's corresponding dynamical mass value ($M_{\rm{dyn}}$).
\tablefoottext{a}{Average contrast value from the two VLT/SPHERE $H2$ 2021 observations.}
\tablefoottext{b}{Average contrast value from the two VLT/SPHERE $H3$ 2021 observations.}}
\label{tab:masses}
\end{table*}

In this paper we present new direct detections of a number of low-mass stellar companions, obtained by combining radial velocities, relative astrometry, and astrometric accelerations to derive precise dynamical masses of these companions. The dynamical masses of stellar companions is fundamental for a number of different reasons:

\begin{itemize}
    \item From an exoplanetary and brown dwarf perspective, it is important to understand the true nature of these massive companions that are detected from their radial velocities with companions that are masquerading as much less massive components, ultimately leading to a different scientific interpretation of these detections.
    \item It is vital for future direct imaging surveys that such systems are thoroughly investigated for such low-mass stellar companions in order to make sure that targeted high-contrast imaging surveys are probing the most promising candidates for exoplanet and brown dwarf studies. This is especially important with expensive telescope facilities, like the \emph{James Webb Space Telescope} (\emph{JWST}), and the upcoming Extremely Large Telescopes (ELTs), where the nature of the detection needs to be  understood better before spending extensive observing time for a comprehensive in-depth characterization, leading to far more efficient direct imaging surveys.
    \item The detection and characterization of low-mass stellar companions is vital in probing and populating stellar mass-luminosity relations, as described in Section~\ref{sec:masses}.
\end{itemize}

As we continue to measure radial velocities and astrometry of our nearby stellar neighbors, it will become increasingly important that we understand the nature of the companions that we are investigating, which all have important astrophysical aspects to learn from, but can be scientifically interpreted very differently if their masses are unknown. By combining these techniques as described in this paper, we are able to detect companions by probing  the giant planet to brown dwarf to low-mass stellar regime, and ultimately it will help us define the blurred boundaries between these different types of companions.

In conclusion, we report the discovery of new directly imaged companions orbiting HD~142234, HD~143616 and HIP~22059. Furthermore, we report the first direct detection of HD~92987~B, and update the orbital parameters and dynamical masses of two previously reported companions, HD~157338~B and HD~195010~B.

For each system we perform an orbital fit using \texttt{orvara} \citep{2021AJ....162..186B} by combining long baseline radial-velocity data from the CORALIE survey for extra-solar planets \citep{2000fepc.conf..571U}, along with long baseline proper motion anomalies from \emph{Hipparcos} data \citep{1997ESASP1200.....E,2007A&A...474..653V,Hipparcos..JAVA} and \emph{Gaia} eDR3 \citep{2021A&A...649A...1G} reported in the Hipparcos-Gaia catalog of accelerations \citep[HGCA;][]{2021ApJS..254...42B}, with high-contrast images taken with VLT/SPHERE, to reveal the orbital properties and precise dynamical masses of each companion. The summary of each system is presented below:

\begin{itemize}
    \item \textbf{HD92987:} We present the first direct detection of HD~92987~B from high-contrast imaging with VLT/SPHERE. By doing a combined orbital fit of CORALIE radial velocities, relative astrometry from direct imaging, and astrometric accelerations, we confirm that HD~92987~B is a low-mass star with a dynamical mass of $276.6^{+17.8}_{-16.8}~M_{\rm{Jup}}$ on a $31.76^{+0.54}_{-0.48}$~year orbital period. \\
    \item \textbf{HD~142234:} We present the CORALIE radial-velocity and direct imaging observations of HD~142234 and discovered that it is a triple stellar system, where HD~142234~B and HD~142234~C are likely to be in a tightly bound orbit, orbiting around the primary star HD~142234~A. For this reason, we performed an orbital fit with the photocenter of the relative astrometry between the two components treated as the barycenter to estimate the combined mass of HD~142234~B+C, and then used the ratio of the contrast values to estimate the ratio of the mass of the individual components. From this we find the mass of HD~142234~B and HD~142234~C to be $487.0^{+40.9}_{-34.1}~M_{\rm{Jup}}$ and $436.2^{+35.0}_{-30.5}~M_{\rm{Jup}}$, respectively.
    \\
    \item \textbf{HD~143616:} We present the CORALIE radial-velocity data for HD~143616, as well as direct imaging observations from VLT/SPHERE and discover a very low-mass star, with a dynamical mass of $291.2^{+12.6}_{-11.5}~M_{\rm{Jup}}$. From an orbital fit combining the radial velocities, relative astrometry, and astrometric accelerations, we find this system to be highly eccentric, with an extreme eccentricity of $e=0.9655^{+0.00058}_{-0.00050}$ and an orbital period of $33.52^{+0.85}_{-0.70}$~years.
    \\
    \item \textbf{HD~157338:} We present new direct images of HD~157338~B with VLT/SPHERE. In addition, we fold in relative astrometry from 2006 VLT/NACO observations reported by \citet{montagnier:tel-00714874}, combined with radial-velocity measurements from HARPS, HIRES, and CORALIE, along with absolute astrometry to determine the dynamical mass of HD~157338~B to be $359.3\pm10.5~M_{\rm{Jup}}$ with an orbital period of $126^{+34}_{-21}$~years.
    \\
    \item \textbf{HD~195010:} We present new direct images of HD~195010~B with VLT/SPHERE. In addition, we fold in relative astrometry from 2006 VLT/NACO observations reported by \citet{montagnier:tel-00714874}, combined with radial-velocity measurements from CORALIE, along with absolute astrometry to determine the dynamical mass of HD~195010~B to be $217.8^{+3.5}_{-3.4}~M_{\rm{Jup}}$ with an orbital period of $35.12^{+0.32}_{-0.31}$~years.
    \\
    \item \textbf{HIP~22059:} We present the new detection of HIP~22059~B with CORALIE and HARPS radial-velocity data, as well as high-contrast imaging from VLT/SPHERE and absolute astrometry from HGCA to determine the dynamical mass of HIP~22059~B to be $248.7^{+4.1}_{-3.9}~M_{\rm{Jup}}$ with an orbital period of $93.1^{+5.9}_{-5.4}$~years.
\end{itemize}

As we enter the next era of instrumentation in direct imaging exoplanets (i.e., JWST and the ELTs), it is crucial that we use the orbital information available to us through radial velocities and astrometry to thoroughly vet targets for low-mass stellar companions, and to make informed choices on target selection to prioritize the most appropriate targets for direct imaging substellar companions and exoplanets. The importance of this is demonstrated through the increasing number of studies that are using this valuable precursor information from RVs and/or astrometry to select informed targets for direct imaging searches, and combining such multiple techniques to improve orbital constraints on previously detected companions \citep[e.g.,][]{2020MNRAS.494.3481B,2021AJ....162..301B,2022arXiv220800334S,2022ApJ...934L..18K,2022arXiv220612266M,2022MNRAS.513.5588B}.

This work also serves as a cautionary tale to warn about the danger of  using a single detection technique to discover giant planets or low-mass brown dwarfs, which can turn out to be stellar companions, like   the cases presented here for HD~92987~B and HD~195010~B. While this contributes to our understanding of low-mass stars, the science case is fundamentally different. Ultimately, this work acts as a precursor observation program as we look toward the future of the direct imaging of exoplanets.

\begin{acknowledgements}
This work has been carried out within the framework of the NCCR PlanetS supported by the Swiss National Science Foundation under grants 51NF40$\_$182901 and 51NF40$\_$205606. The authors acknowledge the financial support of the SNSF.

This publications makes use of the The Data $\&$ Analysis Center for Exoplanets (DACE), which is a facility based at the University of Geneva (CH) dedicated to extrasolar planets data visualisation, exchange and analysis. DACE is a platform of the Swiss National Centre of Competence in Research (NCCR) PlanetS, federating the Swiss expertise in Exoplanet research. The DACE platform is available at \url{https://dace.unige.ch}.

This work has made use of data from the European Space Agency (ESA) mission Gaia (\url{https://www.cosmos.esa.int/gaia}), processed by the Gaia Data Processing and Analysis Consortium (DPAC, \url{https://www.cosmos.esa.int/web/gaia/dpac/consortium}). Funding for the DPAC has been provided by national institutions, in particular the institutions participating in the Gaia Multilateral Agreement.

This research made use of the SIMBAD database and the VizieR Catalogue access tool, both operated at the CDS, Strasbourg, France. The original descriptions of the SIMBAD and VizieR services were published in \citet{2000A&AS..143....9W} and \citet{2000A&AS..143...23O}.

This research has made use of NASA’s Astrophysics Data System Bibliographic Services.

This publication makes use of data products from the Two Micron All Sky Survey, which is a joint project of the University of Massachusetts and the Infrared Processing and Analysis Center/California Institute of Technology, funded by the National Aeronautics and Space Administration and the National Science Foundation.
\end{acknowledgements}

\bibliographystyle{aa} % style aa.bst
\bibliography{bib} % your references Yourfile.bib

\begin{appendix}

\section{DACE links} \label{appendix2}

The radial-velocity measurements and the additional data products discussed in this paper are available in electronic form on the Data Analysis Center for Exoplanets (DACE) web platform for each individual target:

\begin{itemize}
    \item HD~92987: \\
    \url{https://dace.unige.ch/radialVelocities/?pattern=HD92987}
    \item HD~142234: \\
    \url{https://dace.unige.ch/radialVelocities/?pattern=HD142234}
    \item HD~143616: \\
    \url{https://dace.unige.ch/radialVelocities/?pattern=HD143616}
    \item HD~157338: \\
    \url{https://dace.unige.ch/radialVelocities/?pattern=HD157338}
    \item HD~195010: \\
    \url{https://dace.unige.ch/radialVelocities/?pattern=HD195010}
    \item HIP~22059: \\
    \url{https://dace.unige.ch/radialVelocities/?pattern=HIP22059}
\end{itemize}

\section{Posterior distributions of the orbital fits} \label{appendix}

Here we show the corner plots for the posterior distributions of the orbital fits for each system fitted using \texttt{orvara} \citep{2021AJ....162..186B}. For each system the posterior distributions for the primary stellar mass ($M_{\odot}$), the companion mass ($M_{\rm{sec}}$), the semimajor axis ($a$), the eccentricity ($e$), and the orbital inclination ($i$) are shown. In the case of HD~142234 the posterior distributions are shown for HD~1422234~B and HD~142234~C fitted as one single companion, as described in Section~\ref{sec:orbital_solutions}.

\begin{figure*}[ht]
    \centering
    \includegraphics[width=0.8\textwidth]{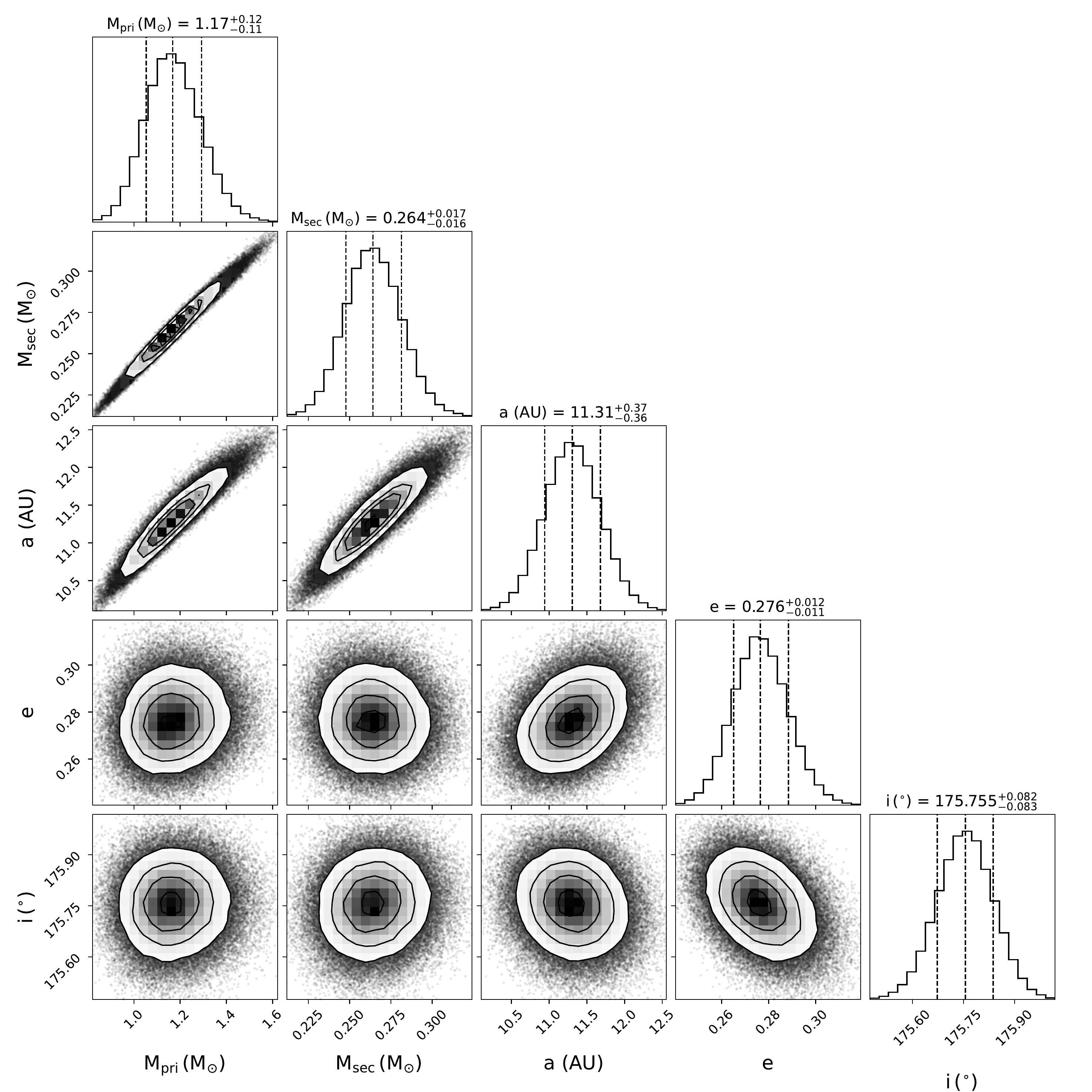}
    \caption{Marginalized 1D and 2D posterior distributions for selected orbital parameters of HD~92987~B corresponding to the fit of the RV, relative astrometry from direct imaging observations, and absolute astrometry from \emph{Hipparcos} and \emph{Gaia} with the use of \texttt{orvara} \citep{2021AJ....162..186B}. Confidence intervals at 15.85\%, 50.0\%, 84.15\% are overplotted on the 1D posterior distributions;   the median $\pm1\sigma$ values are given at the top of each 1D distribution. The  1, 2, and 3$\sigma$ contour levels are overplotted on the 2D posterior distribution.}
    \label{fig:HD92987_corner}
\end{figure*}

\begin{figure*}[ht]
    \centering
    \includegraphics[width=0.8\textwidth]{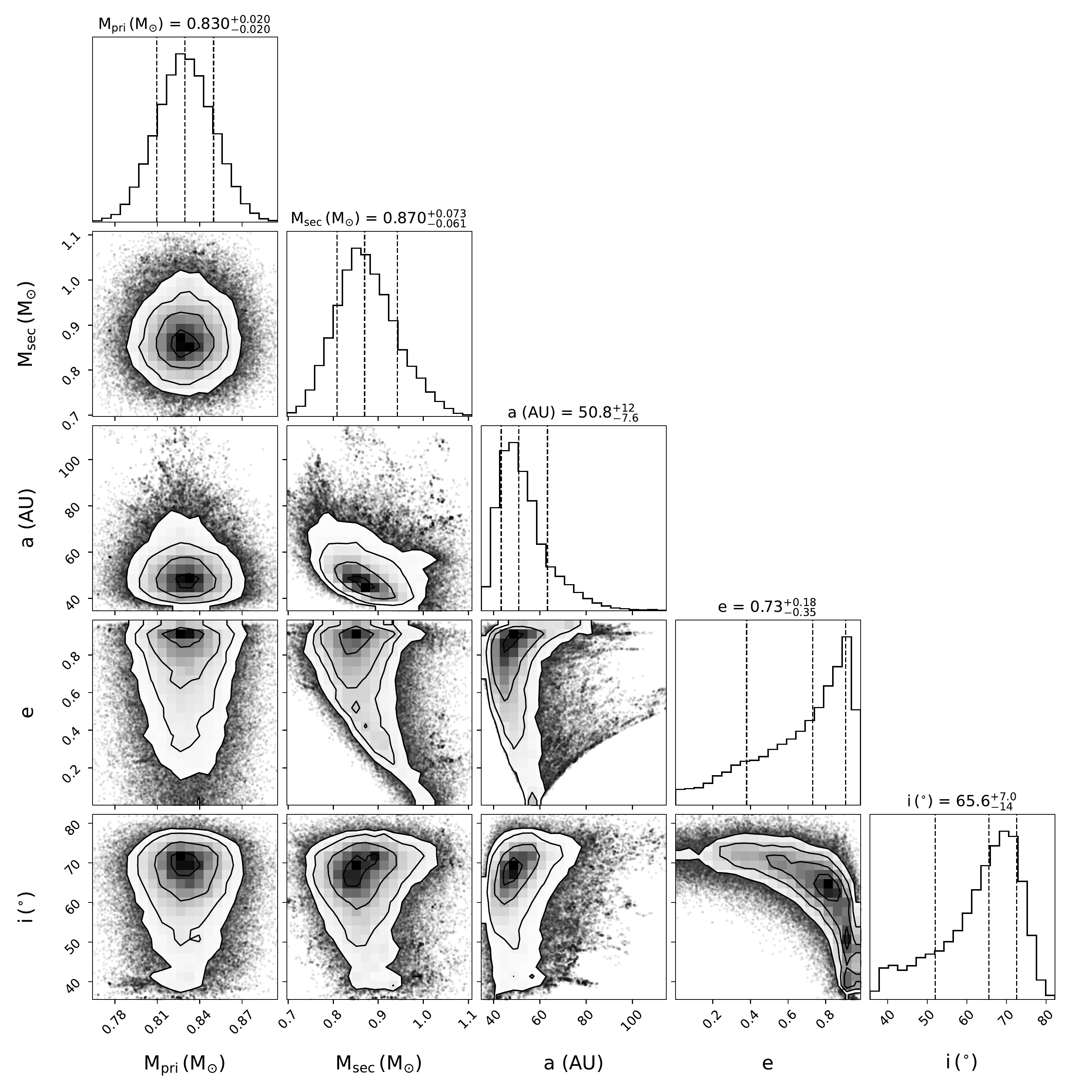}
    \caption{Marginalized 1D and 2D posterior distributions for selected orbital parameters of HD~142234~B and HD~142234~C combined and treated as a single companion. The posterior distributions correspond to the fit of the RV, relative astrometry from direct imaging observations, and absolute astrometry from \emph{Hipparcos} and \emph{Gaia} with the use of \texttt{orvara} \citep{2021AJ....162..186B}. Confidence intervals at 15.85\%, 50.0\%, 84.15\% are overplotted on the 1D posterior distributions;  the median $\pm1\sigma$ values are given at the top of each 1D distribution. Here 1, 2, and 3$\sigma$ contour levels are overplotted on the 2D posterior distribution.}
    \label{fig:HD142234_corner}
\end{figure*}

\begin{figure*}[ht]
    \centering
    \includegraphics[width=0.8\textwidth]{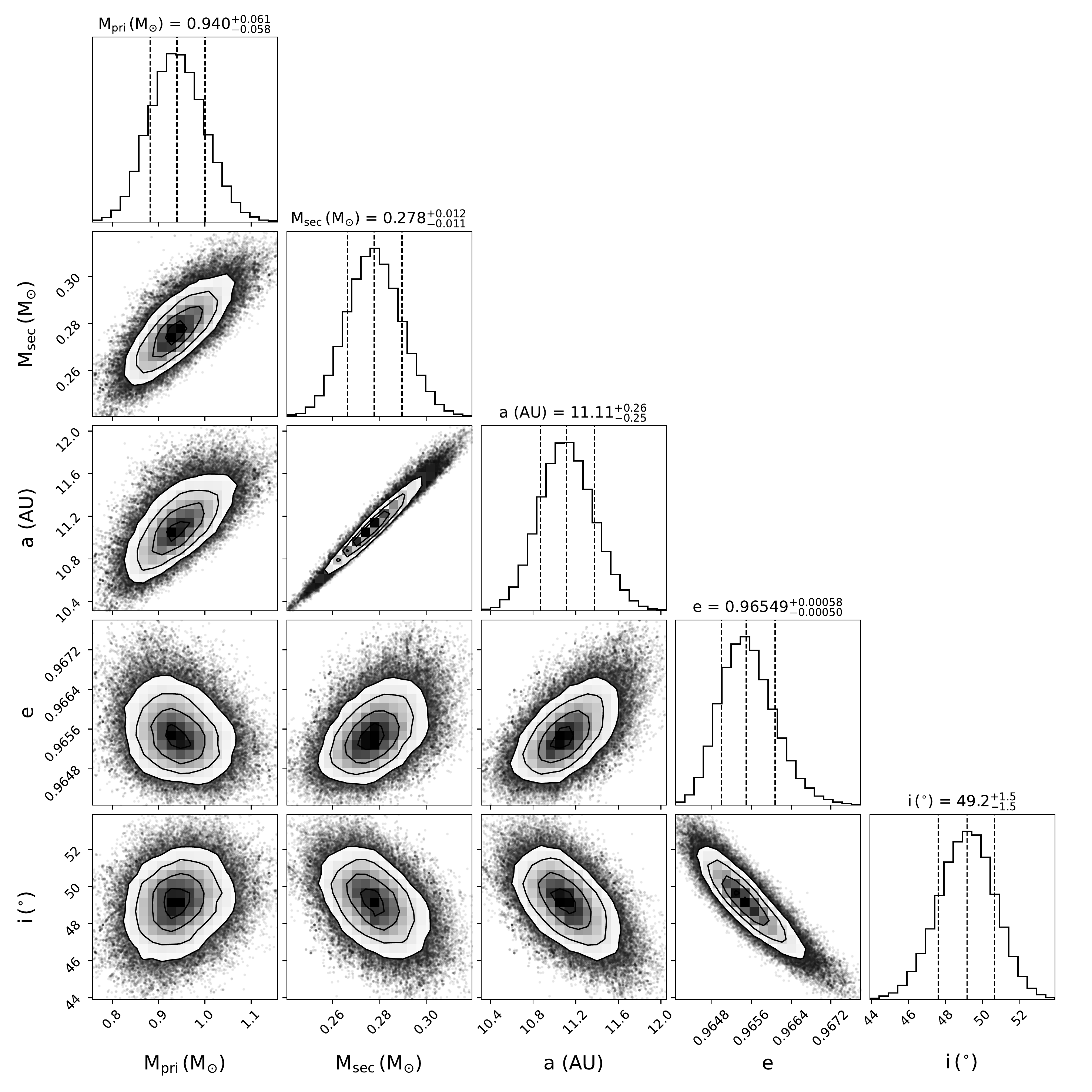}
    \caption{Marginalized 1D and 2D posterior distributions for selected orbital parameters of HD~143616~B corresponding to the fit of the RV, relative astrometry from direct imaging observations, and absolute astrometry from \emph{Hipparcos} and \emph{Gaia} with the use of \texttt{orvara} \citep{2021AJ....162..186B}. Confidence intervals at 15.85\%, 50.0\%, 84.15\% are overplotted on the 1D posterior distributions;  the median $\pm1\sigma$ values are given at the top of each 1D distribution. Here 1, 2, and 3$\sigma$ contour levels are overplotted on the 2D posterior distribution.}
    \label{fig:HD143616_corner}
\end{figure*}

\begin{figure*}[ht]
    \centering
    \includegraphics[width=0.8\textwidth]{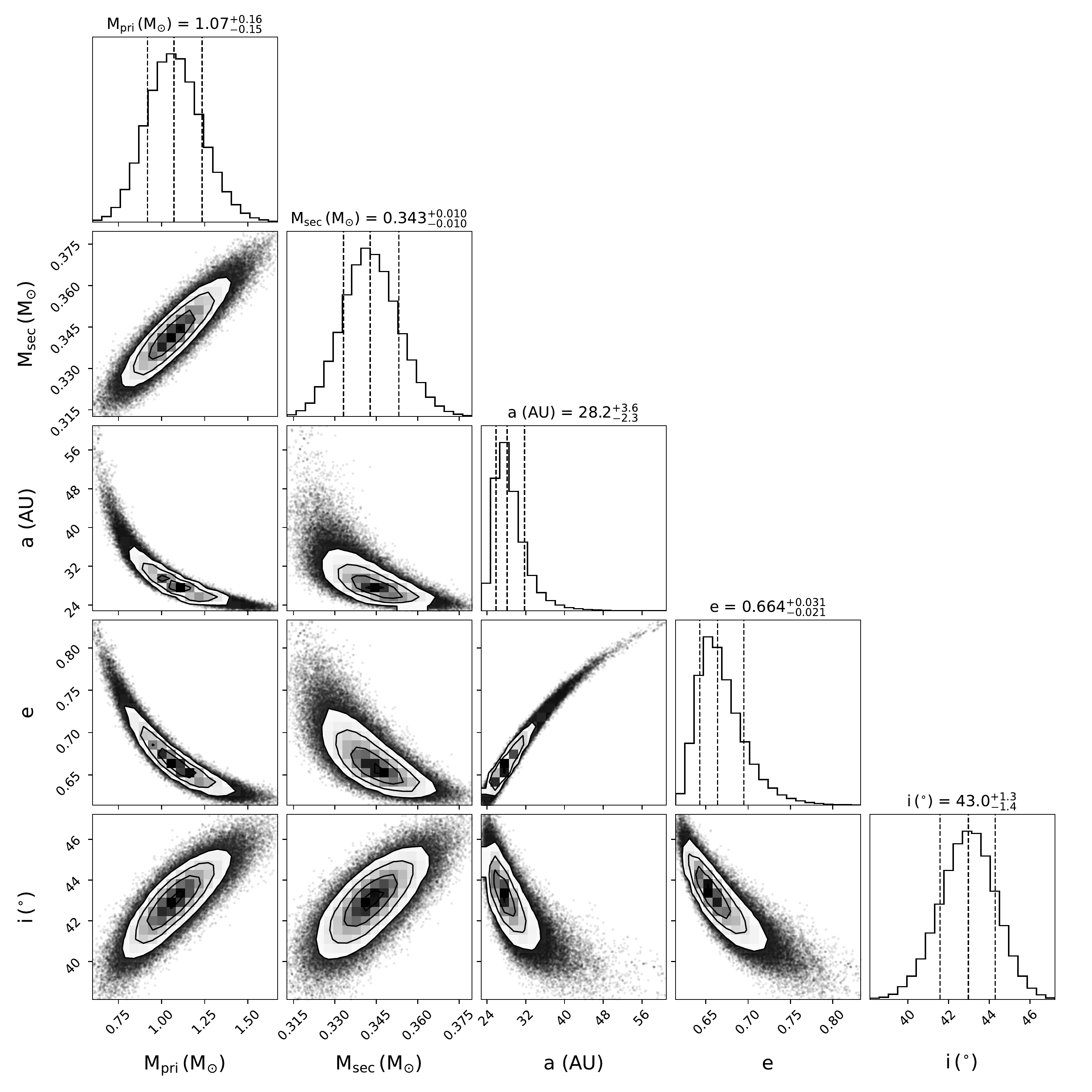}
    \caption{Marginalized 1D and 2D posterior distributions for selected orbital parameters of HD~157338~B corresponding to the fit of the RV, relative astrometry from direct imaging observations, and absolute astrometry from \emph{Hipparcos} and \emph{Gaia} with the use of \texttt{orvara} \citep{2021AJ....162..186B}. Confidence intervals at 15.85\%, 50.0\%, 84.15\% are overplotted on the 1D posterior distributions, with the median $\pm1\sigma$ values are given at the top of each 1D distribution. The 1, 2, and 3 $\sigma$ contour levels are overplotted on the 2D posterior distribution.}
    \label{fig:HD157338_corner}
\end{figure*}

\begin{figure*}[ht]
    \centering
    \includegraphics[width=0.8\textwidth]{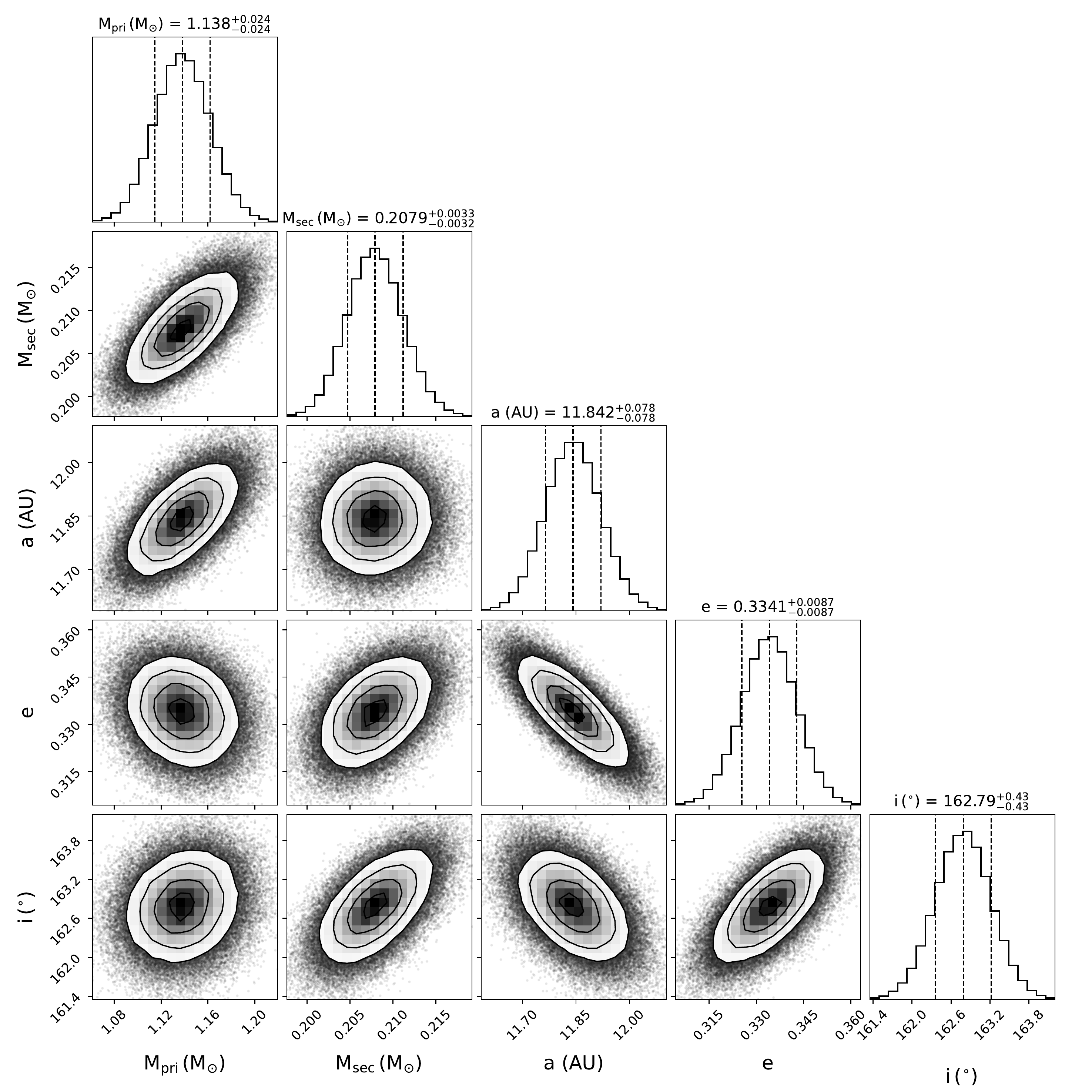}
    \caption{Marginalized 1D and 2D posterior distributions for selected orbital parameters of HD~195010~B corresponding to the fit of the RV, relative astrometry from direct imaging observations, and absolute astrometry from \emph{Hipparcos} and \emph{Gaia} with the use of \texttt{orvara} \citep{2021AJ....162..186B}. Confidence intervals at 15.85\%, 50.0\%, 84.15\% are overplotted on the 1D posterior distributions, with the median $\pm1\sigma$ values are given at the top of each 1D distribution.  The 1, 2, and 3 $\sigma$ contour levels are overplotted on the 2D posterior distribution.}
    \label{fig:HD195010_corner}
\end{figure*}

\begin{figure*}[ht]
    \centering
    \includegraphics[width=0.8\textwidth]{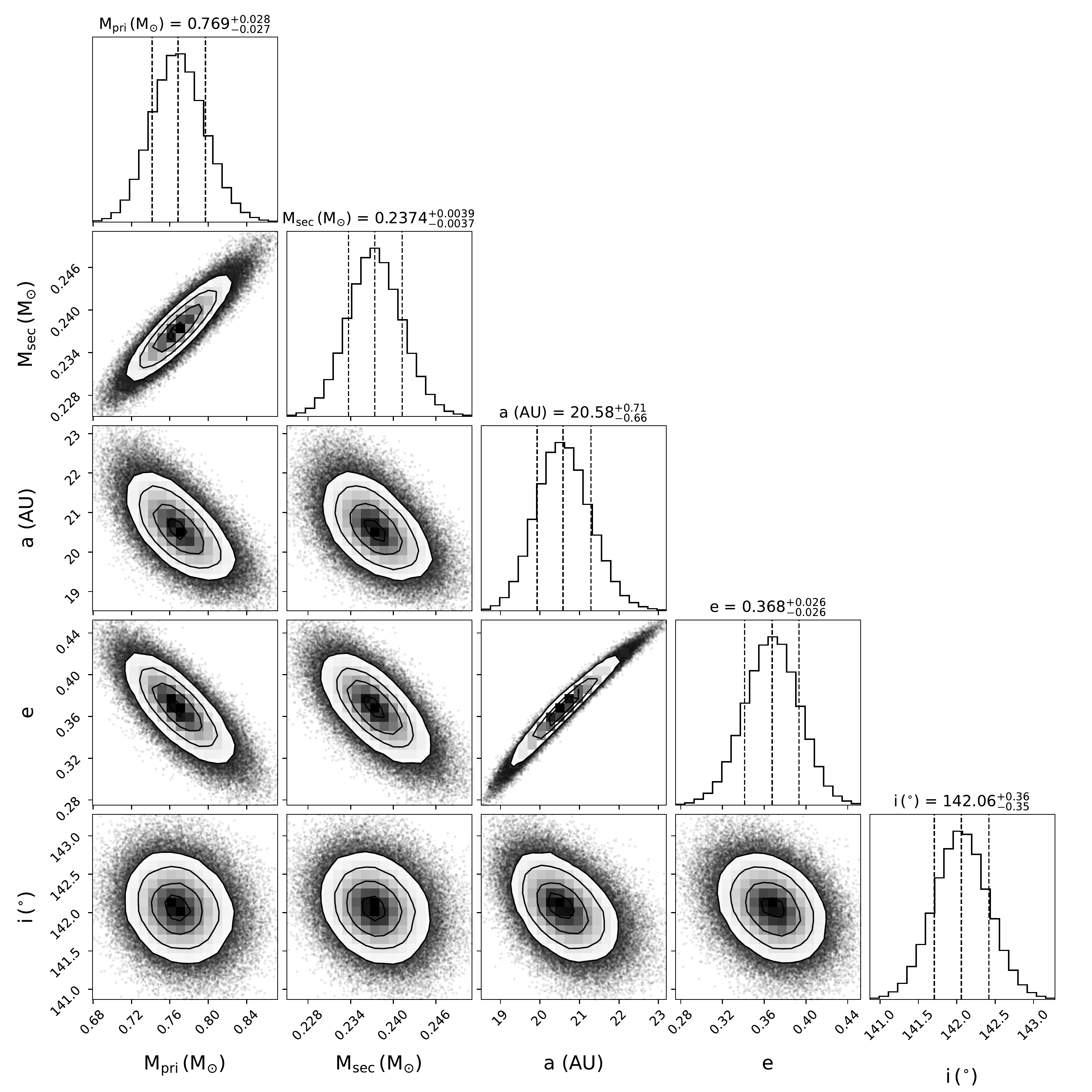}
    \caption{Marginalized 1D and 2D posterior distributions for selected orbital parameters of HIP~22059~B corresponding to the fit of the RV, relative astrometry from direct imaging observations, and absolute astrometry from \emph{Hipparcos} and \emph{Gaia} with the use of \texttt{orvara} \citep{2021AJ....162..186B}. Confidence intervals at 15.85\%, 50.0\%, 84.15\% are overplotted on the 1D posterior distributions;  the median $\pm1\sigma$ values are given at the top of each 1D distribution. The  1, 2, and 3 $\sigma$ contour levels are overplotted on the 2D posterior distribution.}
    \label{fig:HIP22059_corner}
\end{figure*}

\end{appendix}

\end{document}